\documentclass[useAMS,usenatbib]{mn2e}
\title[TNG Model]{Simulating Galaxy Formation with the IllustrisTNG Model}

\author[Pillepich et al.] {Annalisa Pillepich$^1$$^,$$^2$\thanks{E-mail: pillepich@mpia-hd.mpg.de},
Volker Springel$^3$$^,$$^4$,
Dylan Nelson$^5$,
Shy Genel$^6$$^,$$^7$,
\newauthor
Jill Naiman$^2$,
R{\"u}diger Pakmor$^3$,
Lars Hernquist$^2$,
Paul Torrey$^8$,
Mark Vogelsberger$^8$\thanks{Alfred P. Sloan Fellow},
\newauthor
Rainer Weinberger$^3$, and
Federico Marinacci$^8$
\vspace{2mm}
\\
$^1${Max-Planck-Institut f{\"u}r Astronomie, K{\"o}nigstuhl 17, 69117 Heidelberg, Germany}\\
$^2${Harvard--Smithsonian Center for Astrophysics, 60 Garden Street, Cambridge, MA 02138}\\
$^3${Heidelberg Institute for Theoretical Studies, Schloss-Wolfsbrunnenweg 35, D-69118 Heidelberg, Germany}\\
$^4${Zentrum f{\"u}r Astronomie der Universit{\"a}t Heidelberg, ARI, M{\"o}nchhofstr. 12-14, D-69120 Heidelberg, Germany}\\
$^5${Max-Planck-Institut f{\"u}r Astrophysik, Karl-Schwarzschild-Str. 1, 85741 Garching, Germany}\\
$^6${Center for Computational Astrophysics, Flatiron Institute, 162 Fifth Avenue, New York, NY 10010, USA}\\
$^7${Columbia Astrophysics Laboratory, Columbia University, 550 West 120th Street, New York, NY 10027, USA}\\
$^8${Department of Physics, Kavli Institute for Astrophysics and Space Research, Massachusetts Institute of Technology, Cambridge, MA 02139, USA}\\
}

\usepackage[bookmarks,bookmarksnumbered,colorlinks=true, citecolor=blue, linkcolor=cyan]{hyperref}
\usepackage{amsfonts}
\usepackage{amsmath}
\usepackage{amssymb}
\usepackage{color}
\usepackage[american]{babel}
\usepackage{verbatim}
\usepackage{times}
\usepackage{graphicx}
\usepackage{comment}
\usepackage{rotating}
\usepackage{tablefootnote}

\usepackage[normalem]{ulem}

\def \MSUN{\rm M_{\odot}}

\def \epsfr{$\epsilon_f \epsilon_r$}
\def \epsmr{$\epsilon_m \epsilon_r$}
\def \egyw{$\overline{\vphantom{t}e}_{w}~$}

\begin{document}

\maketitle
%%\vspace{-2cm}

\begin{abstract}
We introduce an updated physical model to simulate the formation and evolution of galaxies in cosmological, large-scale gravity+magnetohydrodynamical simulations with the moving mesh code {\sc Arepo}. The overall framework builds upon the successes of the Illustris galaxy formation model, and includes prescriptions for star formation, stellar evolution, chemical enrichment, primordial and metal-line cooling of the gas, stellar feedback with galactic outflows, and black hole formation, growth and multi-mode feedback. In this paper we give a comprehensive description of the physical and numerical advances which form the core of the IllustrisTNG (The Next Generation) framework. We focus on the revised implementation of the galactic winds, of which we modify the directionality, velocity, thermal content, and energy scalings, and explore its effects on the galaxy population. As described in earlier works, the model also includes a new black hole driven kinetic feedback at low accretion rates, magnetohydrodynamics, and improvements to the numerical scheme. Using a suite of (25 Mpc $h^{-1}$)$^3$ cosmological boxes we assess the outcome of the new model at our fiducial resolution. The presence of a self-consistently amplified magnetic field is shown to have an important impact on the stellar content of $10^{12} \,\MSUN$ haloes and above. Finally, we demonstrate that the new galactic winds promise to solve key problems identified in Illustris in matching observational constraints and affecting the stellar content and sizes of the low mass end of the galaxy population.
\end{abstract}


%-------------------------------------------------------------------------------------------------------------------------------------------------------------
% Section 1
%-------------------------------------------------------------------------------------------------------------------------------------------------------------

\section{Introduction}
To model {\it ab initio} the variety of galaxies that are observed across cosmic time is one of the greatest -- and most distant -- challenges for theoretical astrophysics. Galaxies today range in mass from a few thousand to a few trillion solar masses \citep[e.g.][]{Baldry:2008, Baldry:2012, Bernardi:2013, DSouza:2015}, extend in size from a fraction to tens of kilo-parsecs \citep{Shen:2003,Baldry:2012}, and span a diverse morphological mix \citep[e.g.][]{Lintott:2008, Kelvin:2012, Kartaltepe:2015}. Galaxies can reside in heterogeneous environments -- in isolation, or as members of rich groups and clusters \citep[e.g.][]{Ferrarese:2012}. They are expected to be self-gravitating systems of stars and gas embedded in halos of dark matter \citep{Ostriker:1973, White:1978, Mo:1998}, and their distribution throughout space traces a cosmic web defined by filaments, nodes, sheets, and voids of matter \citep{Colless:1999, SDSS:2003, Massey:2007}. 

Within the current $\Lambda$Cold Dark Matter ($\Lambda$CDM) cosmological paradigm \citep{Hinshaw:2013,PlanckXIII:2015}, the highly clustered large-scale structure of the Universe today arose from 13.8 billion years of evolution, starting from a nearly-homogeneous distribution of matter in the early universe \citep{Smoot:1991}. This cosmic topology, emerging through the mass-dominant presence of cold dark matter (CDM), fuels individual galaxies with cosmic gas, imparts gravitational torques and forces, and leads to a bottom-up or ``hierarchical'' growth, with smaller objects collapsing earlier by gravitational-instability and then later merging to form progressively more massive systems. In turn, this cosmological environment and its hierarchical growth are intertwined with the astrophysical processes that define galaxies on scales of kilo-parsecs and below. These include the formation of the dense molecular clouds and stars \citep{Heyer:2015}; the evolution of stellar populations, their winds and energetic feedback \citep{Smith:2014}; the explosions of supernovae \citep{Nomoto:2013}; the formation of supermassive black holes and the interaction of matter and radiation at their accretion disks \citep{Yuan:2014, King:2015}; cooling, heating, radiation, and other microphysical processes in the turbulent, magnetized, multi-phase interstellar medium \citep{McKee:2007, Ostriker:2009, Kennicutt:2012,Naab:2016}; and the driving of galactic outflows of gas, energy, and heavy elements into the circum- and inter-galactic media \citep{Putnam:2012}.

The complex interplay among this diverse ensemble of physical processes, acting across a wide range of spatial and time scales, ultimately governs the formation and transformation of galaxies through cosmic time, regulating their stellar content, star formation activity, gas content, heavy-element composition, morphological structure, and impact on their surroundings. And yet, thus far, advances in numerical astrophysics have proceeded on two parallel tracks: either through targeted modeling of specific phenomena on the scale of a single galaxy or smaller; or by attempting to reproduce the integral properties of entire populations of galaxies by including a large portion of the aforementioned processes within the large-scale structure arising from the now well-constrained initial conditions of the Universe \citep[e.g.][]{Somerville:2014}. Our present efforts follow the latter, {\it top-down} approach. This idea of simulating large cosmological volumes has been pioneered by gravity-only cosmological simulations, which are now a well-understood tool undertaken at massive scale \citep[e.g.][]{Springel:2005,Kuhlen:2012, Kim:2014,Skillman:2014} and coupled with ever-more comprehensive semi-analytical models for galaxy formation (from \citealt{Kauffmann:1993, Springel:2001} to \citealt{Benson:2012,Henriques:2015,Lacey:2016,Croton:2016}).

In the past few years, projects such as Eagle \citep{Schaye:2015}, Horizon-AGN \citep{Dubois:2014}, MassiveBlack-II \citep{Khandai:2015}, and the Illustris simulations \citep{Vogelsberger:2014a,Vogelsberger:2014b,Genel:2014,Sijacki:2015} have demonstrated that hydrodynamical simulations of structure formation at kilo-parsec spatial resolution can reasonably reproduce the basic structural properties and scaling relations of observed galaxies. Together with simulations of individual galaxies at even higher resolutions \citep[e.g.][]{Guedes:2011,Stinson:2013,Hopkins:2014,Wang:2015,Grand:2017,Agertz:2016,Wetzel:2016}, these calculations have provided new insights into the processes underpinning galaxy formation. They open new areas for exploration which are otherwise inaccessible to simpler modeling techniques. At the same time, they have highlighted the need for ongoing improvements in both the numerics and physical models required for cosmological hydrodynamical simulations.

In this paper, we present an updated physical model for simulating the formation and evolution of galaxies in cosmological, large-scale gravity + magnetohydrodynamical simulations with the moving mesh code {\sc Arepo} \citep{Springel:2010}. The overarching philosophy of the current efforts follows directly from the successful Illustris project (www.illustris-project.org). We seek an {\it effective} theory of galaxy formation, which will then allow us to utilize cosmological simulations as laboratories to explore a large variety of astrophysical questions. In practice, we search for an ensemble of numerically approximated physical processes which collectively produce a galaxy population in good agreement with a selected set of observational constraints (e.g. the cosmic star formation rate density and/or the stellar mass content of galaxies at $z=0$). We impose this requirement by simulating thousands or tens of thousands of galaxies across wide ranges of mass, redshift, assembly history, and environment. To do so demands cosmological volumes with at least tens of Mpc per side, as well as the necessity of a subgrid treatment for phenomena acting on scales smaller than $10^2-10^3$ parsecs. Properties of the simulation which have not been directly used to calibrate against observations, and which are not comprised by any simplifications required of the subgrid modeling, are then predictive in nature.

The motivations for an extension of the Illustris galaxy formation model are multifold. 1) We want to bring together under a consistent framework a series of improvements to the numerical techniques implemented in the {\sc Arepo} code, particularly to the hydrodynamical methods. 2) We want to include new physics in large-scale cosmological simulations for galaxy formation, particularly the amplification and evolution of a seed magnetic field \citep{Dolag:2009b, Pakmor:2011, Hopkins:2016, Mocz:2016}
%\citep{Pakmor:2011,Pakmor:2013,Pakmor:2014,Marinacci:2015,Marinacci:2016} 
and the effects of $\gtrsim$ kpc-scale gas outflows triggered by supermassive black holes at the center of galaxies \citep{Dubois:2010,Dubois:2012,Choi:2012,Choi:2014,Choi:2015,Weinberger:2017}. 3) We want to address the most important tensions between the outcome of the Illustris simulation and observational constraints. 

In fact, Illustris has demonstrated good agreement with a broad number of observational scaling relations and galaxy properties at low and intermediate redshifts \citep{Vogelsberger:2014a,Sijacki:2015,Torrey:2015,Snyder:2015,Sales:2015,Genel:2014}. However, important shortcomings have also been identified. These include:
1) a too mild decline in the cosmic star formation rate density at $z\lesssim 1$;
2) a too high stellar mass function at $z<1$ both at the high ($\gtrsim 10^{11.5} \MSUN$) and the low ($\lesssim 10^{10} \MSUN$) mass end with correspondingly too high stellar mass fractions in these mass ranges;
3) a possibly excessive number of galaxies with blue star forming rings;
4) an underestimated total gas fraction within $R_{500c}$ in halos with $M_{500c} \sim  10^{13-14} \MSUN$, and consequently too low bolometric X-ray luminosities in the hot coronae of elliptical galaxies and Sunyaev-Zel'dovich signals from Illustris clusters;
5) too large stellar sizes of galaxies (a factor of a few larger than observed for $M_{\rm stars}\lesssim 10^{10.7} \MSUN$);
and 6) galaxy color distributions without a strong bimodality between red and blue galaxies as observed \citep[see][for a full compendium]{Nelson:2015b}.

In a previous companion paper \citep{Weinberger:2017} we have shown that a new implementation of feedback from black holes (BHs), most importantly a change of feedback injection at low accretion rates in the form of a kinetic, super massive BH driven wind, promises to alleviate many of the discrepancies identified in comparison to observational data at the massive end of the halo mass function ($\gtrsim 10^{13}-10^{14} \MSUN$). 
In this paper, we focus on intermediate and low- mass galaxies (halos of a few $10^{10} - 10^{13}\MSUN$), giving a comprehensive description of all updates and additions since Illustris that constitute the The Next Generation Illustris (hereafter, IllustrisTNG or simply TNG) model. In Section~\ref{sec:tng} we begin with the overall approach of our galaxy physics model (Section~\ref{sec:tng_overview}). We give an overview of the advances in the numerical techniques (Section~\ref{sec:tng_numerics}), and  describe the modifications to the galactic wind feedback and to the stellar evolution and chemical enrichment schemes (Section~\ref{sec:tng_gfm}). In Section~\ref{sec:sims} and in Appendices \ref{sec_appendix1} and \ref{sec_appendix2_winds}, we run a large suite of 25 Mpc $h^{-1}$ cosmological volume simulations in order to discuss the properties of the galaxy population which result from the fiducial TNG model, the dependence of the model on its physical parameters, and the effects of resolution. We conclude and summarize in Section~\ref{sec:summary}. 

%-------------------------------------------------------------------------------------------------------------------------------------------------------------
% Section 2
%-------------------------------------------------------------------------------------------------------------------------------------------------------------

\section{The TNG Model}
\label{sec:tng}

\subsection{Overview}
\label{sec:tng_overview}

\subsubsection{Gravity and Hydrodynamics}

Our galaxy formation model is built upon the cosmological simulation code {\sc Arepo} \citep{Springel:2010}. {\sc Arepo} solves the coupled equations of ideal magneto-hydrodynamics and self-gravity. Gravitational forces are calculated with a Tree-Particle-Mesh (Tree-PM) scheme, while the solution of the magneto-hydrodynamical equations are obtained in a quasi-Lagrangian fashion. For cosmological simulations, gravity is treated in a fully Newtonian framework using periodic boundary conditions. The solution of the General Relativity equations (i.e. the Friedman-Lemaitre-Robertson-Walker equations with null curvature) determine the expansion (or contraction) of space as a function of cosmic time. Spatial quantities and coordinates are expressed in comoving units, where the mapping between the scale factor $a$ and cosmic time depends on the adopted cosmological parameter values.

The Tree-PM scheme is a synthesis of the particle-mesh method and the tree algorithm \citep{Xu:1995, Bode:2000, Bagla:2002}, where the gravitational potential is explicitly split in Fourier space into a long-range and a short-range part. The short-range forces are evaluated using a hierarchical multipole expansion based on an oct-tree \citep{Barnes&Hu:1986, Hernquist&Katz:1989}, modified by a short-range cut-off factor. Long range forces are calculated from the potential obtained with the Fast-Fourier-Transformation (FFT) mesh technique, using cloud-in-cell deposition to construct the mass density field on a uniform Cartesian grid.

For (magneto-)hydrodynamics (MHD), the code employs the finite volume method on an unstructured, moving, Voronoi tessellation of the simulation domain. The Voronoi cells track the conserved quantities of the fluid: mass, momentum, energy, and cell-averaged magnetic field in the case of MHD. They are evolved in time using Godunov's method and a directionally unsplit second order scheme that solves Riemann problems at the cell interfaces \citep{Springel:2010}. Gas cells evolve on a hierarchy of individual timesteps. A fundamental strength of the code is that the generating points of the Voronoi mesh are allowed to move in time, with a velocity close to the local fluid velocity field (corrected to maintain mesh regularity). The {\sc Arepo} code therefore implements an Arbitrary Lagrangian-Eulerian (ALE) scheme. The mesh has no preferred direction or Cartesian grid structure, with a naturally adaptive resolution in both space and time.

\subsubsection{The Model Foundation: the Illustris framework}

Given this numerical basis, an effective theory for galaxy formation and evolution further requires the modelling of several key astrophysical processes. These include star formation, stellar evolution, chemical enrichment, primordial and metal-line gas cooling, stellar feedback driven galactic outflows, and super massive black hole (SMBHs) formation, growth, and feedback. However, the spatial (and mass) resolution of cosmological large-scale simulations remains larger than the typical spatial (and mass) scales of the turbulent inter-stellar medium and, in particular, of the physical sites of star formation and black hole collapse. Therefore, these crucial baryonic processes are implemented in a `subgrid' manner. Inspired by the outcome of observations and small-scale theoretical models, these effective models for processes on unresolved scales are formulated and integrated with the resolved hydrodynamical and gravitational scales and operate within the evolving cosmological context.

The galaxy physics model we present in this paper builds directly upon the framework used for the Illustris simulation \citep{Vogelsberger:2014a,Vogelsberger:2014b, Genel:2014, Sijacki:2015}. Here we briefly summarize the general features of the Illustris setup, before describing the core modifications that define our new model. We refer the reader to the previous methods papers, \cite{Vogelsberger:2013b} and \cite{Torrey:2014a} for implementation details and explorations of the initial model. 

Owing to resolution limits in our simulations, star formation and pressurization of the multi-phase interstellar medium (ISM) are treated following the \cite{Springel:2003} model. Specifically, gas above a star formation density threshold of $n_{\rm H} \simeq 0.1$\,cm$^{-3}$ forms stars stochastically following the empirically defined Kennicutt-Schmidt relation and assuming a Chabrier \citep{Chabrier:2003} initial mass function. Pressurization from unresolved supernovae is included for star forming gas using a two-phase, effective equation of state model. Stellar populations evolve and return mass and metals to their ambient ISM via supernovae of Type Ia and Type II and asymptotic giant branch (AGB) stars according to tabulated mass and metal yields. Here, we do not include the ejection of winds from other stellar evolutionary phases as AGB winds dominate the mass loss. We follow the production and subsequent evolution of nine elements (H, He, C, N, O, Ne, Mg, Si, and Fe). 

Metal enriched gas radiatively cools in the presence of a redshift-dependent, spatially uniform, ionizing UV background radiation field, with corrections for self-shielding in the dense interstellar medium \citep[ISM][]{Katz:1992,Faucher-Giguere:2009}\footnote{We note that in all Illustris and TNG simulations the UV background is instantaneously switched on at $z=6$ instead of following the gradual build up at higher redshifts as prescribed by \cite{Faucher-Giguere:2009}.}. The cooling contribution from metal lines is included using pre-calculated values as a function of density, temperature, metallicity, and redshift \citep{Wiersma:2009a, Smith:2008}, without actually computing line-by-line cooling with individual cell abundance ratios. Cooling is further modulated by the radiation field of nearby AGN, by superimposing the UV background with the radiation field of AGNs, as described in \cite{Vogelsberger:2013b}.

SMBHs form in sufficiently massive haloes, accrete gas from surrounding gas, and inject feedback energy into their environment. In the Illustris model (but differently from the TNG model), for an accretion rate below 5 per cent of the Eddington rate, a radio-mode model injects highly bursty thermal energy into large, $\sim$\,50\,kpc `bubbles' which are displaced away from the central galaxy \citep{Sijacki:2007}. Above this accretion rate, a quasar-mode model injects thermal energy into the immediately surrounding gas, with a lower coupling efficiency and continuously in time \citep{Springel:2005b,DiMatteo:2005}.

Finally, feedback associated with star formation is assumed to drive galactic scale outflows. These outflows are launched directly from star-forming gas, with an assigned wind velocity that scales with the local dark matter velocity dispersion. The wind mass loading is determined from the available SN energy and desired wind speed, while the wind metal content is taken as a constant fraction of the ISM value. In practice, this model is implemented with a kinetic wind scheme where wind particles are stochastically spawned and hydrodynamically decoupled until they leave the local ISM. Outside of the dense ISM wind particles hydrodynamically recouple, allowing them to deposit their mass, momentum, metals, and thermal energy content.

%-------------------------------------------------------------------------------------------------------------------------------------------------------------

\subsection{Improvements to the Numerical Framework}
\label{sec:tng_numerics}

Several key numerical improvements that underpin the TNG model have been implemented in the {\sc Arepo} code \citep{Springel:2010} since the Illustris simulation. We briefly summarize them here.

\subsubsection{Magnetohydrodynamics}
\label{sec:mhd}

A notable addition in the TNG model is a the inclusion of magnetic fields. The TNG model employs cell-centered magnetic fields to solve the equations of ideal magneto-hydrodynamics \citep{Pakmor:2011} combined with an approximate HLLD (Harten-Lax-van Leer with contact and Alfv\'en mode) Riemann solver \citep{Miyoshi&Kusano:2005}. It uses the 8-wave Powell cleaning scheme to maintain the divergence constraint for the magnetic field \citep{Pakmor:2013} and introduces an additional timestep criterion that limits the size of the Powell source terms to increase the accuracy and robustness of the scheme. This approach has been tested and benchmarked on idealized test cases \citep{Pakmor:2013} as well as cosmological galaxy simulations \citep{Pakmor:2014} and cosmological volumes \citep{Marinacci:2015}. In simulations with magnetic fields, a small spatially homogeneous initial seed field is applied with a uniform field strength in an arbitrary orientation. Redshift zero results have been shown to be invariant to both the strength and direction of the seed field provided the seed is sufficiently small \citep[][and Section~\ref{sec:sims_variations}]{Pakmor:2014,Marinacci:2015}.

\subsubsection{Spatial Gradient Estimator}

Each Voronoi gas cell has a set of primitive variables, e.g. density, velocity, pressure, and magnetic field. By construction, the primitive variables are volume-averaged quantities: they can be calculated from the corresponding conserved quantities given the volume of the cell. In addition, the spatial gradients of the primitive variables are needed to linearly reconstruct their values at the cell faces in order to obtain a spatially second order scheme. However, the accuracy of the original Green-Gauss method to estimate gradients can decline for irregular geometries, i.e. whenever the center of mass of a cell is significantly offset from the position of its mesh-generating point. For this reason, the Green-Gauss method has been replaced with a least-squares fit (LSF) method \citep{Pakmor:2016a}. This LSF method improves the accuracy of estimated gradients and guarantees that the gradient estimate is always first order accurate, even for extreme mesh geometries.

\subsubsection{Time Integration}

In both the gravity and hydrodynamics calculations {\sc Arepo} employs a highly adaptive time integration scheme. All cells and particles are evolved on individual timesteps which are organized in a nested binary hierarchy of factors of two. The original \textsc{arepo} code, as well as the Illustris simulation employed a MUSCL-Hancock time integration scheme \citep{VanLeer:1977,Springel:2010}. However, this scheme does not take changes of the mesh geometry during a timestep fully into account and therefore is in general only first order accurate for moving meshes. Therefore, it has been replaced with an approach following Heun'€s method, a second-order Runge-Kutta scheme \citep{Pakmor:2016a}. Together with the improved gradient estimates, the new time integration scheme especially improves the conservation of angular momentum (although the latter is mainly relevant in idealized test problems rather than the significantly more complex cosmological simulations presented here). In doing so it avoids an expensive additional reconstruction of the Voronoi mesh at the mid-timestep point, which would otherwise be required to achieve second order convergence in time. The new scheme also retains individual time-stepping for all particles and cells.

Still, it is challenging to maintain high efficiency of the code during the smallest timesteps, when only a few particles or cells are active. In the context of the tree-gravity solver in {\sc Arepo}, this has motivated the development of a new method using a particle set decomposition based on a recursive splitting of the N-body Hamiltonian into short and long timescale systems \citep[inspired by][]{Pelupessy:2012}. Consequently, the forces required to integrate the rapidly evolving particles can be computed without constructing the full gravity tree, leading to significantly improved parallel efficiency. A full description of this scheme is outside the present scope and will be presented in future work (\textcolor{blue}{Springel et al. in prep}).

\subsubsection{Advection of Passive Scalars (Metal Abundances)}

Our model follows the total mass fraction of elements heavier than Helium in each gas cell and stellar particle -- the metallicity, $Z$. In the gas this field acts as a passive scalar which is advected between cells with the mass. However, this advection was treated entirely independently from the individual elemental abundances, leading to various inconsistencies as detailed in \citealt{Nelson:2015b} (Sec 6.1). Therefore, in addition to the evolution of the previous nine individual species (H, He, C, N, O, Ne, Mg, Si, and Fe) we now also follow a tenth pseudo `other metals' element which represents the sum of all metals that are {\it not} otherwise individually tracked. We use this field to guarantee that the sum of the mass fractions always adds up to unity for all flux exchanges and dynamically calculate the total metallicity from the mass fractions of all elements heavier than He. This is done by normalizing individually the extrapolated abundances at cell interfaces before they enter the Riemann solver \citep{Fryxell:1989}.

Similarly, we previously initialized only the hydrogen and helium mass fractions to a nonzero value, 0.76 and 0.24 respectively, at the start of the simulation. This choice, together with small numerical errors in individual element advection, led to a small fraction of gas cells and stars with unrealistically small or even negative metal abundance values. To avoid this, we now initialize the cells with a metallicity floor for all individual metals (and `other metals') at a mass fraction of $10^{-10}$. The galaxy population statistics studied in this paper (see next Sections) are robust to the precise value to which the metals are initialized.
 
\subsubsection{Resolution and Softening Choices}
\label{sec:resolution_choices}

In a hydrodynamical galaxy formation simulation such as Illustris we include several types of resolution elements: collisionless dark-matter and stellar particles, gas cells, black-hole sink particles, and short-lived wind particles. They all interact gravitationally, each softened independently on potentially different spatial scales. For dark matter, star, and wind particles we adopt a fixed comoving softening value until $z$\,=\,1, after which it is fixed to its physical redshift one value. We note that in Illustris, this did {\it not} apply to DM particles, such that at $z=0$ dark matter had twice the softening length as stars. For the comoving value we roughly take, in every simulation, $\epsilon_{\rm DM,stars} = L_{\rm box} / N_{\rm DM}^{1/3} / 40$. This is the smallest choice that avoids significant collisional effects in small dark matter halos and a spurious heating of the gas by dark matter particles. For the cosmological volumes presented here, these values are listed in Table \ref{tab:res}. The quoted values, $\epsilon$, are Plummer-equivalent spline softening, where the gravitational potential at zero lag for a test mass $m$ is $-G\times m / \epsilon$ and the gravitational forces become Newtonian at a separation  $2.8\times\epsilon$.

We treat Voronoi cells as point masses at the position of their centers of mass. They use an adaptive gravitational softening of $\epsilon_{\rm gas} = 2.5 r_{\rm cell}$ where the effective cell radius $r_{\rm cell}$ is derived from the total volume of the Voronoi cell approximating it as a sphere. We then enforce a minimum gas softening $\epsilon_{\rm gas}^{\rm min}$ that is 8 times smaller than the comoving value of the collisionless component (in Illustris this was taken as 2 times smaller than the stellar softening). For technical reasons, the adaptive gas softenings are chosen from a discrete spectrum increasing from this minimum by factors of 1.2 up to 64 total values. The softening for each black hole is chosen as $\epsilon_{\rm BH} = \epsilon_{\rm DM,stars} (m_{\rm BH}/m_{\rm DM})^{1/3}$, similarly restricted to a discrete list spaced by factors of 1.7 and increasing from the collisionless softening.
%\footnote{As we cannot in any case well resolve the regions where the gravity of SMBHs dominate over the background, choosing relatively large softening lengths for the BHs allows somewhat larger time steps and overall a better computational efficiency.}.

Besides the softenings, two additional numerical parameters may be scaled with resolution. 
The first is the black hole kernel-weighted neighbor number $n_{\rm ngb}$ of \cite{Weinberger:2017}, that defines the BH accretion and feedback region. We now scale this as $\propto m_{\rm baryon}^{-1/3}$, whereas in Illustris this was scaled as $\propto m_{\rm baryon}^{-1}$. 
The second is the stellar enrichment neighbor number $N_{\rm enrich}$ (see \citealt{Vogelsberger:2013b}) for the return of metals from the stars to the ISM. We now keep this fixed with resolution, whereas in Illustris this was also scaled as $\propto m_{\rm baryon}^{-1}$. The scaling choices of Illustris imply that the baryon mass - and so the approximate spatial extent - impacted by the associated physics is constant with resolution. The new choices are primarily made to avoid extremely low neighbor numbers for low resolution simulations, and we have verified that these scaling changes have no relevant impact on our results\footnote{We have verified that no galaxy integral properties, including all those explored herein, are affected in any measurable way by different values of enrichment neighbor number (32, 64, 128) at fixed resolution (L25n512).}.
All other parameters of the TNG model are kept constant with respect to changes in the numerical resolution.

%-------------------------------------------------------------------------------------------------------------------------------------------------------------

\begin{table*}
  \centering
  \begin{tabular}{r|c|ll} 
  \hline 
  Illustris  &  Model features  &  TNG  &  Technical Reference \\ 
  \hline\hline
  &&&\\[-1ex]
    
  & \textbf{MHD} & &\\
  &&&\\
   no & magnetohydrodynamics (MHD)   & yes: Powell $\nabla \cdot B$ cleaning     	& \citealt{Pakmor:2011} \\
   -  & Seed B field strength        & $1.6 \times 10^{-10}$ phys Gauss at $z=127$  & \citealt{Pakmor:2013} \\ % $10^{14}$ "comoving Gauss"
   -  & Seed B field configuration   & uniform in random direction               	& \citealt{Pakmor:2013} \\
   
  &&&\\[-1ex]
  \hline
  &&&\\[-1ex]
  
  & \textbf{BHs and BH Feedback} & &\\
  &&&\\
  $1\times10^{5} h^{-1} \MSUN$        & BH Seed Mass                 & $8\times10^{5} h^{-1} \MSUN$            & \citealt{Weinberger:2017} \\
  $5\times10^{10} h^{-1} \MSUN$       & FoF Halo Mass for BH seeding & $5\times10^{10} h^{-1} \MSUN$           & \citealt{Vogelsberger:2013b} \\
  $\alpha=100$ Boosted Bondi-Hoyle    & BH Accretion                 & Un-boosted Bondi-Hoyle (w/ $v_{\rm A}$) & \citealt{Weinberger:2017} \\
  parent gas cell, Eddington limited  & BH Accretion                 & nearby cells, Eddington limited         & \citealt{Weinberger:2017} \\
  fixed to local potential minimum     & BH Positioning               & fixed to local potential minimum         & \citealt{Vogelsberger:2013b} \\
   &&&\\
  Two: ``Quasar/Radio''            & BH Feedback Modes                      & Two: ``High/Low Accretion State''  & \citealt{Weinberger:2017} \\
  Thermal Injection around BHs 	   & High-Accr-Rate Feedback 	            & Thermal Injection around BHs       & \citealt{Weinberger:2017} \\
  Thermal `Bubbles' in the ICM     & Low-Accr-Rate Feedback                 & BH-driven kinetic wind             & \citealt{Weinberger:2017} \\
  constant: 0.05                    & Low/High Accretion Transition: $\chi$  & BH-mass dependent, $\le 0.1$       & \citealt{Weinberger:2017} \\
  0.2                              & Radiative efficiency: $\epsilon_r$     & 0.2                                & \citealt{Weinberger:2017} \\
  \epsfr, with $\epsilon_f=0.05$   & High-Accr-Rate Feedback Factor         & \epsfr, with $\epsilon_f=0.1$      & \citealt{Weinberger:2017} \\
  \epsmr, with $\epsilon_m=0.35$   & Low-Accr-Rate Feedback Factor          & $\epsilon_{\rm f, kin} \le 0.2$    & \citealt{Weinberger:2017} \\
  yes                              & Radiative BH Feedback                  & yes                                & \citealt{Vogelsberger:2013b} \\
 
  &&&\\[-1ex]
  \hline
  &&&\\[-1ex]
  
  & \textbf{Galactic Winds} & &\\
  &&&\\
  non local, from sf-ing gas        & General Approach        & non local, from sf-ing gas                           & \citealt{Vogelsberger:2013b} \\
  bipolar                           & Directionality          & isotropic                                            & this paper \\
  cold                              & Thermal Content         & warm                                                 & this paper \\
  $\propto$ local $\sigma_{\rm DM}$ & Injection Velocity      & $\propto$ local $\sigma_{\rm DM}$ with $H(z)$ scaling  & this paper \\
  -                                 & Injection Mass Loading  & gas-metallicity ($Z$) dependent                        & this paper \\
  &&&\\
  no   & Injection Velocity Floor                       & yes: 350 ${\rm km\,s^{-1}}$ & this paper \\
  3.7  & Wind Velocity Factor: $\kappa_w$               & 7.4           & this paper \\
  1.09 & Wind Energy Factor: \egyw                      & 3.6           & this paper \\
  -    & Thermal Fraction: $\tau_w$              & 0.1           & this paper \\
  -    & $Z$-dependence Reduction Factor: f$_{w, Z}$      & 0.25          & this paper \\
  -    & $Z$-dependence Reference Metallicity: $Z_{w, Z}$ & 0.002         & this paper \\
  -    & $Z$-dependence Reduction Power: $\gamma_{w, Z}$  & 2             & this paper \\
  0.4  & Metal loading of wind particles: $\gamma_w$    & 0.4           & \citealt{Vogelsberger:2013b} \\
  
  &&&\\[-1ex]
  \hline
  &&&\\[-1ex]
  
  & \textbf{Stellar Evolution} & &\\
  &&&\\
  \citealt{Chabrier:2003}     & IMF                           & \citealt{Chabrier:2003}    & \citealt{Vogelsberger:2013b} \\
  $[6, 100]~ \MSUN$           & [min, max] SNII Mass          & $[8, 100] ~\MSUN$          & this paper \\
  see Table \ref{tab:yields}  & Yield Tables                  & see Table \ref{tab:yields} & this paper \\
  at every star timestep                  & ISM Chemical Enrichment       & time/stellar mass discrete & this paper \\
  
  &&&\\[-1ex]
  \hline
  &&&\\[-1ex]
  
  & \textbf{Metal Advection} & &\\
  &&&\\
  gradient extrapolation          & Advection Scheme                & same + renormalization              & this paper \\
  0                               & Initialization Metal Fractions  & $10^{-10}$ at $z=127$ 	          & this paper \\
  H, He, C, N, O, Ne, Mg, Si, Fe  & Tracked Element Scalars         & same 9 + other metals               & this paper \\
  -                               & Metal Tagging                   & from SNIa, SNII, AGB separately     & \cite{Naiman:2017}\\
  -                               & Iron Tagging                    & from SNIa and SNII separately       & \cite{Naiman:2017}\\
  -                               & r-processes                     & from NS-NS mergers                  & \cite{Naiman:2017}\\
  &&&\\[-1ex]
\hline
  
  \end{tabular}
  \caption{Overview of the differences between the new TNG galaxy formation physics model and the original Illustris simulation model, including the fiducial values of all parameters (see text for details). The references for the new model are the present paper, together with \citealt{Weinberger:2017}, while the main references for the original Illustris model are \citealt{Vogelsberger:2013b} and \citealt{Torrey:2014a}. References in the last column provide a technical description of the feature implementation and (ideally/often/if possible) also a study of its effects on cosmological galaxy populations. Prescriptions for star formation and the treatment of the unresolved ISM in TNG remain unchanged with respect to Illustris \citep{Springel:2003}.}
  \label{tab:illvstng}
\end{table*}

%-------------------------------------------------------------------------------------------------------------------------------------------------------------

\subsection{Improvements of the TNG Galaxy Formation Model}
\label{sec:tng_gfm}

The main modifications to the Illustris galaxy physics model are concentrated in three areas:

\begin{itemize}  
\item the growth and feedback of supermassive black holes,
\item the galactic winds,
\item and the stellar evolution and gas chemical enrichment.
\end{itemize}

\noindent In the following subsections we give a full description of the implemented changes. Table \ref{tab:illvstng} provides a concise summary of the main new features, with fiducial values of the model parameters as described in Section~\ref{sec:sims_fiducial}.

\subsubsection{Formation, Growth and Feedback of Black Holes}

In the low-accretion state, the new approach employs a kinetic AGN feedback model which produces black hole-driven winds. This replaces the previous feedback model in the low-accretion mode known as the {\it bubble model} \citep{Sijacki:2007}. The BH-driven wind is motivated by recent theoretical arguments for the inflow/outflow solutions of advection dominated accretion flows (ADAFs) in this regime \cite[see][for a review]{Yuan:2014}. At high accretion rates, the TNG model invokes a thermal feedback that heats gas surrounding the BH, as was also employed in Illustris \citep{Springel:2005b,DiMatteo:2005}.
In the new model, we remove the $\alpha$ boost factor in the formulation of the Bondi-Hoyle accretion rate for the BH sinks. To guard against a slow early growth, we modify the choice of the initial black hole seed mass, which is increased to $8\times 10^5~ h^{-1} \MSUN$. 
%which is largely unmotivated and not required. 
All details related to the seeding, growth and feedback of black holes in the TNG model, including their impact on the galaxy population, are given in the companion paper by \cite{Weinberger:2017}. We also summarize the key details in Table \ref{tab:illvstng}. As a reminder, BH positioning and merging are implemented similarly as in Illustris. Namely, BHs are (re)positioned to their local potential minima, i.e.  searched within a sphere of a kernel-weighted number of neighboring gas cells, $n$, with $n$ fixed to 1000 in TNG (instead of $n_{\rm ngb}$ in Illustris). A BH merges with another if it is found within the accretion/feedback region of the other. As discussed in \cite{Weinberger:2017}, the new AGN feedback model is responsible for the quenching of galaxies residing in intermediate and high mass haloes ($\sim 10^{12} - 10^{14}\MSUN $), and for the production of a population of red and passive galaxies at late times. 

\subsubsection{Galactic Winds}
\label{sec:tng_winds}

The TNG model for galactic-scale, star formation-driven, kinetic winds is refined in several ways with respect to the approach described in \cite{Vogelsberger:2013b} and \cite{Torrey:2014a}. As in Illustris, wind particles inherit the properties of the gas cell from which they are launched (in proportion to the fraction of mass converted into wind), including their thermal energy.

Differently from Illustris, winds are now injected isotropically. In the Illustris model, the initial direction of the wind particles was a random orientation along the direction given by $\vec{v} \times \nabla \Phi$, where $\vec{v}$ and $\nabla\Phi$ are the velocity and acceleration of the gas cell of origin in the rest frame of the hosting friends-of-friends halo. In this `bipolar winds' approach wind particles are ejected preferentially along e.g. the rotation axis of a spinning object. In the new model, for simplicity, we abandon this directionality and assign wind particles an initial velocity in random directions: `isotropic winds'. Even in this case, winds will naturally propagate along the direction of least resistance and thereby show characteristic patterns of gas motion, such as large-scale galactic fountains around massive star-forming disks \citep[see Section~\ref{sec:sims_fiducial} as well as][]{Springel:2003}.

The initial speed of wind particles is also modified. As in the Illustris model, the TNG model launches wind particles with an initial speed that scales with the the local, one-dimensional dark matter velocity dispersion $\sigma_{\rm DM}$ \citep[as in Eq. (14) of][]{Oppenheimer:2006, Oppenheimer:2008,Vogelsberger:2013b}, measured with a weighted kernel over the $N=64$ nearest DM particles. In addition, the TNG model now (i) introduces an additional redshift-dependent factor for the wind velocities and (ii) sets a minimum wind velocity, $v_{w, \rm min}$, such that:

\begin{equation}
 v_w = {\rm max} \left[ \kappa_w ~\sigma_{\rm DM} \left( \frac{H_0}{H(z)} \right)^{1/3} , ~ v_{w, \rm min} \right]
\label{eq:winds_vel}
\end{equation}

\noindent where $\kappa_w$ is a multiplicative dimensionless factor. The chosen redshift dependence of the wind velocity implies that the wind velocity and the growth of the virial halo mass have the same scaling with redshift (we demonstrate this in Section~\ref{sec:sims_fiducial_winds}). This choice of redshift-independent wind velocities at fixed halo mass is inspired by the results of semi-analytical models, where a similar approach was needed to reproduce observed stellar mass functions and rest-frame B- and K-band luminosity functions across redshift \citep{Henriques:2013}.

The wind velocity floor \citep[whose effects on the gaseous haloes have already been investigated by][within the Illustris model]{Bird:2014, Suresh:2015} prevents wind mass loading factors in low-mass haloes from becoming unphysically large \citep[e.g.][for a recent compilation of observations, that however pertain outflow measurements at some radial distance from the galaxy centers and hence are not directly comparable to our choices at injection]{Zahid:2014, Schroetter:2016}. Indeed, it is physically plausible that there is an upper limit to the amount of mass that can be entrained by a supernova (or a group of neighbouring supernovae in a superbubble), and hence the wind speed cannot become arbitrarily small in low-mass haloes. We note that, by fixing the injection velocity at high redshift, our Hubble factor scaling implies larger wind velocities at recent times and across all halo masses. It thereby increases the effectiveness of the galactic winds at suppressing low-redshift star formation. Simultaneously, because $v_{w, \rm min}$ imposes a floor almost exclusively in low-mass galaxies, the TNG stellar feedback is also more effective at high redshift (we show both aspects in Section~\ref{sec:sims_variations_winds_1}). Some mechanism to create more effective wind feedback at higher redshifts has been advocated in many galaxy formation models and implemented in a variety of ways \citep{Murray:2010, Hopkins:2011e, Stinson:2013, Sales:2014}. Finally, a wind velocity floor impacting low-mass galaxies has been introduced recently also by \citealt{Dave:2016} in their implementation of launched outflows.

Once the wind injection velocity is determined, the wind mass loading factor in turn depends on the specific energy available for wind generation, $e_w$ \citep[or $\rm egy_w$ in the formalism of][]{Vogelsberger:2013b}, which in our framework is tied to the energy released by SNII per formed stellar mass. The TNG model contains two modifications wrt Illustris affecting the available wind energy: (i) some given fraction of this energy is thermal, via a parameter $\tau_w$; and (ii) the wind energy depends on the metallicity of the star-forming gas cell, such that galactic winds are weaker in higher-metallicity environments. Taken together, the mass loading factor at injection is:

\begin{equation}
 \eta_w \equiv  \frac{\dot{M}_w}{\dot{M}_{\rm SFR}} = \frac{2}{v_w^2}~ e_w  (1- \tau_w)
\label{eq:winds_eta}
\end{equation}
\noindent where $\dot{M}_w$ denotes the rate of gas mass to be converted into wind particles and $\dot{M}_{\rm SFR}$ the instantaneous, local star-formation rate.

The wind energy available to a star-forming gas cell with metallicity $Z$ is given by:
\begin{eqnarray}
e_w &=& \overline{\vphantom{t} e}_{w}~ 
                   \left[ f_{w,Z} + \frac{1-f_{w,Z}}{1 + (Z/Z_{w, \rm ref})^{\gamma_{w,Z}}} \right] \nonumber \\[2ex]
            && \times ~ N_{\rm SNII} ~E_{\rm SNII,51} ~10^{51} ~\rm{erg}~ \MSUN^{-1},
\label{eq:wind_energy}
\end{eqnarray}

\noindent similar to the parameterization by \cite{Schaye:2015}.

Here, $E_{\rm SNII,51}$ denotes the available energy per core collapse supernovae in units of $10^{51}$ erg, which we take equal to unity (as in the Illustris model). $N_{\rm SNII}$ is the number of SNII per formed stellar mass (in solar mass units) and depends on the shape of the IMF and the mass limits for core-collapse supernovae (see Section~\ref{sec:tng_stars}). The wind energy factor \egyw  is a dimensionless free parameter of the model, while the metallicity dependence implies that

\begin{eqnarray}
\nonumber
\eta_w &\propto& \frac{2}{v_w^2}~\overline{\vphantom{t} e}_{w}~ f_{w,Z} ~(1- \tau_w)  ~{\rm for }~ Z \gg Z_{w, \rm ref}\\
\nonumber
\eta_w &\propto& \frac{2}{v_w^2}~\overline{\vphantom{t} e}_{w} ~(1- \tau_w)  ~{\rm for }~ Z \ll Z_{w, \rm ref}.\\
\label{eq:winds_eta_scaling}
\end{eqnarray}

\noindent Namely, the new scaling reduces by a factor $f_{w,Z}$ the energy at injection for gas cells with metallicities much larger than a reference value $Z_{w, \rm ref}$ (see Figure~\ref{fig:galprop_winds_Z} for a visual representation of the effect). For the TNG fiducial choice of the parameter values ($Z_{w, \rm ref}=0.002, \gamma_{w,Z}=2, f_{w,Z}=0.25$, and \egyw $=3.6$), the effective wind mass loading factor at injection is $\eta_w = (2/v_w^2)~ 0.9~ (1- \tau_w) N_{\rm SNII} ~E_{\rm SNII,51} 10^{51} \rm{erg}~ \MSUN^{-1}$ for gas metallicities similar to solar ($Z\sim 0.02$). In comparison, the Illustris model gives $\eta_w = (2/v_w^2)~ 1.09 ~N_{\rm SNII} ~E_{\rm SNII,51} 10^{51} \rm{erg}~ \MSUN^{-1}$. In practice, the wind energy in Eq.~(\ref{eq:winds_eta}) varies in the range  $e_w = [3.6, 0.9] \times N_{\rm SNII} 10^{51}  \rm{erg}~ \MSUN^{-1}$, according to the gas metallicity. Because the majority of the stars in the Universe is made in MW-like galaxies and given that the average metallicity of star-forming gas in ($z=0$) TNG MW-like galaxies is about $Z = 0.015-0.02$, in our model the bulk of the galactic winds is launched with energies very close to $e_w = 1 ~ N_{\rm SNII} 10^{51}  \rm{erg}~ \MSUN^{-1}$ (see again Figure~\ref{fig:galprop_winds_Z} for reference), in agreement with the qualitative expectations for the energy available from core-collapse SNe.  Eq.~(\ref{eq:winds_eta}) here follows directly from Eq.~(15) of \cite{Vogelsberger:2013b} assuming no momentum-driven winds. 

By allowing the galactic winds to carry some (additional) thermal energy, we avoid spurious star formation and other artifacts where the wind particles hydrodynamically recouple, as already done by \cite{Marinacci:2014,Grand:2017}. We also allow wind particles to radiatively cool as they travel, their effective density adopted from the nearest gas cell and including self-shielding corrections. As we demonstrate later, a $\tau_w > 0$ fraction of thermal energy can significantly impact the overall stellar mass content of galaxies at all times. 

The reduction of the available wind energy with metallicity is motivated by the idea that higher metallicity galaxies may imply larger radiative cooling losses of the supernova energy \citep[see also][]{Schaye:2015,Martizzi:2015}. Additionally, the evolution, mass-loss, and end states of SNII progenitors may themselves depend on metallicity \citep{Smartt:2009}, in turn affecting the amount of energy released into the ISM per core collapse. Given our choice not to change the effective, overall wind energy at injection in metallicity environments typical of $L^{\star}$ galaxies, the metallicity modulation again implies that winds in lower-mass galaxies are stronger in the new model than in Illustris.

The conditions for wind re-coupling are invariant with respect to the Illustris model, i.e. a wind particle is donated to the gas cell in which it is currently located once either it falls below a certain density ($0.05\times$ the density threshold for star formation) or a maximum travel time has elapsed ($0.025\times$ the current Hubble time). In practice, the time condition is almost never exercised and the typical travel length of the wind while being decoupled is at most a few kpc. Finally, we have not modified any choice compared to Illustris in relation to the metal loading of the winds. 

To summarize, the total energy release rate available to drive galactic winds depends on the instantaneous star-formation rate as:

\begin{equation}
\dot{E}_w  = \frac{ \dot{M}_w~ v_w^2}{2 ~(1- \tau_w)} = e_w~ \dot{M}_{\rm SFR},
\end{equation}
where all terms depend on spatial location within a galaxy and on time. However, even at fixed wind energy rate, the actual effectiveness of the winds at preventing star formation across a galaxy life time depends on the properties of the wind particles at injection, i.e. their velocity, thermal content, specific energy, or even the number of wind particles spawned per given mass loading factor (within the limits of the cells (de-)refinement scheme). Such properties ultimately determine the characteristics of the galactic outflows, including time scales for gas recycling and gas mass return to the galaxy. While not the main focus of this paper, in Appendix \ref{sec_appendix2_winds} we quantify the effects of different galactic wind parameterizations on the galaxy populations. 

%-------------------------------------------------------------------------------------------------------------------------------------------------------------

\begin{table*}
\centering
\begin{tabular}{l| l l} 
\hline
		& Illustris Tables & TNG Tables \\
\hline
AGB 	& \cite{Karakas:2010} 	        & \cite{Karakas:2010} \\
		& $[1-6]~\MSUN, \mathit{Z}\in [0.0001, 0.004, 0.008, 0.02]$	& $[1-6]~\MSUN, \mathit{Z}\in [0.0001, 0.004, 0.008, 0.02]$\\
		& & \cite{Doherty:2014} \\
        & & $[7.0, 7.5]~\MSUN, \mathit{Z}\in [0.004, 0.008, 0.02]$ \\
        & & \cite{Fishlock:2014} \\
        & & $[7.0]~\MSUN, \mathit{Z}\in [0.001]$\\
        &&\\
SNII 	& \cite{Portinari:1998}		& \cite{Kobayashi:2006}\\
		& $[6-120]~ \MSUN, \mathit{Z}\in [0.0004, 0.004, 0.008, 0.02, 0.05]$ 
        & $[13-40]~ \MSUN, \mathit{Z}\in [0, 0.001, 0.004, 0.02]$\\ 		
        & & \cite{Portinari:1998}\\
        & & $[6-13, 40-120]~ \MSUN, \mathit{Z}\in [0.0004, 0.004, 0.008, 0.02, 0.05]$ \\ 		
		&&\\
SNIa 	& \cite{Travaglio:2004}		& \cite{Nomoto:1997a}, W7\\
		& \cite{Thielemann:2003}		& 					\\
\hline
\end{tabular}
\caption{Overview of the choices for the stellar yield tables compiled from the literature in the Illustris versus the TNG model. In the new model, the minimum mass for SNII is raised to 8$\MSUN$. To simultaneously use the yields proposed by \textcolor{blue}{Kobayashi et al 2006} and \textcolor{blue}{Portinari et al. 1998}, SNII yields are renormalized such that the IMF-weighted yield ratios at each metallicity are equal to those from the Kobayashi mass range models alone (see text for details).}
\label{tab:yields}
\end{table*}

\subsubsection{Stellar Evolution and Chemical Enrichment}
\label{sec:tng_stars}

\begin{figure*}
\centering
\includegraphics[width=17cm]{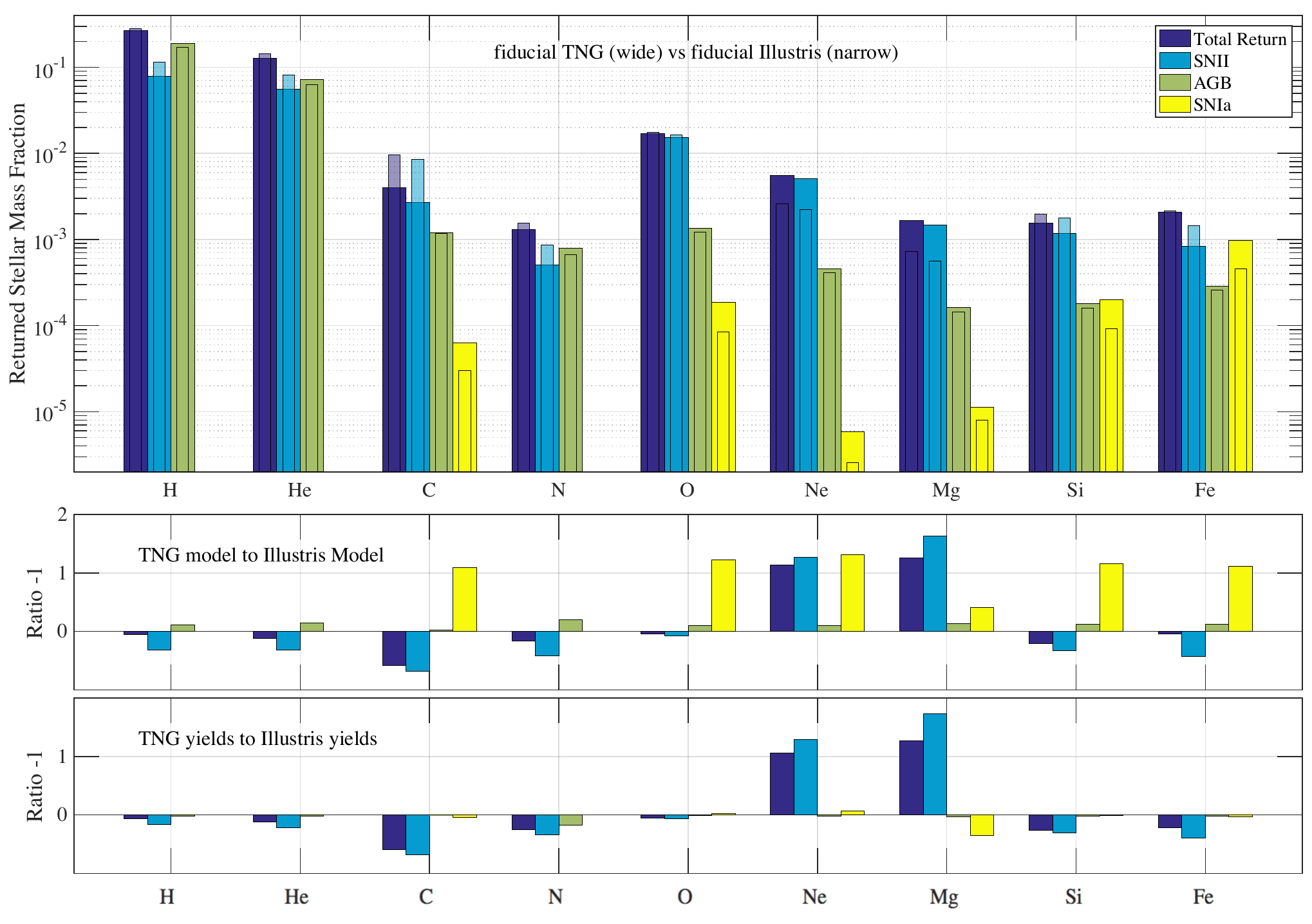}
\caption{Top: Fraction of mass returned to the ISM in a Hubble time per stellar mass formed at Solar metallicity. Two sets of bars compare results from the TNG model (wide bars) and the Illustris model (thin bars). Hydrogen and Helium dominate the stellar return in total mass. SNII dominate the metal return of essentially all elements considered here, but for Nitrogen and Iron, for which AGB and SNIa are, respectively, at par. Middle: ratio between the two models in linear scale to highlight the differences. These comparisons take into accounts \textit{all} changes to the enrichment process. Namely, the different yield tables, the increased minimum mass for SNII, and the updated SNIa normalization factor. The differences from the yield tables alone are reported in the bottom panel, which account for most of the changes in metal return for SNII. Modifications from Illustris to TNG for SNIa are due to the more consistent normalization factor to the SNIa DTD, for all species but Magnesium.}
\label{fig:yields}
\end{figure*}

Stellar particles represent a population of stars with the same birth time (single-age stellar population) and with an initial mass function given by \cite{Chabrier:2003}. These populations are allowed to evolve and age, with a return of mass and individual elements to the surrounding inter-stellar medium. As described by \cite{Vogelsberger:2013b} and following \cite{Wiersma:2009b}, the mass and metal return of a stellar particle is tracked as a function of time, by following the evolution of stars through three stellar phases: asymptotic giant branch (AGB) stars, core collapse supernovae (SNII) and supernovae Type-Ia (SNIa). In the TNG model, we assume stars pass through an AGB phase in the mass range $1-8\MSUN$, and core collapse supernovae are the fate of stars between 8 and 100 $\MSUN$, while in Illustris the minimum mass for SNII was 6 $\MSUN$. This was in practice dictated by the lack of available elemental yields for massive AGB stars. Although the minimum mass for core collapse supernovae is not known precisely, it is typically thought to be roughly 8 $\MSUN$ or slightly higher \citep{Smartt:2009,Ibeling:2013,Woosley:2015}, making this update important. Given the \citealt{Chabrier:2003} IMF, a minimum mass limit for SNII of 8 $\MSUN$ implies 0.0118 stars going off as core collapse supernova per formed stellar mass in $\MSUN$, a decrease by 30 per cent in the factor $N_{\rm SNII}$ of Eq. (\ref{eq:wind_energy}) with respect to the choice with 6 $\MSUN$ (0.0173 $\MSUN^{-1}$). The upper mass limit of SNII progenitors is also uncertain \citep{Jennings:2014}, although the precise value is largely unimportant for our purposes.

\label{sec:tng_sf_enrichment}
For reasons of computational efficiency, we change the time integration of mass and metal return from stars to their surrounding gas. Previously, every star at every timestep searched for nearby gas cells and then donate a minute amount of mass and metals. Now, we discretize this process more coarsely in time, such that this only occurs if the total mass fraction relative to the star mass exceeds 0.0001 or if the star age is less than 100 Myr, to ensure we capture quickly evolving young stellar populations. Therefore, the yields are integrated over a time interval corresponding to the time since the last enrichment event, rather than the dynamical timestep itself. We have verified that this change makes no qualitative or quantitative impact on the simulation results, as desired, except for saving computation time.

\subsubsection{Updated Yield Tables}
\label{sec:tng_yields}

In the TNG model we update our choices for the elemental mass yields of stars from the literature. Tabulated stellar yields are uncertain, often incomplete at needed stellar mass or metallicity ranges, and continuously updated in the literature: here we opt for a series of choices which are overall more consistent with observed abundance ratios, specifically for Milky Way-like galaxies. In practice, construction of the yield tables requires a combination of data from several different sources, in diverse ranges of stellar masses and metallicities, and based on a variety of methods, all summarized in Table \ref{tab:yields}. 

We extend the AGB tables from their original $1-6 ~\MSUN$ mass range to $1-7.5 ~\MSUN$. We do this by using the \cite{Karakas:2010} yields for the metallicity values $Z \in [0.0001, 0.004, 0.008, 0.02]$ and for masses between 1 and 6 $\MSUN$. For masses of 7 and 7.5 $\MSUN$ we use the \cite{Doherty:2014} yields with $Z \in [0.004, 0.008, 0.02]$, and supplement this with the \cite{Fishlock:2014} data at 7 $\MSUN$ and $Z=0.001$, extrapolated down to $Z=0.0001$. An interpolation across masses gives us the final values for 7.5 $\MSUN$ and $Z=0.0001$ AGB metal return.

With respect to the Illustris choices, the largest update is to the SNII tables, which now extend from 8 to 120 $\MSUN$, over $Z \in [0, 0.001, 0.004, 0.02]$. For the mass range $13-40~\MSUN$, we use yields from \cite{Kobayashi:2006}. To extend this mass range to lower and higher masses we supplement the Kobayashi models with those by \cite{Portinari:1998}, extrapolated from their metallicity range of [0.0004, 0.004, 0.008, 0.02, 0.05] to the metallicity range of the Kobayashi data. To obtain yield ratios consistent with Kobayashi, we renormalize all our models such that the IMF-weighted yield ratios at each metallicity are equal to those from the Kobayashi mass range models alone. For consistency with previous work, we also renormalize our yields such that the total ejected mass is the same as in the Illustris yield tables which were taken fully from \cite{Portinari:1998}.

The SNIa yields have been changed to the ``W7'' model from \cite{Nomoto:1997a} to be consistent with other groups, however this leads to only minor changes in the yields. In fact, all SNIa yields are independent of mass and metallicity, and are injected with a simple delay time distribution (DTD). We also note that, solving an issue identified by \cite{Marinacci:2014b}, we have corrected the upper limit of the normalization integral for the SNIa delay time distribution from infinite to the Hubble time. This makes the total number of SNIa consistent with the prescriptions given by \cite{Maoz:2012} and with the choice for the SNIa rate normalization constant of $N_0=1.3\times10^{-3} \MSUN^{-1}$. In the TNG model, the effective total number of SNIa per stellar mass formed is larger by about a factor of 2 than in Illustris, if all other parameters are kept as in the Illustris model. 

In Figure~\ref{fig:yields} we diagnose the net yield changes, showing the stellar mass fraction by chemical species returned to the ISM in a Hubble time, per stellar mass formed, at Solar metallicity. Wide bars represent the new TNG model, while thin bars show the fiducial Illustris model. Different columns refer to different enrichment channels: all together, or separated by SNII, AGB, and SNIa alone. This comparison includes \textit{all} changes to the enrichment (shown through ratios in the middle panel): the different yield input as per Table \ref{tab:yields}, the different minimum mass for SNII, and the different SNIa normalization factor. The last is responsible for the increased metal fractions in the SNIa channel from Illustris to the TNG model. On the other hand, changes due solely to the updated yields (shown through ratios in the bottom panel) are small for SNIa, of the order of a few percent, with the notable exception of Magnesium. The different metal mass fractions returned by core-collapse supernovae between the two models are mostly due to the changes in the underlying yield tables: the enrichment of Neon and Magnesium can be different by up to a factor of 2.5 because of the modifications detailed in Table \ref{tab:yields}. Finally, the $\sim$ 10 per cent-level increase in the mass returned by AGB stars is due to the fact that in TNG AGB stars can be as massive as 8 $\MSUN$ instead of 6 $\MSUN$. 

\subsubsection{Star Formation in the Presence of Magnetic Fields}
\label{sec:sf_mhd}

As a final point of clarification, we comment on the treatment of the magnetic field for star-forming gas. Four distinct events can take place: (i) a gas cell can convert entirely into a star, (ii) a gas cell can convert entirely into a wind, (iii) a gas cell can spawn a star with a fraction of its mass, and (iv) a gas cell can spawn a wind with a fraction of its mass. In practice, for the types of simulations presented here, the former two cases dominate over the latter two.

In the first two cases, the originating cell is removed from the Voronoi mesh. The neighboring cell volumes then fill the evacuated space and -- although their conservative variables remain the same -- their primitive variables, including magnetic field strength, decrease in proportion to the volume increase. We are therefore assuming that the corresponding magnetic field flux is locked up into the star (or wind) and removed from the gas. This follows \cite{Pakmor:2014} who found that, compared to not removing the B-field \citep[as done instead in][]{Pakmor:2013}, this choice gives somewhat smaller magnetic field strengths in galaxies, by at most a few tens of percent. 

In the latter two cases, when only a fraction of the originating gas cell mass is used to spawn a new star or wind particle, we leave the magnetic field as well as the magnetic energy of the gas cell unchanged \citep[in the same manner as e.g.][]{Dubois:2010,Pakmor:2013}. This has the advantage of not introducing any additional divergence error or change in the local magnetic field topology. The mass, (non-magnetic) total energy, and momentum of the originating gas cell is reduced by the corresponding mass fraction of the newly spawned particle.\footnote{In fact, a final complication exists for case (iv), where a gas cell spawns a wind with only a fraction of its mass. Although the magnetic energy of the gas cell is unchanged, the total energy -- including the magnetic component -- is rescaled as the fractional mass loss. This implies a transfer of thermal to magnetic energy, because the total energy decrease is larger than it would have been in the absence of MHD. This inconsistency is resolved going forward, such that case (iv) follows the more obvious behavior of case (iii), although for concreteness we note that this resolution has not been applied to any of the simulations of the present work.}

These aspects of the model are non-trivial and relate directly to unresolved physical effects. For the mass resolution we can afford, each star particle represents approximately one or more stellar populations formed within a molecular cloud. The magnetic field dynamics in such a cloud is complex, and it may be that the magnetic field surrounding or within a molecular cloud is amplified during star formation \citep{Schober:2012}, likely through turbulence \citep{Federrath:2011}. On the other hand, it is possible that star formation depletes the magnetic energy of a molecular cloud. This could occur if part of the magnetic field was locked up in stars, or removed from the cloud through e.g. ambipolar diffusion or turbulent reconnection \citep{Shu:1983,Santos-Lima:2010}. For the present model we therefore adopt the rather simple approach described above.

%-------------------------------------------------------------------------------------------------------------------------------------------------------------

\subsection{Additional Code Features}
\label{sec:tng_features}

The simulation code incorporates a number of on-the-fly analysis routines which produce rich data beyond particle level information. These include the {\sc SUBFIND} and {\sc FOF} algorithms to identify haloes and subhaloes \citep{Springel:2005,Dolag:2009}; a custom output strategy, whereby particle information is ordered on disk according to its hierarchical group membership, allowing for rapid retrieval and query; a Monte Carlo tracer particle scheme \citep{Genel:2013}, which actively follows the flow of a large number of ``tracer particles'' and enables the reconstruction of Lagrangian fluid element histories; and the application of stellar population synthesis models to derive broad-band luminosities for the stellar particles \citep{Vogelsberger:2013b}.
The TNG codebase includes a new, differentiated output approach that optimizes
disk usage by writing snapshots covering different data fields, at different times, and covering different spatial volumes according to
three types: full, mini, and subbox snapshots. The TNG model also includes a series of novel code capabilities:

\subsubsection{Shock Finder}

A cosmological shock finder has been developed for the {\sc Arepo} code \citep{Schaal:2015}: it uses a ray-tracing method in the Voronoi mesh to identify shock surfaces and measure their properties, and it can be used on-the-fly as a simulation runs. Its functioning and application have already been demonstrated on the Illustris simulation data \citep{Schaal:2016} and new non-radiative test cases \citep{Weinberger:2017}. By including this component we measure the dissipated energy rate -- the amount of kinetic energy irreversibly transformed into thermal energy -- and the Mach number of every gas cell at each output.

\subsubsection{Metal Tagging}

In addition to following the production and evolution of the 9 individual species, we now \textit{separately} follow the production and mixing of the {\it sum of all metals} produced by AGB stars, SNIa and SNII, respectively. These are recorded as mass fractions in both gas and stars, in the exact same way as the 9 individual species. In addition, we now also keep track of the iron mass separately produced by SNIa and SNII, which is then also followed as it advects throughout the gas over time. More details on the implementation details, and demonstration of their scientific application, will be shown in an upcoming paper \citep{Naiman:2017}.

\subsubsection{Neutron-Star Mergers and r-processes}

We include the injection of r-process material from neutron stars - neutron star (NSNS) mergers by modeling such infrequent events in a similar fashion as with  SNIa. Following the method of \cite{Shen:2015}, we use a power-law delay time distribution \citep{Piran:1992, Kalogera:2001} with the same DTD index (1.12) but with a different cut off time scale and normalization as compared to the SNIa DTD \citep[see][for more details]{Naiman:2017}. The delay time distribution is used to probabilistically determine the number of NSNS mergers (if any) which occur in the stellar particles. These merger events would then return mass and metals to the surrounding gas. To maintain consistency with the previous method, any NSNS material is treated as a tracer and does not add to the dynamical mass (total nor by species) of the receiving gas cells. This is not an issue because the NSNS merger ejecta mass is negligible. In practice, we derive the total metal mass produced by NSNS by using yields re-scaled from the SNIa tables, and track this as a single additional property for both gas cells and stars. The details of this implementation and the methods to extract an estimate of the production of Europium, as a specimen of r-process material, is presented by \cite{Naiman:2017}.

%-------------------------------------------------------------------------------------------------------------------------------------------------------------
% Section 3
%-------------------------------------------------------------------------------------------------------------------------------------------------------------

\begin{figure*}
\centering
\includegraphics[width=15.9cm]{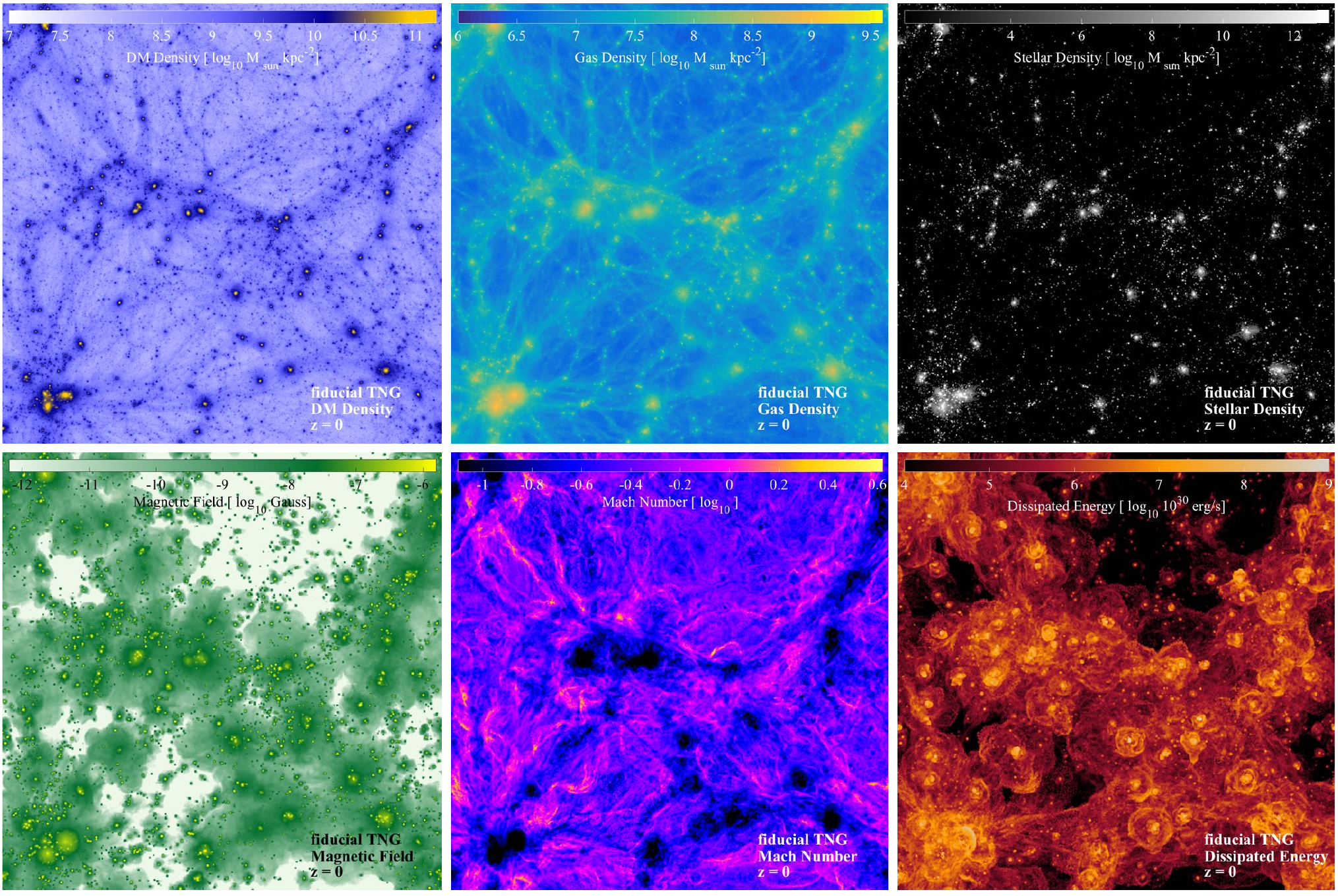} 
\caption{Qualitative inspection of the L25n512 volume run with the fiducial TNG model showing the large scale structure at $z=0$. Top row: column density of dark matter, gas and stellar mass. Bottom row: mass-weighted projected average of magnetic field strength, gas Mach number, and kinetic energy dissipation rate via shocks, to highlight the new diagnostic capabilities.}
\label{fig:L25n512_box_1} 
\end{figure*}

\section{Performance of the TNG Model}
\label{sec:sims}

We demonstrate the effectiveness of the model introduced in the previous Sections, as well as its dependence on the associated parameters, with a suite of cosmological simulations in uniform boxes. In the following we show the outcome of the model and its variations on a handful of galaxy population observables which solely depend on the integral stellar content of galaxies: the star formation rate density as a function of cosmic time (SFRD), the galaxy stellar mass function at $z=0$ (GSMF) and the present-day stellar-to-halo mass relation (SMHM). These are the quantities which were used to calibrate the parameters of the Illustris model. However, in order to improve upon the identified issues with the original model, we also now evaluate: the BH mass to galaxy or halo mass relation, the halo gas fraction, and the galaxy stellar sizes. In all cases we include observational constraints intended only as rough guidelines. We postpone to future work a quantitatively robust and consistent comparison between simulated galaxies and observations.

\begin{figure*}
\begin{center}
\includegraphics[width=15.9cm]{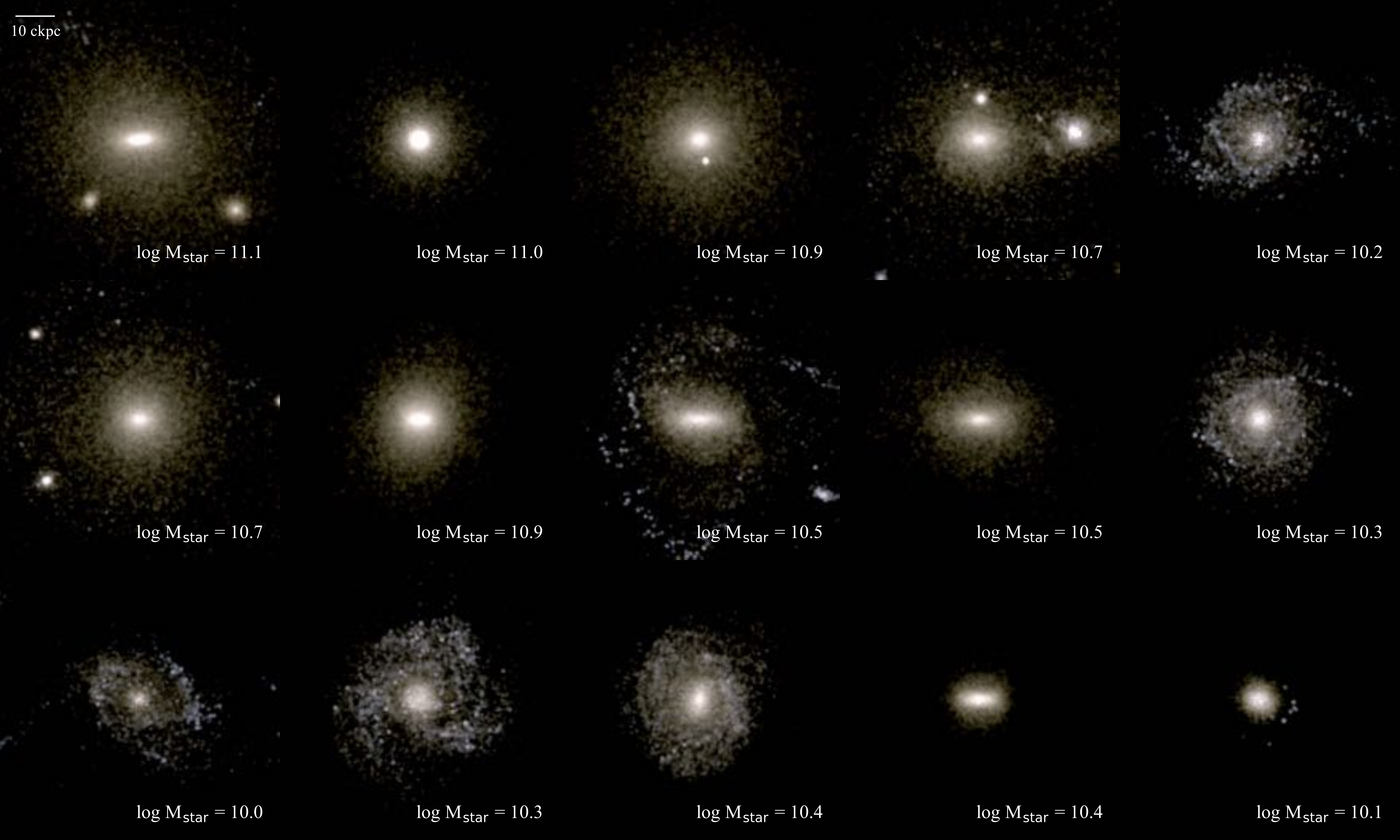}
\includegraphics[width=15.9cm]{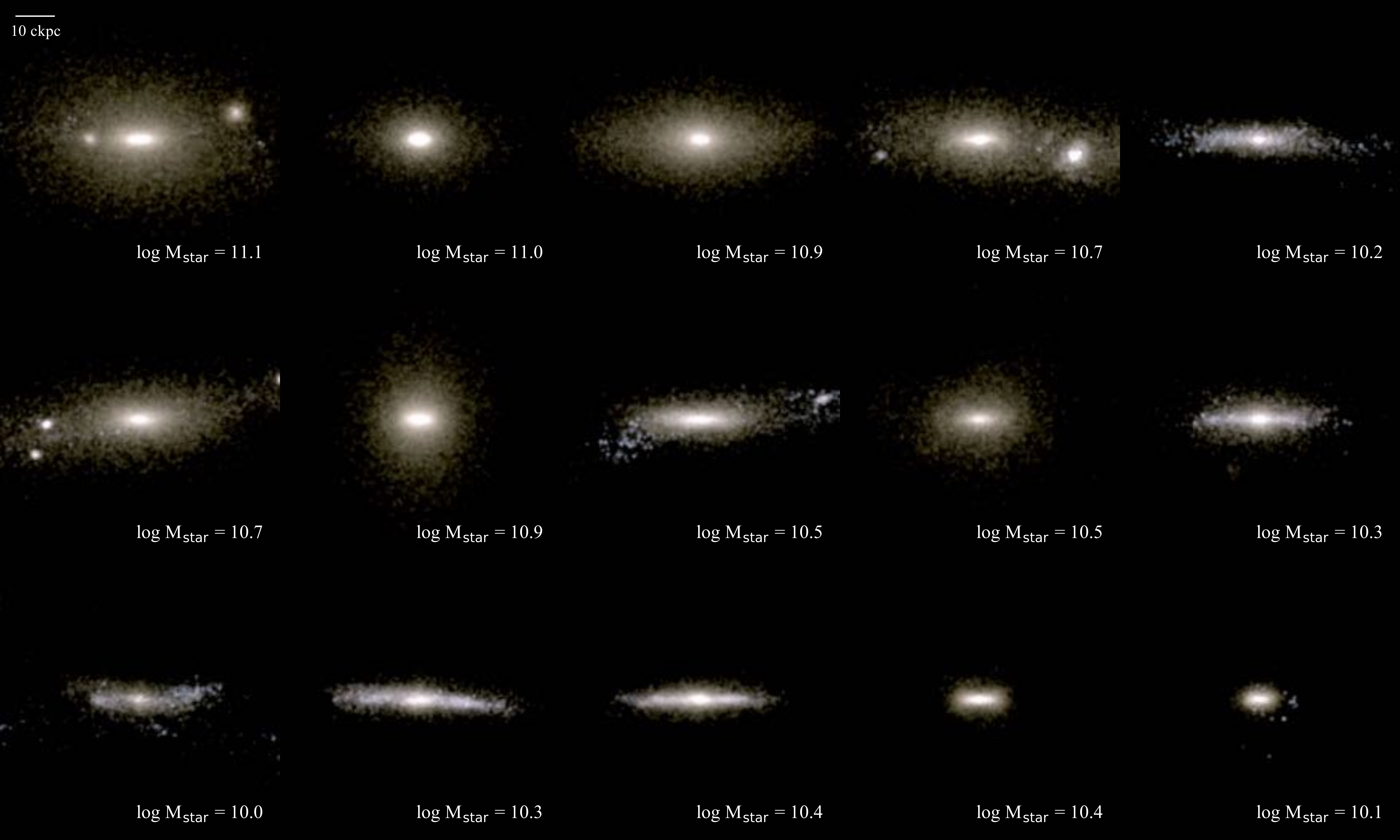}
\caption{Qualitative inspection of the L25n512 volume run with the fiducial TNG model: a random sample of fifteen $z=0$ galaxies selected to have halo mass greater than $10^{12}\,\MSUN$. These include a mix of spheroid-type and disk-type systems, where the top fifteen panels are face-on, and the bottom fifteen panels are the same galaxies edge-on. Each is shown in projected stellar light combining three wide optical NIRCam filters (f200w, f115w, and f070w) from JWST. The TNG model still reproduces a diverse galaxy population, which is a basic requirement for any theoretical model for galaxy formation. Stellar masses are in $\MSUN$ units.}
\label{fig:L25n512_box_2} 
\end{center}
\end{figure*}

%-------------------------------------------------------------------------------------------------------------------------------------------------------------

\subsection{Cosmological Test Simulations: Setup}
\label{sec:sims_setup}

We run a number of cosmological simulations of a periodic box 25 $h^{-1} {\rm Mpc}$ on a side with the {\sc Arepo} code (d203ec8) and cosmological parameters consistent with the most recent Planck results \citep[][ $\Omega_m = 0.31, \Omega_L = 0.69, \Omega_b = 0.0486, h = 0.677, \sigma_8=0.8159, n_s = 0.97$]{PlanckXIII:2015}. Variations of the model are carried out on a realization with (512)$^3$ dark-matter particles and initially (512)$^3$ gas cells. This resolution corresponds to modern large-scale cosmological simulations such as Horizon-AGN, Illustris, Eagle \citep{Dubois:2014,Vogelsberger:2014a,Schaye:2015}: an average gas-cell and stellar particle mass of $\simeq 10^6\,\MSUN$, DM particle mass of $\simeq 10^7\,\MSUN$, and stellar/DM softening lengths of $\sim 500 - 1000$\,pc at $z=0$. The exact values are given in Appendix \ref{sec_appendix1}, specifically Table \ref{tab:res}, where we also consider the effects of resolution and the convergence behavior of our model. The simulations are evolved to the present epoch from $z=127$ initial conditions. These have been obtained with the code {\sc N-GenIC} \citep{Springel:2005} by applying the Zel'dovich approximation on a glass distribution of particles with a linear matter transfer function computed using the CAMB code \citep{Lewis:2000}.

In order to simulate a representative volume and minimize sample variance, we have selected the realization with care. Namely, we have drawn 10 random realizations of the initial density field and evolved low-resolution gravity-only versions to the current epoch. We have chosen the realization exhibiting a cumulative dark matter halo mass function at $z=0$ that is closest to the average one across realizations, via a $\chi^2$-minimization across the widest halo mass range allowed by resolution. In fact, we note (but do not show here) that such choice does not guarantee that the resulting galaxy population is unaffected by sample variance nor, therefore, that it is a representative sample. However, it at least minimizes usually large variations arising from the underlying DM density field. 

The simulation box we study in this paper (L25n512) contains at the current epoch 9 haloes with total mass ($M_{\rm 200c}$\footnote{$M_{\Delta {\rm c}}$ denotes the mass enclosed in a sphere whose mean density is $\Delta$  times the critical density of the universe at the time the halo is considered.}) exceeding $10^{13}\MSUN$, about 40 Milky-Way like central haloes ($6\times10^{11}\MSUN \le M_{\rm 200c} \le 2\times10^{12} \MSUN$), and about 1500 luminous galaxies (among centrals and satellites) resolved with at least 100 stellar particles. While the most massive object does not reach $\sim5\times10^{13}\MSUN$, the chosen volume size is still well suited to study the galaxy population at the low-mass end of the galaxy mass function, which is the main scope of this paper.

\begin{figure*}
\centering
\includegraphics[width=8.8cm]{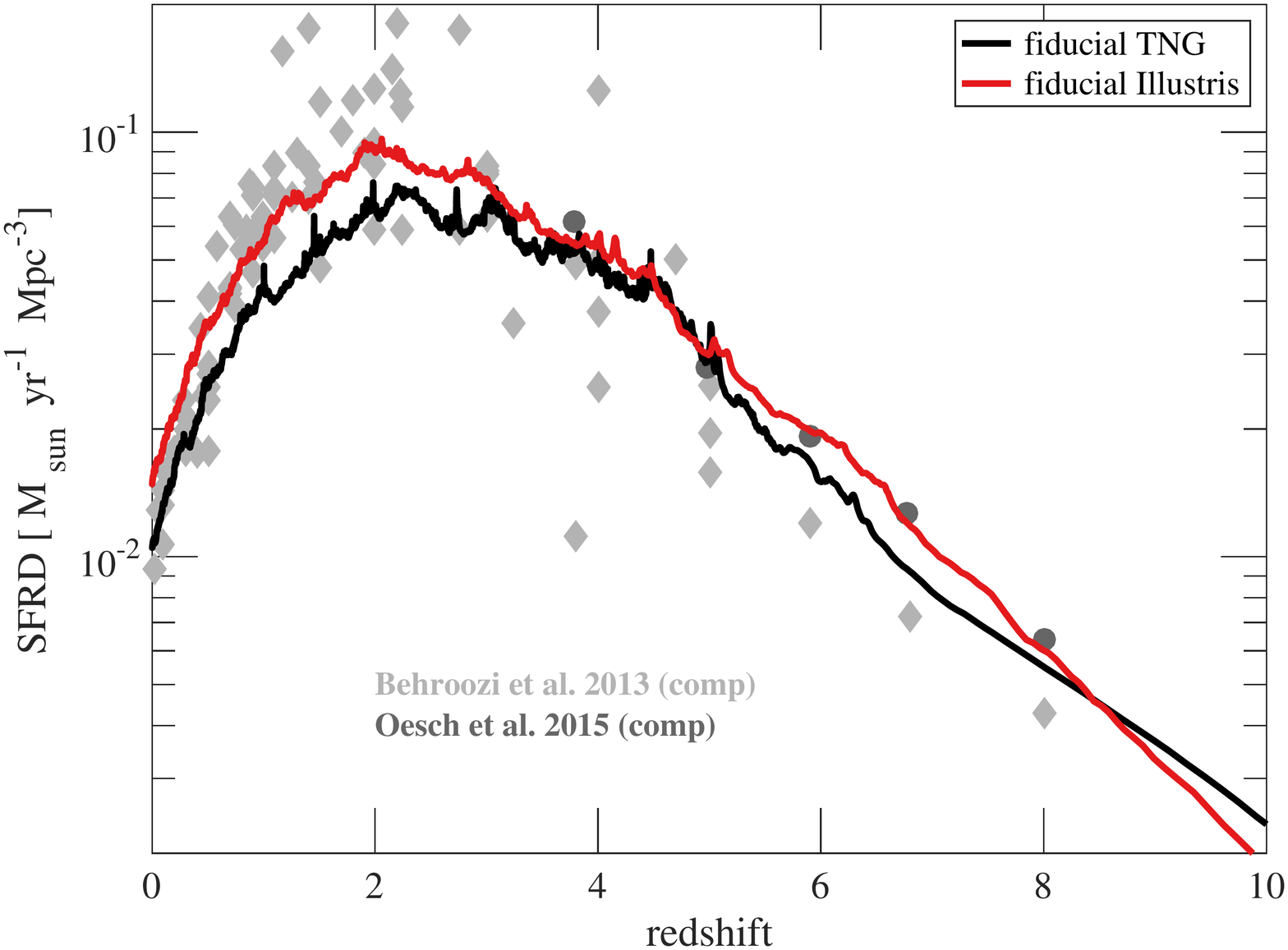}
\includegraphics[width=8.8cm]{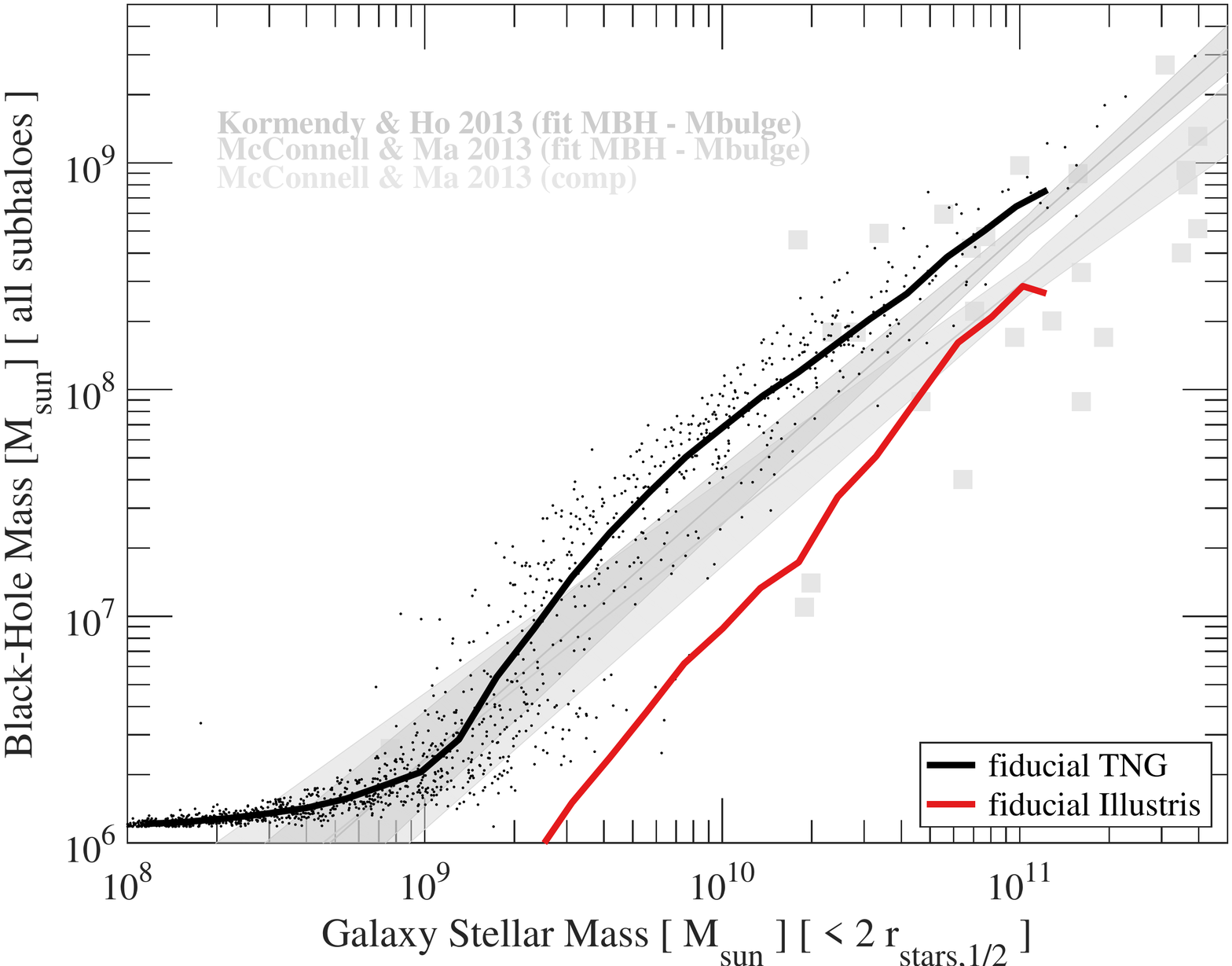}
\includegraphics[width=8.8cm]{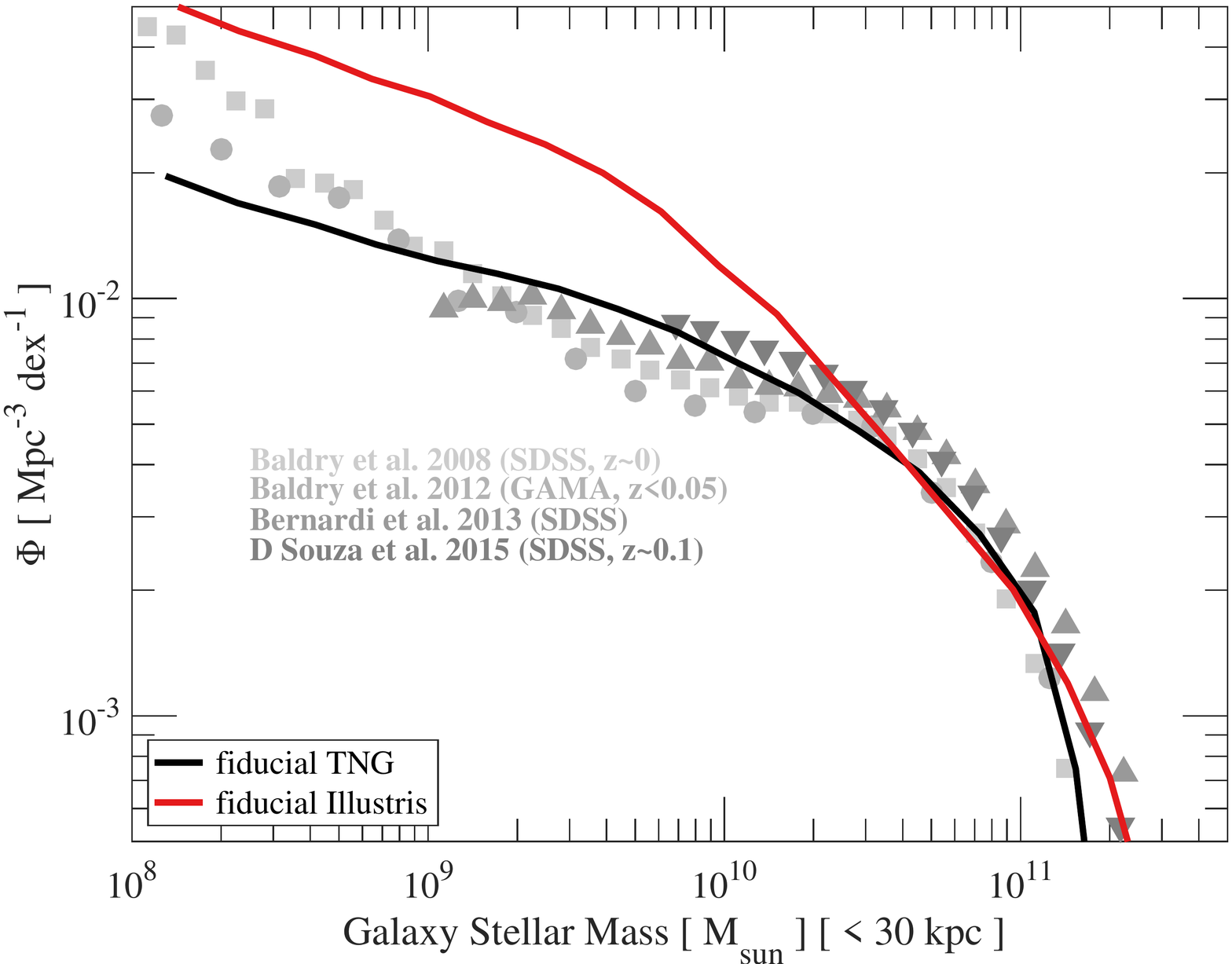}
\includegraphics[width=8.8cm]{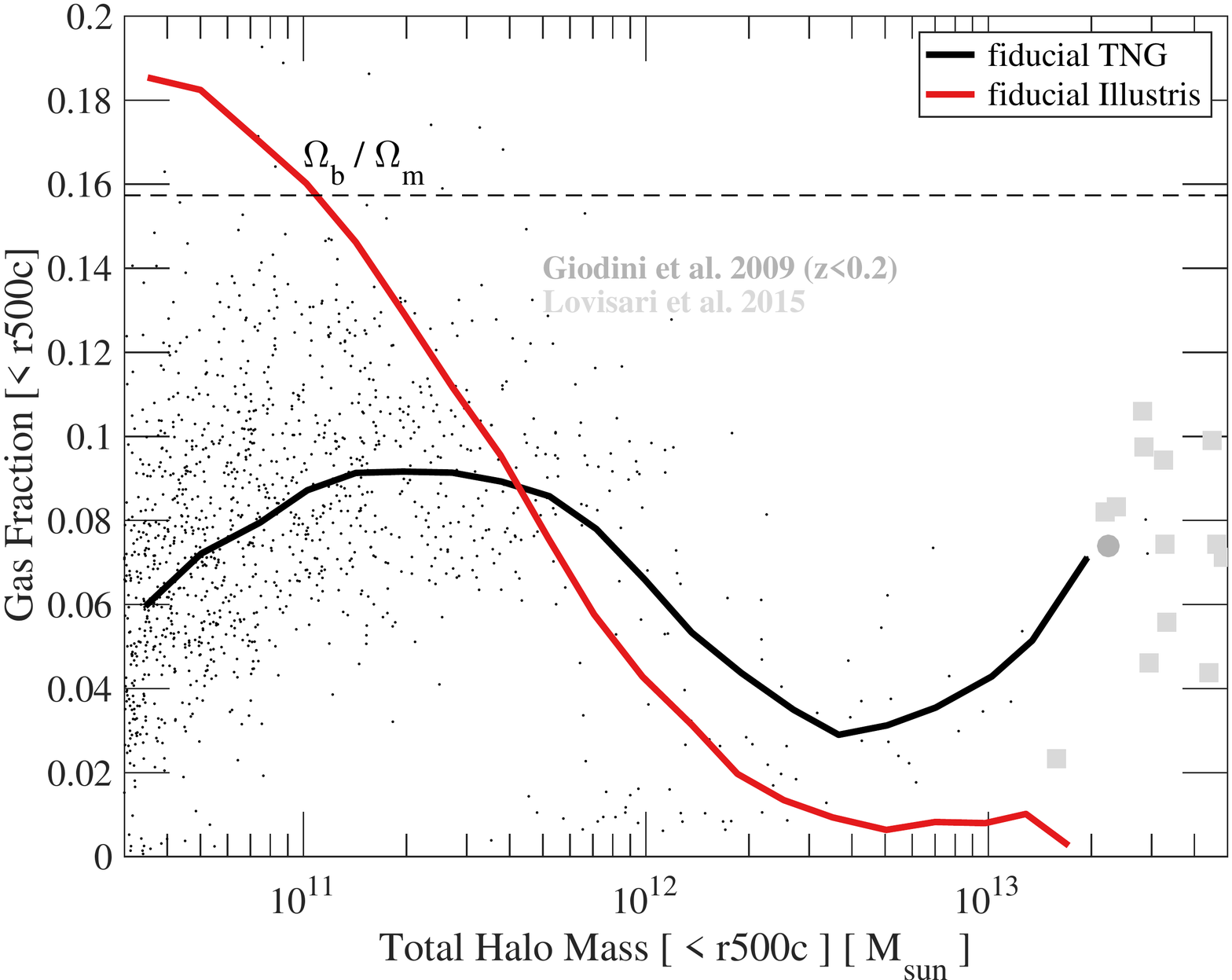}
\includegraphics[width=8.8cm]{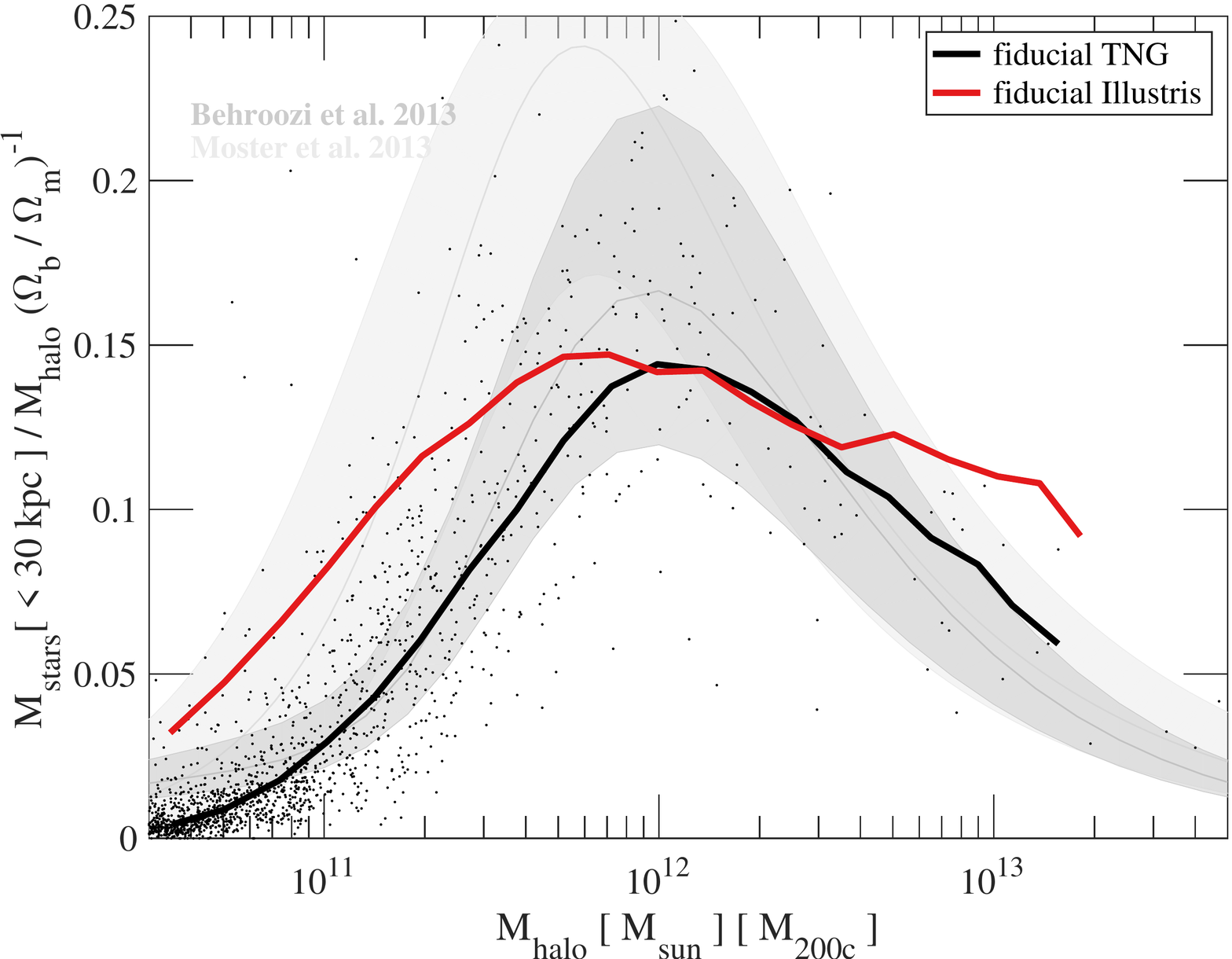}
\includegraphics[width=8.8cm]{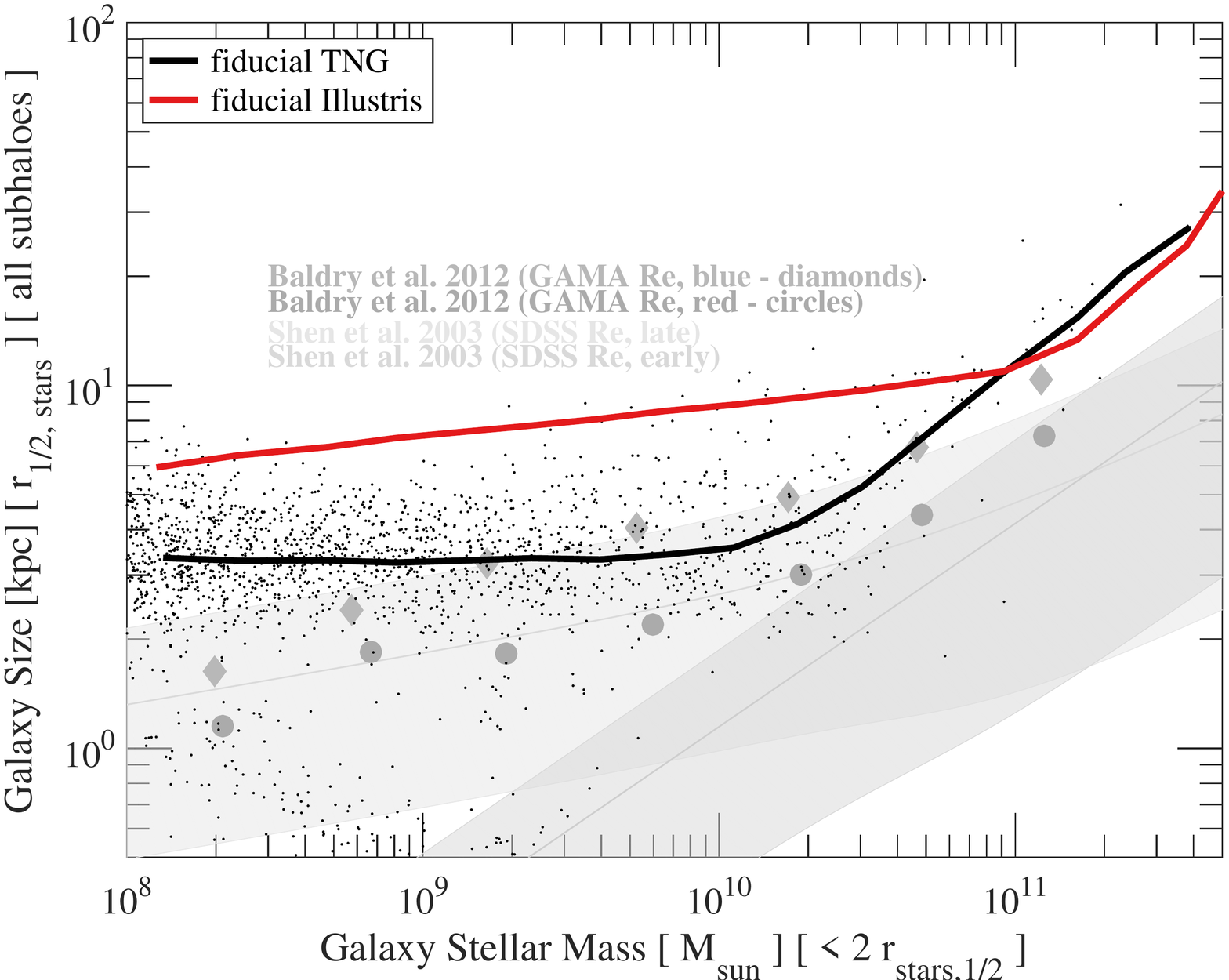}
\caption{Quantitative properties of the fiducial model via galaxy population statistics at $z=0$ (unless otherwise stated). The black line shows the result of the fiducial TNG L25n512 simulation, while the red line shows the original Illustris model outcome on the same volume. We always give running medians (but for the cosmic star-formation rate density as a function of redshift - top left panel). Individual galaxies are shown as data points only for the fiducial TNG run L25n512. When aperture definitions are needed to measure e.g. stellar masses, they are denoted in each panel. Gray curves, shaded areas, and filled large symbols represent observational data or empirical constraints: \citealt{Behroozi:2013,Oesch:2015,Baldry:2008,Baldry:2012,Bernardi:2013,DSouza:2015,Moster:2013,Kormendy:2013,McConnell:2013,Giodini:2009,Lovisari:2015,Shen:2003}. We note that the comparison to observational data is only intended as rough guideline, as we are not applying any observational mock post processing to our simulated galaxies.}
\label{fig:galprop_fiducial}
\end{figure*}

%-------------------------------------------------------------------------------------------------------------------------------------------------------------

\subsection{Fiducial Implementation}
\label{sec:sims_fiducial}

In this Section we show the outcome of our TNG model in its fiducial implementation at our nominal resolution (L25n512). The choices and parameter values which define the TNG fiducial implementation are summarized in Table \ref{tab:illvstng}, where they are shown in comparison to the Illustris fiducial setup. In the following sections we explore the parameter choices and dependencies in depth. Comparisons to the outcome at lower resolution are given and discussed in Appendix \ref{sec_appendix1}.
\begin{figure*}
\centering
\includegraphics[width=15cm]{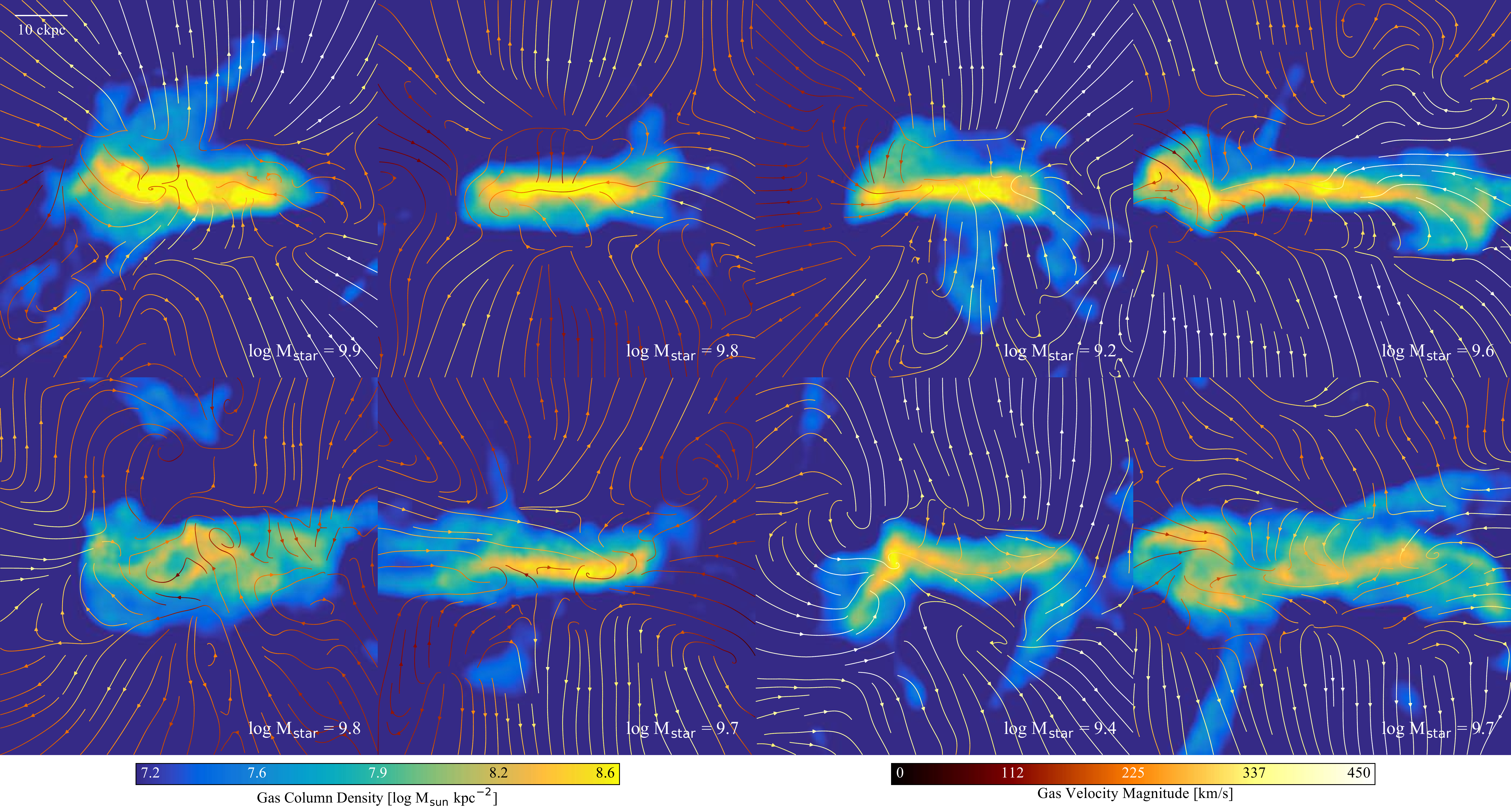}
\caption{Galactic wind morphologies: gas mass density projection with overlaid gas velocity streamlines, with colour indicating velocity. We show a random selection of four galaxies from L25n512 having halo masses $\simeq 10^{11.5}\,\MSUN$. Each is seen edge-on at $z=2$. Top row: gas patterns in the TNG fiducial model where wind particles are launched with random directions from the star forming gas cells (isotropic winds); Bottom row: matching galaxies simulated with the Illustris model, where the directionality of the winds is instead bipolar.}
\label{fig:winds_patterns}
\end{figure*}
A visual presentation of the richness of the model outcome is shown in Figures \ref{fig:L25n512_box_1} and \ref{fig:L25n512_box_2}. In the first, projections of various fields across the whole simulated box $\sim 37$ Mpc on a side are given at $z=0$. The top panels depict DM, gas, and stellar mass density projections. As a demonstration of new information, in the lower panels we show intensity of the magnetic field, the gas Mach number, and the kinetic energy dissipation via shocks. Figure~\ref{fig:L25n512_box_2}, on the other hand, shows a random collection of galaxies, depicted with stellar luminosity projections in stamps of 74\,kpc per side. Broadly speaking, the TNG model continues to reproduce a diverse galaxy population in terms of stellar morphologies and galaxy types, which is a basic requirement for any theoretical model for galaxy formation. 

A more quantitative assessment of the galaxy population demographics is provided in Figure~\ref{fig:galprop_fiducial}, for the Illustris-like resolution box L25n512. Black curves and data points refer to the fiducial TNG implementation. Red curves denote the outcome of the Illustris model, also realized on the same L25n512 box, with the Planck cosmology, and according to prescriptions in the left column of Table \ref{tab:illvstng}. The left panels show observables related to the stellar content of galaxies: 1) the global star-formation rate density as a function of time (SFRD); 2) the current galaxy stellar mass function; and 3) the stellar-to-halo mass relation at $z=0$, in terms of $M_{200c}$. These are the quantities we have used to {\it calibrate} the TNG model, as done by \cite{Vogelsberger:2013b}. The three quantities on the right column are: 1) the BH mass vs stellar or halo mass relation; 2) the halo gas fraction within $R_{500c}$; and 3) the stellar half-mass radii of galaxies. During our implementation of the TNG model, we have additionally kept these three quantities under consideration, in order to discriminate among model realizations which appear similarly acceptable at reproducing the stellar content of galaxies, and in order to solve some of the issues identified in the Illustris simulation.

In these as well as all subsequent galaxy/halo property plots, gray shaded areas and data points represent a selection of currently available observations: we show them here as visual guidelines, not as robust quantitative comparisons. 

Broadly speaking, we observe the following relative changes in TNG with respect to Illustris: (i) the SFRD is lower at late times and the peak shifts to higher redshifts; (ii) the GSMF is significantly suppressed at the low mass end and slightly at the high mass end; (iii) the SMHM relation is lower overall and has a more well defined peak; (iv) the BH to galaxy mass relation is somewhat higher and essentially a powerlaw in the median; (v) the halo gas fractions are lower for low mass halos and higher for high mass halos; and (vi) the galaxy sizes for low mass objects are much smaller and the relation becomes flat for galaxies below $10^{10}\MSUN$. In broad comparison to the observations, keeping in mind the limited volume and the tensions identified for the Illustris model in the full Illustris volume \citep{Nelson:2015b}, we find that: (i) the $z=0$ SFRD is improved, while the $z=2$ SFRD may be slightly low; (ii) the GSMF at the low and high mass ends is much improved, and essentially unchanged at the knee; (iii) the SMHM shape and normalization is improved, although the value at the peak decreases slightly; (iv) black hole masses are larger at fixed stellar mass and the relation between black hole and galaxy masses is a power law over a larger mass range, although we do not here model bulge masses in any way \citep[for the differences resulting in modeling the bulge see][]{Volonteri:2016, Weinberger:2017}; (v) the halo gas fractions at the group scale are in much better agreement; and (vi) the low-mass galaxy sizes are also now well within the rough observational constraints.

Note that in deciding the TNG model parameters we have not performed a parameter search via fitting of the simulated outcome to such observational curves. Instead, the approach we have adopted to calibrate the TNG model was intentionally simple, primarily because several issues would complicate any actual fine tuning methodology. First, it has been demonstrated that in order to consistently compare simulated to observed astronomical data, the simulated raw data of synthetic galaxies must be transformed into realistic mock observations \citep{Overzier:2013,Torrey:2015, Snyder:2015, ZuHone:2016, Camps:2016, Bottrell:2017a, Bottrell:2017b}. Second, as demonstrated by \cite{Genel:2014} and mentioned in Section~\ref{sec:sims_setup}, the sample variance of the galaxy population due to the limited volume of a simulation test box is not negligible and could potentially offset the result of a rigorous fine tuning to the observational data. Finally, the outcome of our method is {\it not} fully resolution independent at the resolutions reached here (see Appendix \ref{sec_appendix1}). All these considerations, together with the computational costs of simulating large volumes at fixed resolution, make it conceptually challenging and practically prohibitive to identify the best model parameter values via a fit to a selection of galaxy observables. 

Instead, we have proceeded as follows. First, we have chosen the most striking observational tensions with observations identified in the Illustris simulation (see Introduction). We have then made educated guesses as to how e.g. the velocity and energy wind scaling needed to be modified in order to alleviate such tensions (see Section~\ref{sec:tng_winds}). Next, we have run the same test cosmological volumes with the fiducial Illustris model and with a series of perturbation of the TNG model. Finally, we have chosen an implementation and parameter set that simultaneously alleviates the largest number of targeted tensions while demonstrating good overall agreement with benchmark observables.

%-------------------------------------------------------------------------------------------------------------------------------------------------------------

\subsubsection{Characteristics of the galactic winds}
\label{sec:sims_fiducial_winds}

As described in Section~\ref{sec:tng_winds} the TNG wind particles are launched isotropically. As we show in Figure~\ref{fig:winds_patterns} for a selection of galaxies extracted from the fiducial realization at $z=2$ of the L25n512 box, this in practice still corresponds to non spherically-symmetric gas patterns around disk-like galaxies. The wind mass, although launched from the star forming gas with random orientations, still propagates along the trajectories of least resistance. Large-scale galactic fountains naturally emerge both with isotropic winds (TNG fiducial choice, top panels) as well as with bipolar winds (Illustris fiducial choice, bottom panels). In fact, irregular gas flow patterns can also be found in both cases, especially in regimes where outflows also depend on the central BH activity.

\begin{figure*}
\centering
\includegraphics[width=8.7cm]{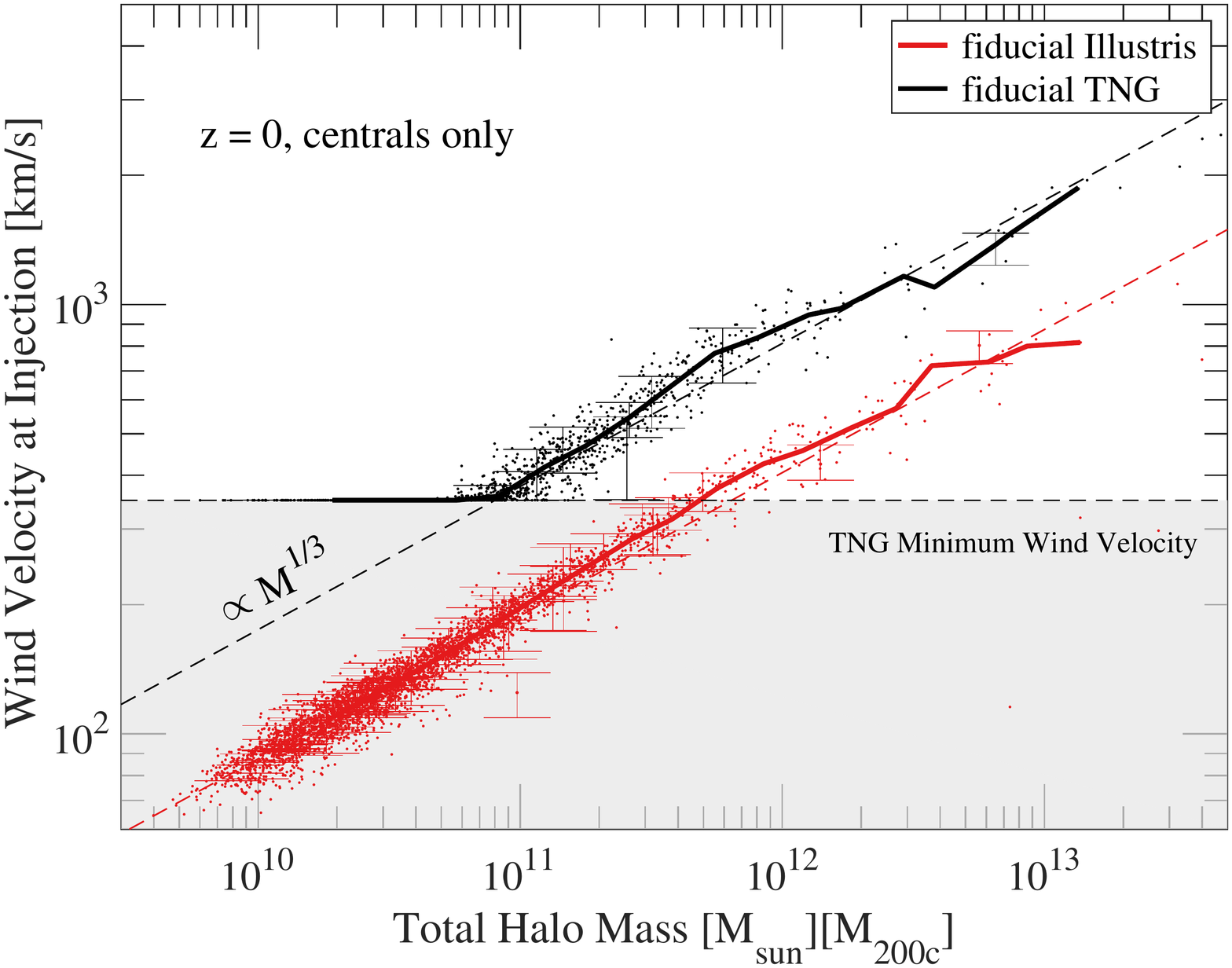}
\includegraphics[width=8.7cm]{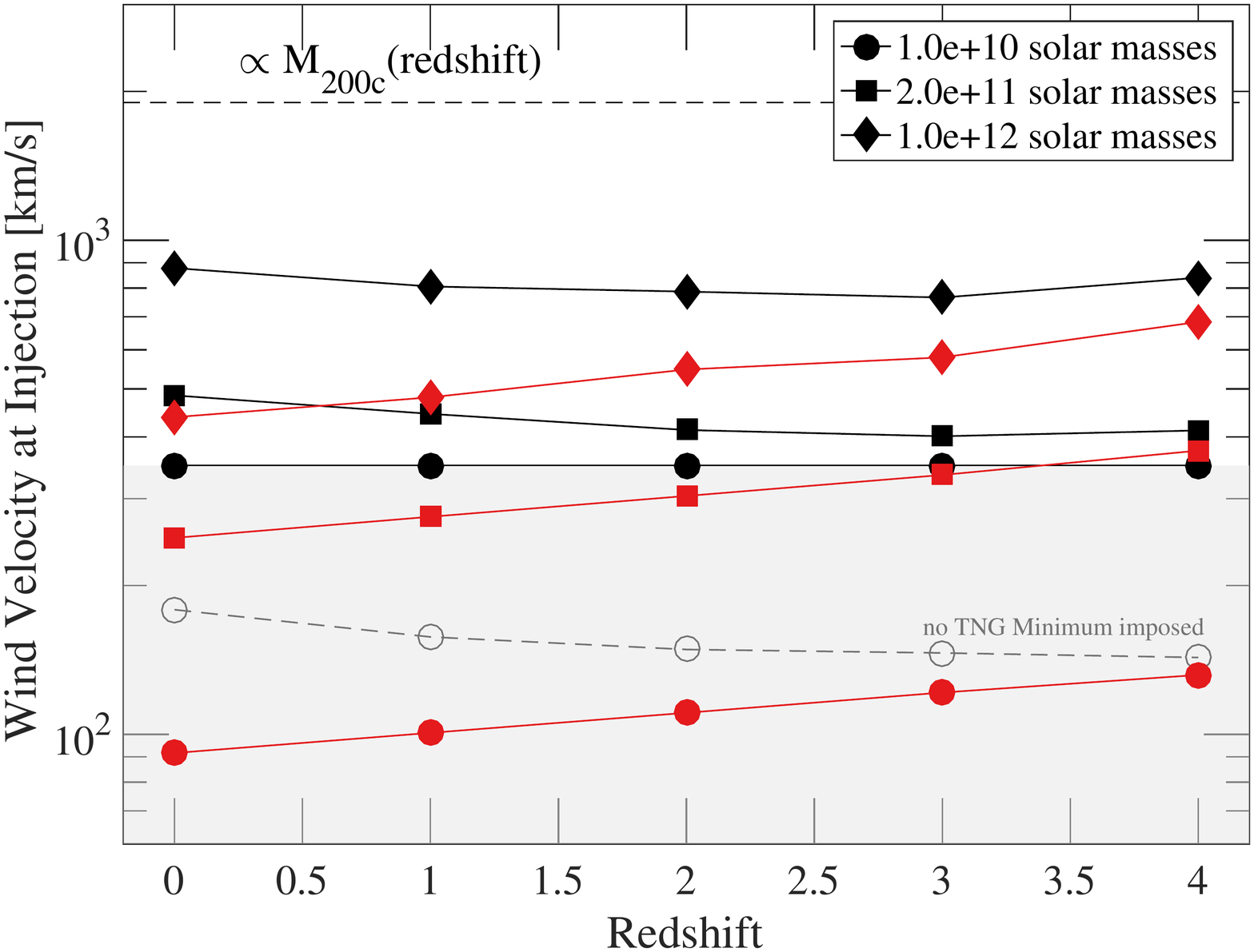}
\caption{Characteristics of the galactic winds at injection: initial velocities of wind particles at launch, as a function of halo mass at $z=0$ (left) and as a function of redshift in bins of halo mass (right). The black (red) color shows the fiducial TNG (Illustris) model performed on the L25n512 box. In the left panel, errorbars denote the 5th and 95th percentiles of the wind velocity distribution {\it within} individual galaxies. 
%Black symbols are shown {\it without} having imposed the minimum velocity constraint, to highlight its effects. 
The solid curves are running medians as a function of halo mass of the galaxy average wind velocities. TNG winds are faster than Illustris winds at injection (at all $z\lesssim 4-5$): $\kappa_w$ = 7.4 in TNG vs 3.6 in Illustris, in addition to the minimum velocity floor. In the right panel, symbols show the average wind velocities as a function of time, averaged across galaxies in three bins of total halo mass (within a factor of 2 around $10^{10}$, $2\times10^{11}$, and $10^{12} \MSUN$). The TNG velocity parameters of Eq.~(\ref{eq:winds_vel}) are chosen such that they match the Illustris model at $z\sim5$. The Hubble scaling forces the evolution of the wind velocities in haloes to be similar to the evolution of halo mass (black curves are essentially horizontal). The TNG velocity floor raises the wind velocity of galaxies in haloes $\lesssim 10^{11}\MSUN$ essentially across all times. Note that these wind velocities do {\it not} equal the velocity of actual gas outflows at larger radius, and therefore {\it cannot} be compared directly to  observations.}
\label{fig:winds_vel}
\end{figure*}

The wind velocity in the TNG model is determined by the local one-dimensional DM velocity dispersion with a normalization factor $\kappa_w = 7.4$ and the velocity floor imposed at $v_{w, min} = 350 {\rm km\,s^{-1}}$. The wind velocity is modulated in time via the Hubble factor, and the choice of $\kappa_w = 7.4$ corresponds to a velocity multiplicative factor of about $\kappa_w = 3.7$ at $z\sim5$, which is the Illustris fiducial choice across time (see Table \ref{tab:illvstng}). In Figure~\ref{fig:winds_vel} (left panel) we show the ranges of velocity values at which the wind particles are ejected at $z=0$ as a function of total halo mass. Each data point gives the average velocity of wind particles at injection for a single galaxy, excluding satellites. Error bars denote the 5th and 95th percentiles of the wind velocity distribution {\it within} a random selection of individual galaxies. Solid curves show running medians as a function of halo mass. Black symbols and curves represent the TNG model for the L25n512 run, red symbols the same measurements for the Illustris model. 

First, we see that in both models, galactic winds within a given galaxy launch with a range of velocities, which can span a couple of hundreds ${\rm km\,s^{-1}}$ for MW-like galaxies at $z=0$ (5th - 95th percentiles of the velocity distribution). The scatter within galaxies is similar to the scatter around the average galaxy wind velocity, however it can be much larger in the case of non central galaxies (not shown). Secondly, the average wind velocity at fixed redshift scales roughly like a power of the halo mass, $v_w \propto$ (halo mass)$^{1/3}$, as for Illustris and above halo masses of $\times 10^{11}\MSUN$. 
Dashed thin lines in Figure~\ref{fig:winds_vel} give best fits (fixed power factor $=1/3$) to the galaxy average wind properties, 
\begin{equation}
v_w = \kappa_w \times 110 ~{\rm km\,s^{-1}} ~ \left( \frac{M_{200c}}{10^{12}\MSUN}\right)^{1/3},
\label{eq:wind_vel_scaling}
\end{equation}
\noindent where $110~ {\rm km\,s^{-1}}$ is approximately the ambient 1D velocity dispersion around {\it star forming} gas in Milky Way-like haloes at $z=0$.
%Thirdly, the minimum wind velocity floor affects only the slow tail of the wind particle distributions at $z=0$ and overall the typical wind speed of low-mass galaxies at higher redshifts. While the average DM velocity dispersion in haloes is larger at higher redshifts, the redshift-dependent wind-speed-scaling ($\kappa_w (H_0/H(z))^{1/3}$) is smaller prior to $z=0$. The combination of the two effects makes the importance of the minimum velocity floor at higher redshifts greater than what can be seen directly in left panel of Figure~\ref{fig:winds_vel}, at least for low mass galaxies. In fact, at all redshifts, the average wind velocity at injection flattens towards lower mass. This is because it depends on the velocity dispersion of material around {\it star-forming} gas, and the equation of state pressure provides a floor to the local pressure and temperature regardless of the virial scaling of the velocity dispersion. 
Thirdly, the minimum wind velocity floor affects mostly the typical wind speed of low-mass galaxies, both at $z=0$ and at higher redshifts. While the average DM velocity dispersion in haloes is larger at higher redshifts, the redshift-dependent wind-speed-scaling ($\kappa_w (H_0/H(z))^{1/3}$) is smaller prior to $z=0$. The combination of the two effects makes the importance of the minimum velocity floor at higher redshifts greater than what can be seen directly in left panel of Figure~\ref{fig:winds_vel}, at least for low mass galaxies. 
%In fact, at all redshifts, the average wind velocity at injection flattens towards lower mass. This is because it depends on the velocity dispersion of material around {\it star-forming} gas, and the equation of state pressure provides a floor to the local pressure and temperature regardless of the virial scaling of the velocity dispersion. 
%
\begin{figure*}
\centering
\includegraphics[width=8.7cm]{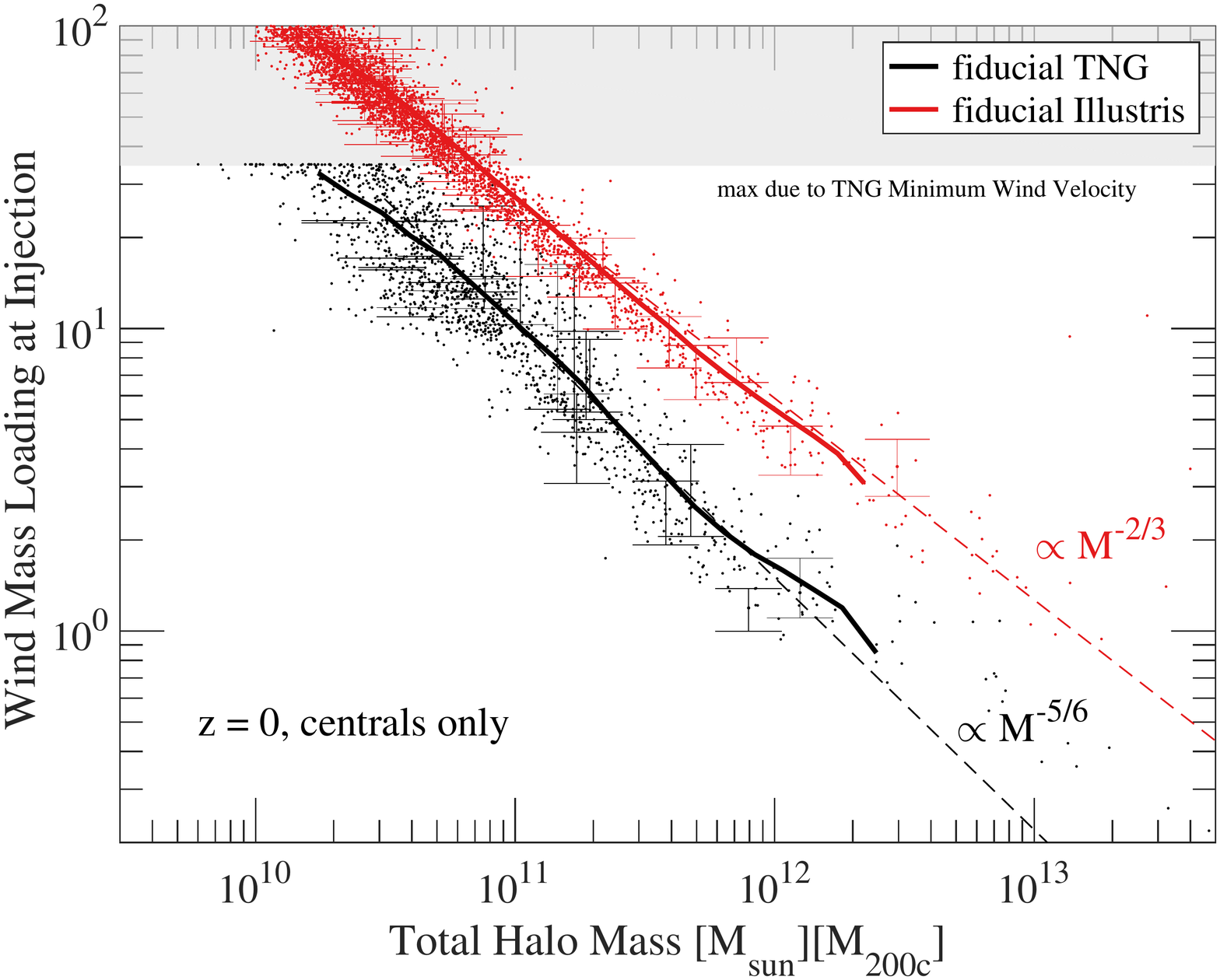}
\includegraphics[width=8.7cm]{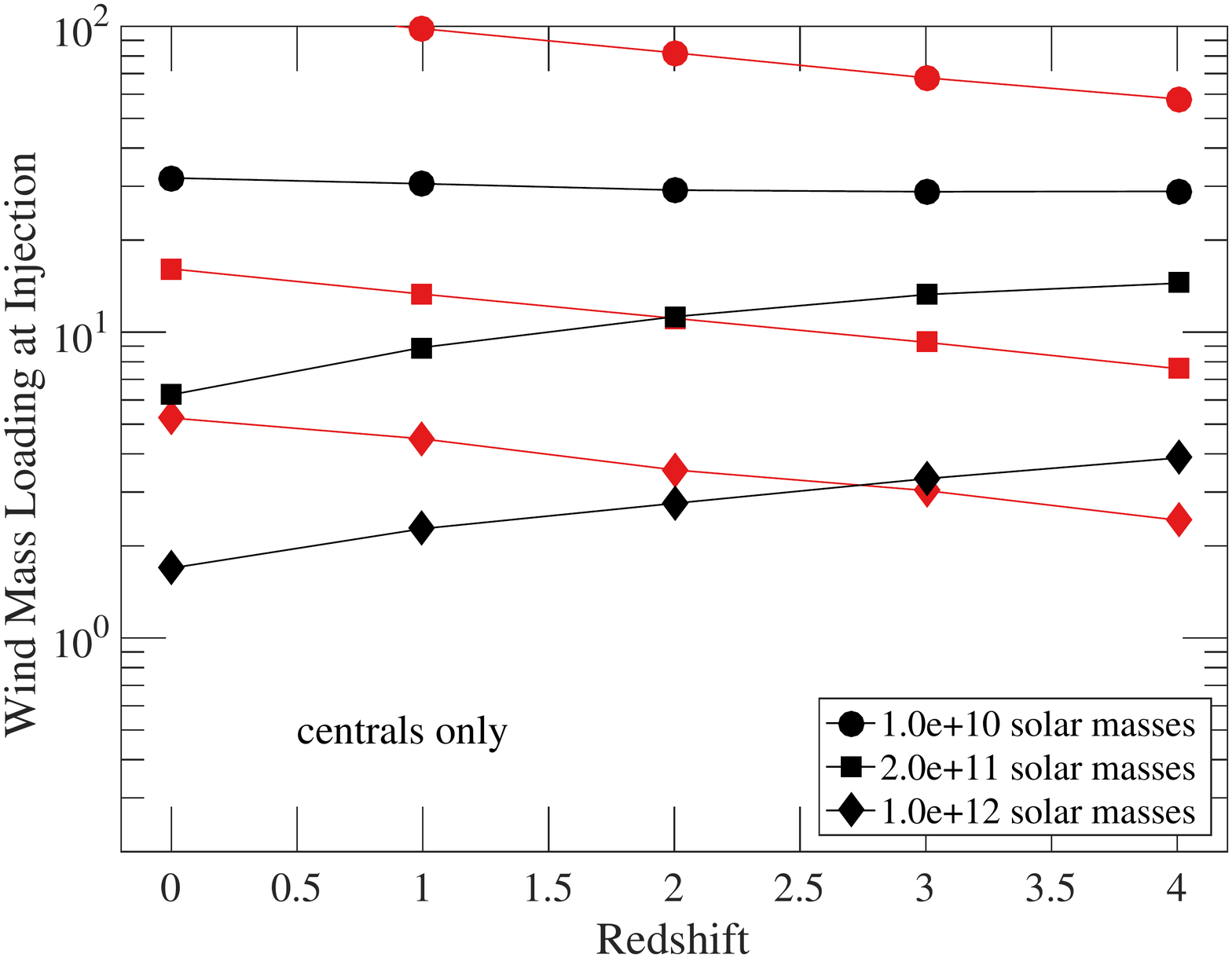}
\caption{Characteristics of the galactic winds at injection: effective wind mass loading factors in the TNG (black) and Illustris (red) fiducial models as a function of halo mass at $z=0$ (left) and as a function of redshift in bins of halo mass (right). Symbols are as in Figure~\ref{fig:winds_vel}. The wind mass loading factor is generally defined as $\eta = \dot{M}_{\rm out}/{\rm SFR}$, i.e. it is the ratio between the outward flux of gas and the galaxy star formation rate. Here instead we show the effective mass loadings of the wind particles at injection as defined in Eq. (\ref{eq:winds_eta}), but without accounting for the fraction of wind thermal energy: in TNG, the thermal fraction $\tau_w$ is 10 per cent (in Illustris winds are cold), and therefore the black dots are in fact 10 per cent higher than the actual TNG fiducial case. The plotted values do {\it not} correspond to a measurable mass loading factor at any larger radius, and therefore {\it cannot} be compared directly to observations.}
\label{fig:winds_eta}
\end{figure*}

From the right panel of Figure~\ref{fig:winds_vel}, we see that in the new TNG scheme winds are faster than in Illustris at all times up to $z\sim5$ and at all halo masses. The mean galaxy wind velocity, averaged across galaxies in the same halo mass bin (large symbols), exhibits by construction a trend with redshift which is similar to the time evolution of the virial halo mass -- that is, the black curves are more or less flat, and this makes the TNG winds faster than Illustris at all times. The imposed wind velocity floor results in wind velocities at injection of 350 ${\rm km\,s^{-1}}$ for haloes
smaller than $\sim 10^{11}\MSUN$ at essentially all redshifts.

At $z=0$, the effective mass loading factor of the winds in the TNG model is lower than in Illustris. Figure~\ref{fig:winds_eta}, left panel, shows the measurement of the wind mass loading as per Eq. (\ref{eq:winds_eta}), averaged across the star-forming gas cells of galaxies as a function of halo mass, but assuming cold winds. In the TNG model, however, winds are given a 10 per cent thermal energy content, such that the actual TNG wind mass loading is lower by an additional ten per cent than the black symbols reported in Figure~\ref{fig:winds_eta}. We have chosen the parameters for the wind energy, \egyw in Eq. \ref{eq:wind_energy}, and metallicity-dependent scaling from Eq. \ref{eq:winds_eta} so that the wind energy in galaxies with average Solar gas metallicity is within 20 per cent of the typical wind energy factor of Illustris winds in similar galaxies (see Figure~\ref{fig:galprop_winds_Z} for clarity). However, the larger TNG wind velocity factor $\kappa_w$ makes the effective TNG wind mass loading lower at equal wind energy, while the metallicity dependence introduces a modulation with halo mass. Taken together, these choices result in a slightly steeper trend of the effective wind mass loading at $z=0$ as a function of halo mass in TNG than in Illustris. The minimum wind velocity imposes a maximum to the wind mass loading at injection that at $z=0$ occurs at about $\eta$ = 30-40. At $z=0$, dashed thin lines denote best fits to the Illustris and TNG
wind loading at injection (two-parameters fit): 
\[
\eta_w \simeq 5.9 ~ \left( \frac{M_{200c}}{10^{12}\MSUN}\right)^{-2/3} \rm{for ~ Illustris}
\]
\[
\eta_w \simeq 1.5 ~ \left( \frac{M_{200c}}{10^{12}\MSUN}\right)^{-5/6} \rm{for~  TNG}.
\]

The redshift evolution at fixed halo mass (Figure~\ref{fig:winds_eta}, right panel) is such that for haloes below about a few $10^{10}\MSUN$, the TNG effective wind loading are lower than in Illustris at essentially all redshifts, even though winds in those haloes are faster. For galaxies residing in more massive haloes, TNG winds are weaker at lower redshifts and stronger at earlier times, with a transition depending on halo mass.

We emphasize that the input choices for the wind particles at injection cannot easily be mapped into properties of gas outflows around galaxies. The latter are the ones to be adopted as a natural test bed of the model against observational constraints, not the characteristics of the wind particles once they are spawned from the star-forming gas from the innermost regions of galaxies. Furthermore, even deriving the effective wind properties at injection corresponding to a given model is not straightforward, and can further lead to ill-posed comparisons.
%\footnote{For example, Figure 3 of \cite{Zahid:2014} is inconsistent with the mass loading we present here for the Illustris model. This is due primarily to an assumed relationship between dark matter velocity dispersion and other halo properties, which should instead be taken directly from the simulation around star-forming gas. Figure 10 of \cite{Schroetter:2016} suffers from the same issue, although they have corrected for the factor of 3 parameter error in \egyw noted in the erratum of \cite{Vogelsberger:erratum}.}. 
Effective wind properties should instead be directly measured in the simulation, as we have done here. We postpone to future work the task of measuring the physical state and properties of gas outflows in order to make a robust connection to observations \citep[e.g.][for a review]{Veilleux:2005} and other simulated galaxies based on different implementations of the stellar feedback \citep[e.g.][]{Muratov:2015, Christensen:2016}.

\begin{figure*}
\centering
\includegraphics[width=8.3cm]{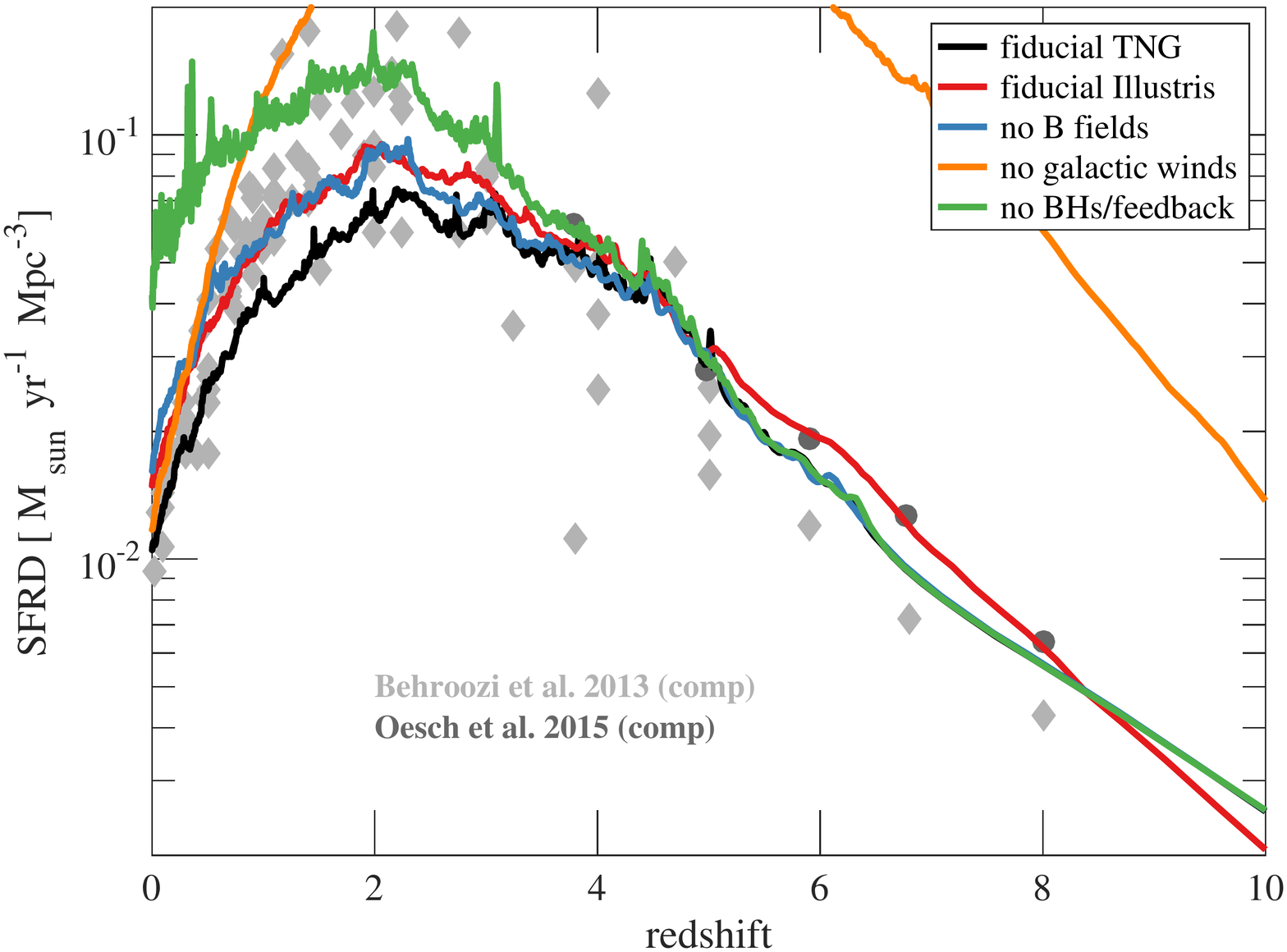}
\includegraphics[width=8.3cm]{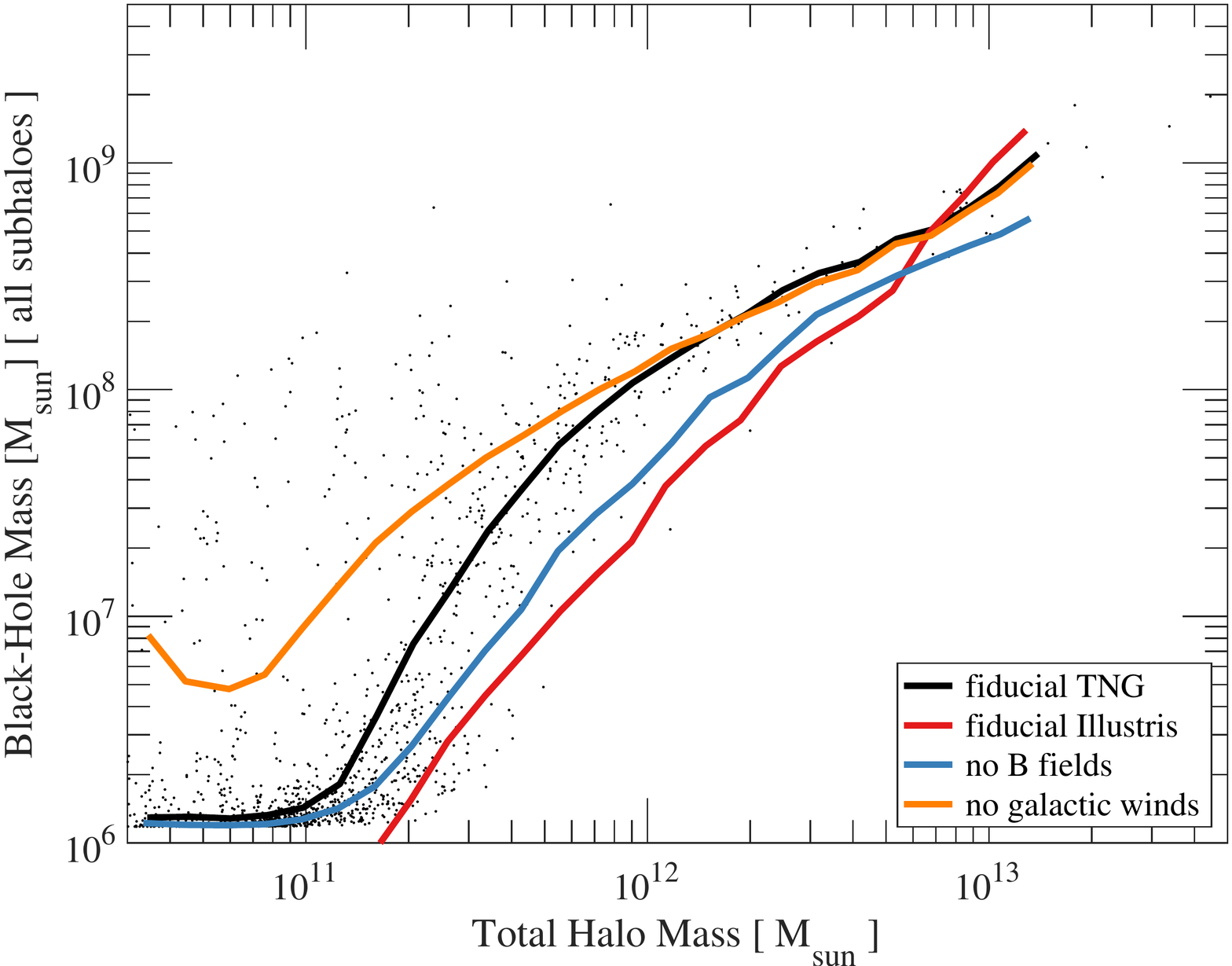}
\includegraphics[width=8.3cm]{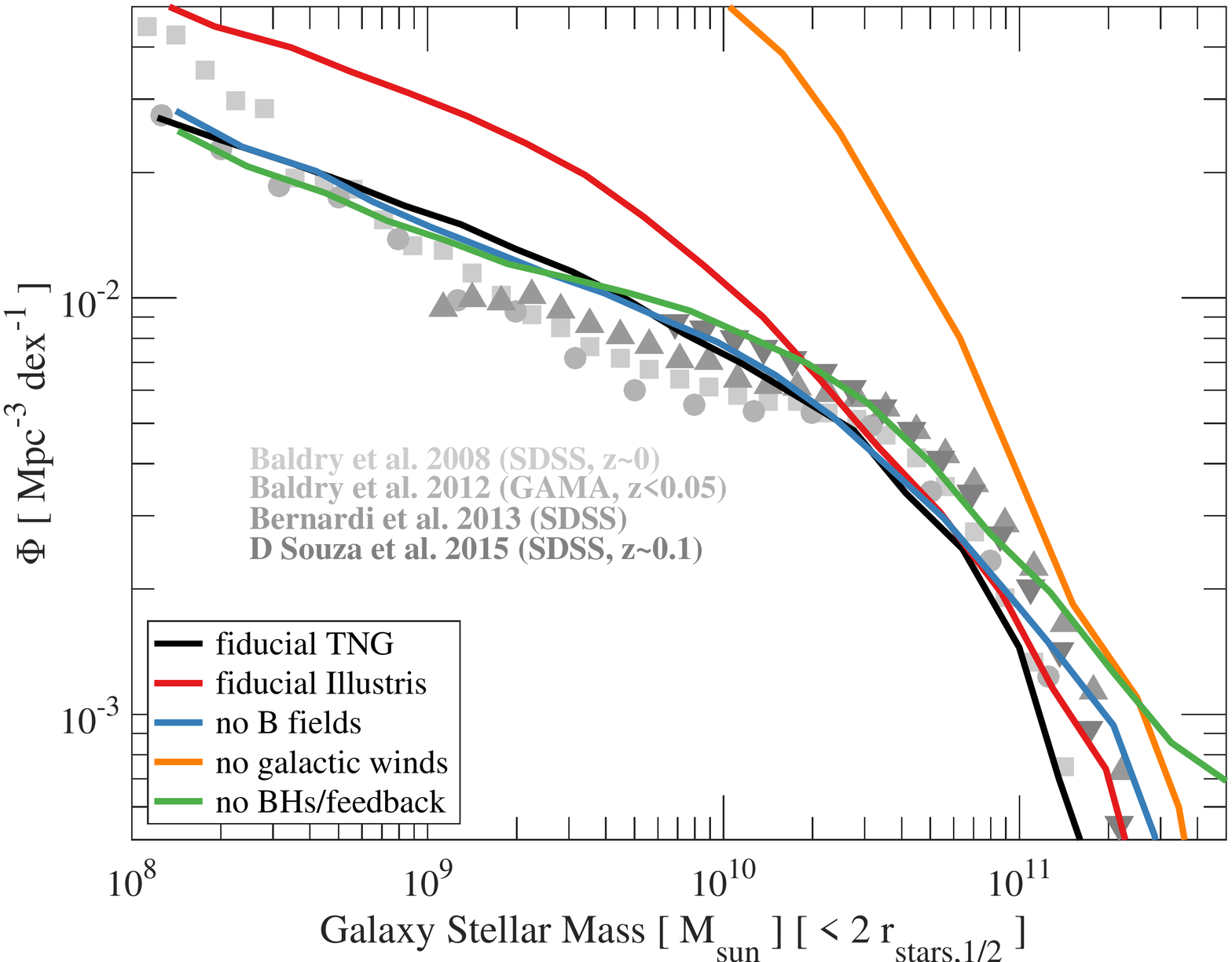}
\includegraphics[width=8.3cm]{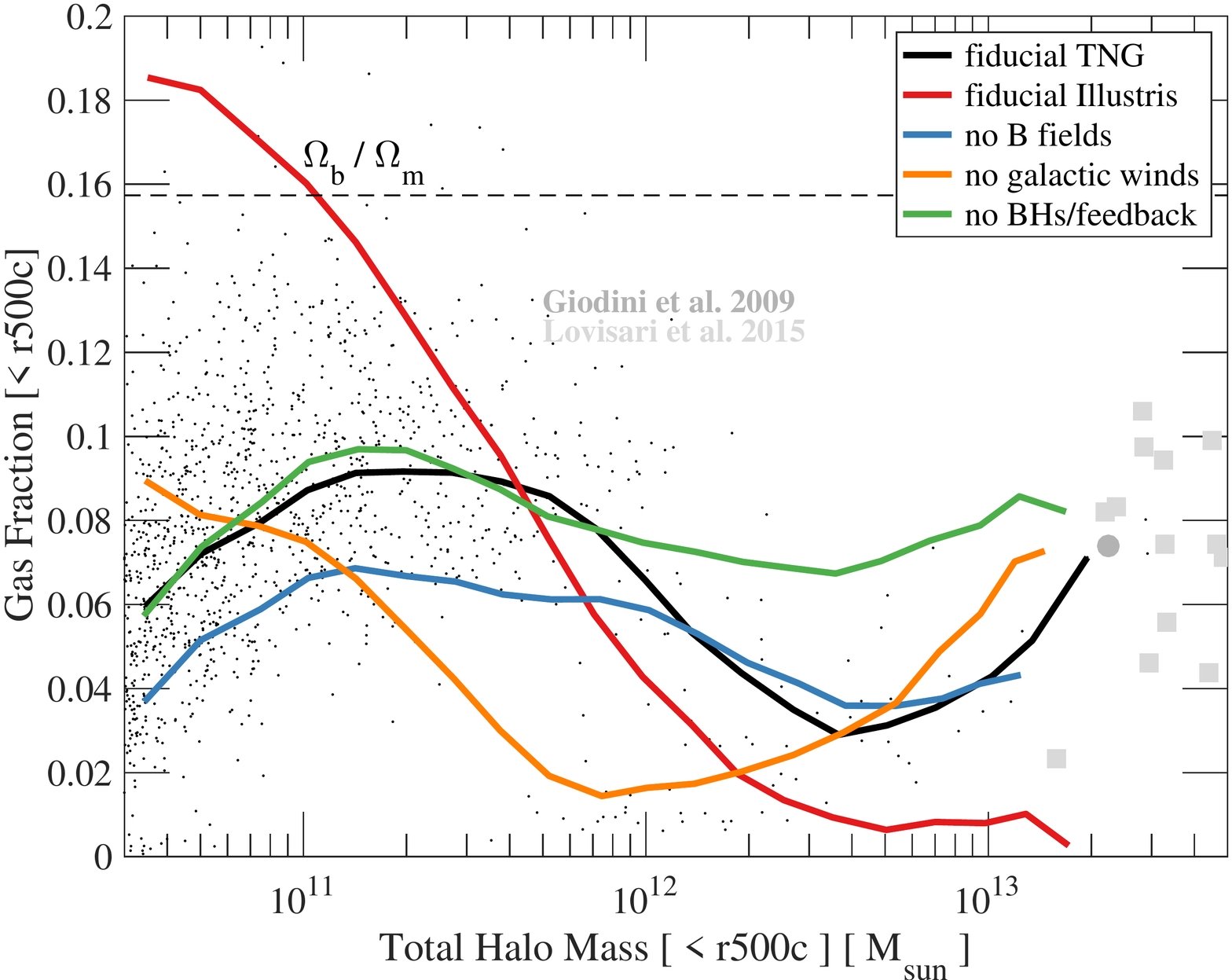}
\includegraphics[width=8.3cm]{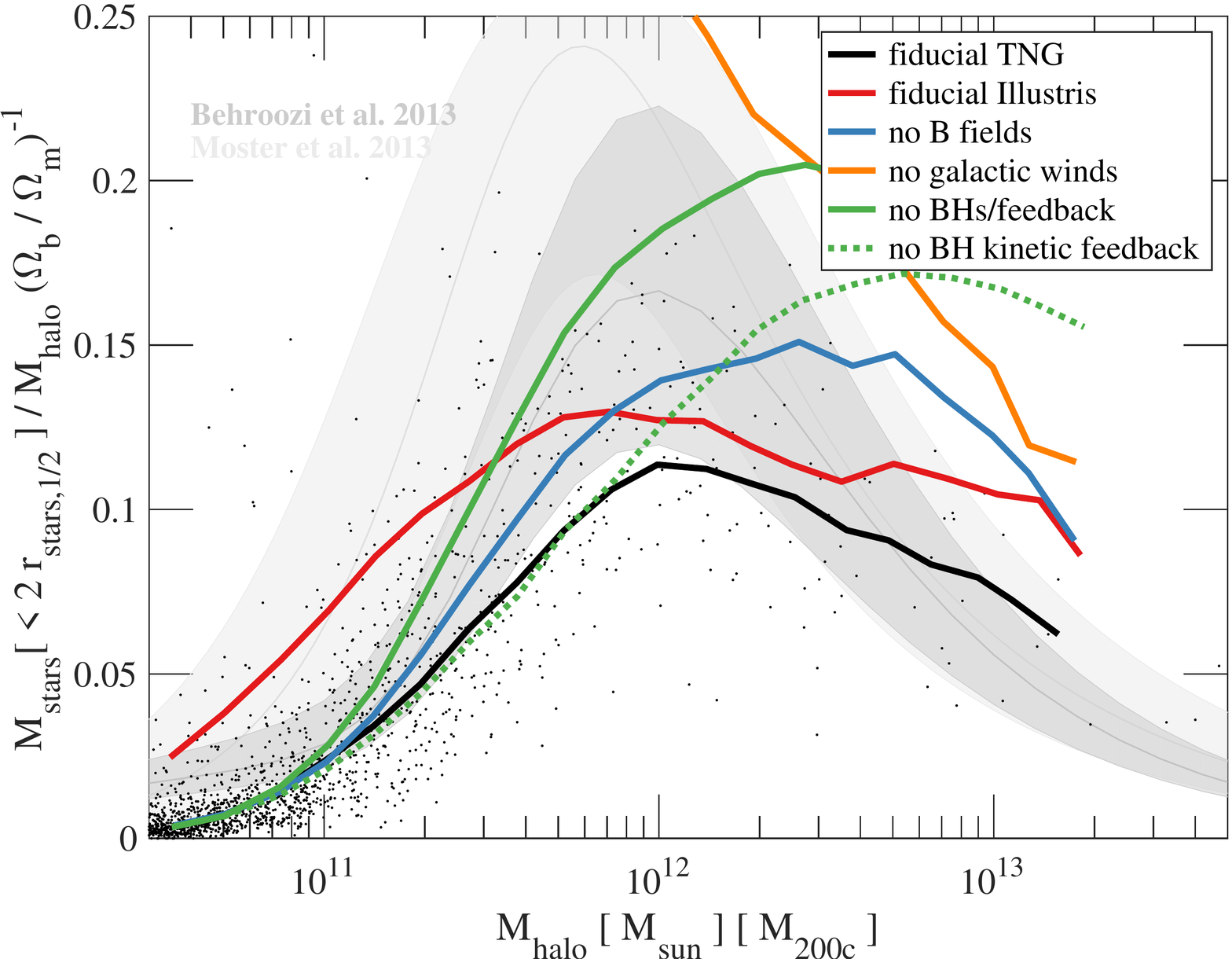}
\includegraphics[width=8.3cm]{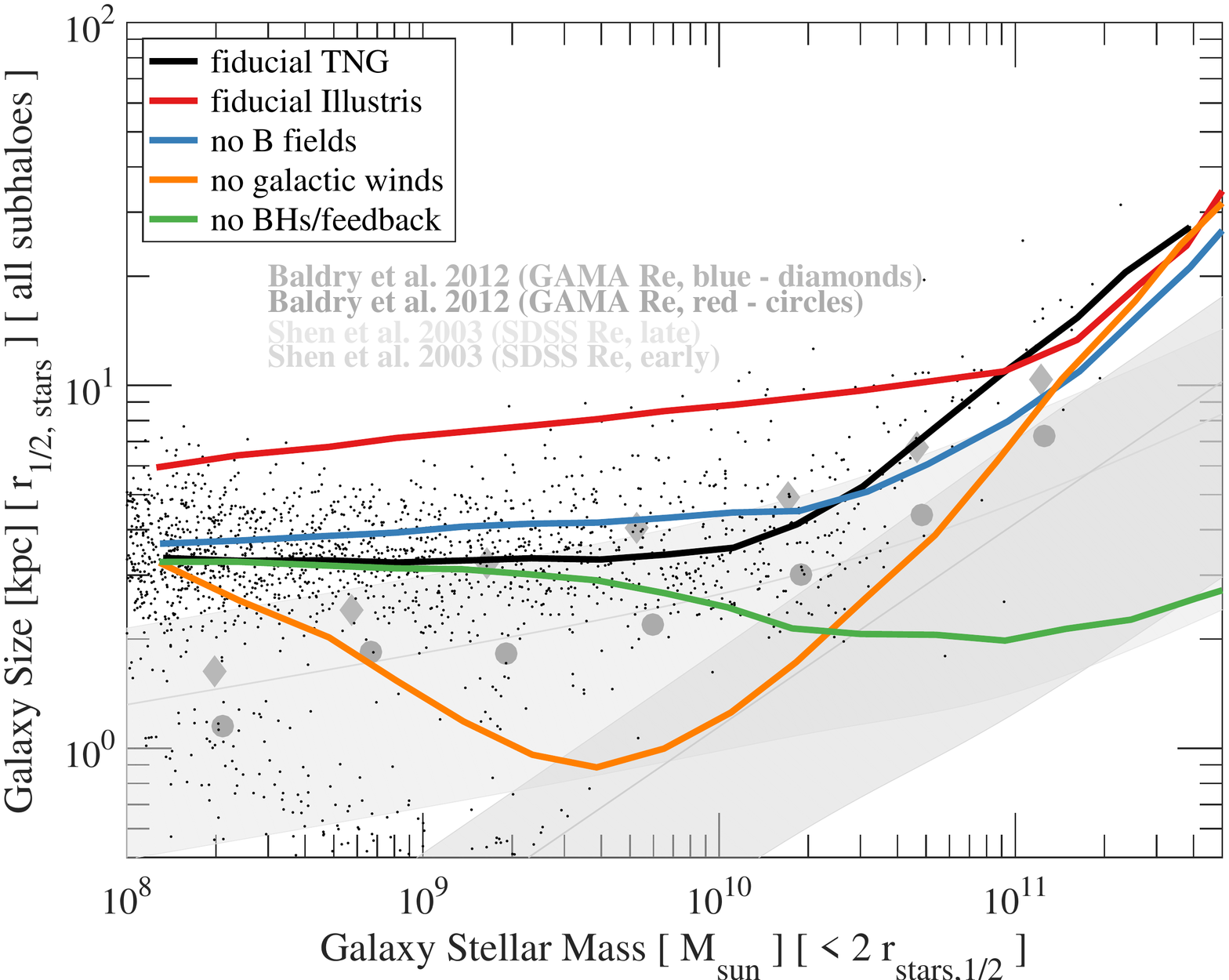}
\caption{The galaxy population at L25n512 resolution for different choices in the galaxy physics model. The labels ``no B fields'' (blue), ``no galactic winds'' (orange), and ``no BHs/feedback'' (green) all refer to runs identical to the TNG fiducial setup but for the absence of the indicated mechanism. The red curves denote the Illustris model outcome in the same cosmological volume and with the same cosmology as TNG fiducial. For reference, and only in the bottom left panel, the dotted green curve shows the outcome of a run where only the low-accretion BH feedback is turned off (dotted green). Symbols and annotations are as in Figure~\ref{fig:galprop_fiducial}, but stellar masses are measured within twice the stellar half-mass radius in every panel.}
\label{fig:galprop_misc}
\end{figure*}

%-------------------------------------------------------------------------------------------------------------------------------------------------------------

\subsection{The model at various levels of complexity}

Before quantifying the sensitivity of the TNG model to its most important parameter values, we give an overview of the effects of the key model ingredients: the presence or absence of magnetic fields, galactic winds, and BH feedback.  The galaxy properties are shown in Figure~\ref{fig:galprop_misc}. The labels ``no B fields'' (blue), ``no galactic winds'' (orange), and ``no BHs/feedback'' (green) all refer to runs identical to the TNG fiducial setup but for the absence of the indicated mechanism.\footnote{In the case of no galactic winds, we also switch off metal-line cooling to avoid runaway star formation: including the metal-line cooling contribution does not alter the qualitative statements we provide.} Black and red curves refer to the fiducial TNG and Illustris runs, as from Figure~\ref{fig:galprop_fiducial}. For reference, and in the bottom left panel only, we also provide the outcome of a run where the low-accretion BH-driven wind is switched off (green dotted curve), i.e. when only the thermal injection feedback around the BH occurs.

The case without magnetic fields is implemented by adopting a vanishing initial magnetic field seed.\footnote{We have checked, but do not show, that the results are unchanged if the code is run disabling MHD functionality altogether. In this case the exact Riemann solver (instead of the HLLD solver) is used, as adopted in the Illustris model.} Noticeably, magnetic fields clearly have an effect on haloes $\gtrsim 10^{12}\MSUN$ at $z=0$ and in general at low redshifts ($z\lesssim 3$). This is not in contradiction with the findings of \cite{Pakmor:2013,Pakmor:2014,Pakmor:2017} who have not detected a significant effect in their sample of MW-like galaxies realized with a slightly different galaxy physics model. While the effect on this halo mass scale is small, the large statistical sample available for our TNG model shows that the presence of cosmic magnetic fields suppresses the total star formation at low redshifts (top left panel), as well as for large halo masses at the current epoch (middle and lower left panels).
%This is not strictly in contradiction with the findings of \cite{Pakmor:2013,Pakmor:2014,Pakmor:2017} who have not detected a significant effect and who have inspected only a handful of MW-like galaxies realized with a slightly different galaxy physics model. 
%It does not contradict the quantitative findings of \cite{Marinacci:2016} or \cite{Pakmor:2017} either, as there no results are given with null magnetic field seed and in both cases different subgrid implementations are adopted, especially for the BH feedback (see Discussion). 
%In the TNG model, our large statistical sample shows that the presence of cosmic magnetic fields suppresses the total star formation at low redshifts (top left panel), as well as for large halo masses at the current epoch (middle and lower left panels), by relative amounts that can be larger than the differences between the fiducial Illustris and TNG realizations. 
Including MHD increases the stellar sizes for galaxies above $2 \times 10^{10}\MSUN$, although this is driven by the lower stellar masses in the TNG model at fixed halo mass. On the other hand, the presence of the magnetic fields affects the total gas content within low mass haloes, raising the gas fraction below $M_{\rm halo}~\sim 10^{12}\MSUN$.

By comparing to the runs {\it without} galactic winds and BH feedback (orange and green lines, respectively), it is clear that the galactic winds have a significant impact on stellar mass content across all halo mass scales we probe here, and at all times. The effects of the BH feedback are noticeable in the stellar masses at the current epoch above a threshold halo mass scale of roughly $1-2 \times 10^{11}\MSUN$: the suppression of the central stellar mass due to the BH feedback as a whole reaches a factor larger than two for haloes with a total mass of a few $10^{12}\MSUN$. Beyond $2\times 10^{12}\MSUN$ this suppression at $z=0$ is predominantly due to the effects of the BH-driven wind feedback in the low-accretion states (solid green vs dotted green vs black curves, bottom left panel), in agreement with what already shown by \cite{Weinberger:2017}. In the TNG fiducial model, the BH-driven wind has no effect on the $z=0$ stellar fractions below the peak of SF efficiency ( $M_{\rm halo} \sim 10^{12}~ \MSUN$ ), whereas this is not the case for the BH high-accretion thermal feedback mode, which continues to play a role down to $\sim 10^{11} \MSUN$.
In order to reconcile the stellar content of simulated galaxies with observational constraints, it has been commonly proposed that stellar feedback is needed below $L^*$ galaxies, while feedback from the central SMBH is needed for more massive haloes \citep[e.g.][]{Croton:2006}. In our model, the ultimate stellar content of galaxies depends on a {\it combination} of the two feedback mechanisms, not only at $L^*$ but also towards both lower and higher masses. As we discuss later, the mass scales over which stellar feedback dominates BH feedback (and low-accretion BH feedback dominates high-accretion BH feedback) are model dependent, and indeed differ between the Illustris and TNG models. In the TNG model, stellar feedback is responsible for regulating the global star formation rate density down to the $z\sim 2-3$ peak; and only at redshifts $z \lesssim 4$ does BH feedback lead to a noticeable reduction of the global SFRD. On the other hand, BH feedback is needed to establish an appropriately peaked and non-monotonic stellar-to-halo mass relation, as well as to shape the knee in the galaxy stellar mass function. The location of such a peak and knee, and how pronounced they in turn appear, depends on specific choices for the combined model parameters. 

In the top right panel of Figure~\ref{fig:galprop_misc} we see that the presence of galactic winds slows down the growth of black holes in halos below $10^{12}\MSUN$. We show the relation as a function of halo mass in order to exclude the effects of the large stellar mass discrepancies across models. A somewhat similar effect has been seen in the Eagle model \citep{Bower:2017} where it was attributed to the removal of gas from small galaxies due to stellar feedback. However, in our model, a steep transition to the median BH-halo mass relation is still present, and it is simply shifted to lower masses. Conversely, the presence of magnetic fields generally leads to more massive black holes for a given halo mass. We reiterate that in the TNG model we modify the Bondi-Hoyle-Lyttleton formula for the gas accretion onto the BH in the presence of magnetic fields. Specifically, the accretion rate inversely depends on an effective gas sound speed that takes into account both thermal and magnetic signal propagation, the latter expressed in terms of the Alfv\'en speed (see Section 2.1 of \citealt{Weinberger:2017}). We note, but do not show here, that removing this dependence of the BH accretion on the magnetic Alfv\'en speed also reduces the {\it growth lag} of BHs in haloes around $10^{11}\MSUN$, similarly to the removal of galactic winds. This is the only other perturbation to the model we find to exhibit such an effect, although it is very small and does not correspond to any appreciable change in stellar masses.

The trend of gas mass fraction within $R_{500c}$ as a function of halo mass strongly depends on the presence and characteristics of all feedback mechanisms (center right panel).  In the TNG model, the AGN feedback is responsible for setting up a transition just below $10^{12}\MSUN$, above which the gas fraction suddenly drops, before rising again towards group and cluster-size objects. The Illustris and Eagle models exhibit very different trends of the gas fraction within haloes, the former producing a non-monotonic function at masses $\gtrsim 10^{11}\MSUN$ (red curve in Figure~\ref{fig:galprop_misc}; also \citealt{Genel:2014}) and the latter having a monotonically rising gas fraction across the whole mass range between $10^{11}\MSUN$ and a few $\times 10^{14}\MSUN$ (Figure 3 of \citealt{Schaller:2015}). %The precise location of this gas-fraction transition mass depends on the combination of involved processes and choices for the model parameters. 
Given the strikingly different trends for the different model perturbations, we argue that the amount of gas mass (cold and hot, ionized and neutral) retained within the virial radius ($R_{\rm 500c}$) of haloes as a function of halo mass could be one of the most powerful observational constraints to discriminate among competing galaxy formation
feedback models that appear degenerate with respect to approximately recovering the observed galaxy stellar mass function. Altering the gas fraction in group and cluster-size haloes was one of the main motivations to modify the AGN feedback mechanism from Illustris (red
curve) to TNG (black curve). However, an even more useful constraint may be the gas fraction of MW-mass haloes and below.

Finally, the galaxy sizes reveal a clear and intriguing phenomenon (lower right panel). Galactic winds predominantly determine the spatial extent of galaxies in haloes below $10^{12}\MSUN$, making them significantly more extended. On the other hand, it is the BH feedback which enlarges the half mass radii of galaxies at the high mass end, leading to the steep slope above Milky Way masses \citep[see also e.g.][]{Crain:2015}. We find that high- and low-state accretion modes contribute in a similar fashion to the increase of galaxy sizes at the high mass end, and without each of them or BH feedback altogether the size-mass relation is nearly flat. However, in Appendix \ref{sec_appendix2_winds}, Figure~\ref{fig:galprop_winds_2}, we show how changes to individual parameters of the galactic winds can still impact the exact shape of the galaxy size-mass relation, as well as the location of the size upturn.

%-------------------------------------------------------------------------------------------------------------------------------------------------------------

\subsection{Effects of Model Parameters and Choices}
\label{sec:sims_variations}

Before exploring the sensitivity of our model to variations of the most important physical parameters, we state a number of broad conclusions that our tests have demonstrated but that we do not explicitly show in this paper:

\begin{enumerate}
\item The changes implemented to the stellar yields (Section~\ref{sec:tng_yields}, Table \ref{tab:yields} and Figure~\ref{fig:yields}, bottom panel) do not lead to any noticeable difference in the galaxy/halo properties examined thus far. The new yield tables have been constructed so that the total mass ejected by SNII is unchanged, with SNII dominating the cumulative metal production and resulting in about just 20  per cent difference between the new and Illustris tables. Therefore, the new yield choices do not impact significantly the overall gas cooling and subsequent star formation. However the abundances of individual species as well as their spatial gradients will differ because of the different yields, and possibly even more so because of the differences in the SNII mass limits (see below) and SNIa normalization.  \\

\item The discrete enrichment scheme (Section~\ref{sec:tng_sf_enrichment}), as opposed to a return of metals at every timestep from stars to their surrounding gas, has no relevant effect on the galaxy and/or halo properties we have studied far. \\

\item The increase of the minimum mass for core-collapse supernovae from 6 (Illustris) to 8 $\MSUN$ (TNG) has on the other hand a non negligible impact, and indeed the TNG fiducial parameters -- chiefly of the galactic wind feedback -- have been chosen to take this into account. In particular, the smaller minimum mass of SNII in Illustris implies a larger number of stars going off as core-collapse supernovae per formed stellar mas ($N_{\rm SNII} = 0.0173 $ vs $ 0.0118$ SNII $\MSUN^{-1}$). This results in $\sim$46 per cent larger effective wind loading factors and hence more efficient galactic winds at all times and masses, in Illustris vs TNG for the same choice of the other parameters. Consequently, with 6 $\MSUN$ as minimum SNII mass instead of 8 $\MSUN$, the star-formation rate density is lower at $z\gtrsim1$ (up to 50 per cent more suppressed) and galaxies in haloes $\lesssim 10^{12}\MSUN$ are 20-30 per cent less massive at $z=0$. \\

\item Finally, varying the initial magnetic field seed in the range $10^{-8} - 10^{-21}$ physical Gauss does not have any relevant effect on the galaxy population at $z=0$ \cite[in agreement with][]{Marinacci:2015}. We recall that in the fiducial setup, the magnetic field is seeded at $z=127$ with a strength of $10^{-14}$ comoving Gauss, equivalent to about $10^{-10}$ physical Gauss. This is within observational constraints on the primordial magnetic field strength \citep{PlanckXIX:2015}.
\end{enumerate}

\begin{figure}
\centering
\includegraphics[width=8.4cm]{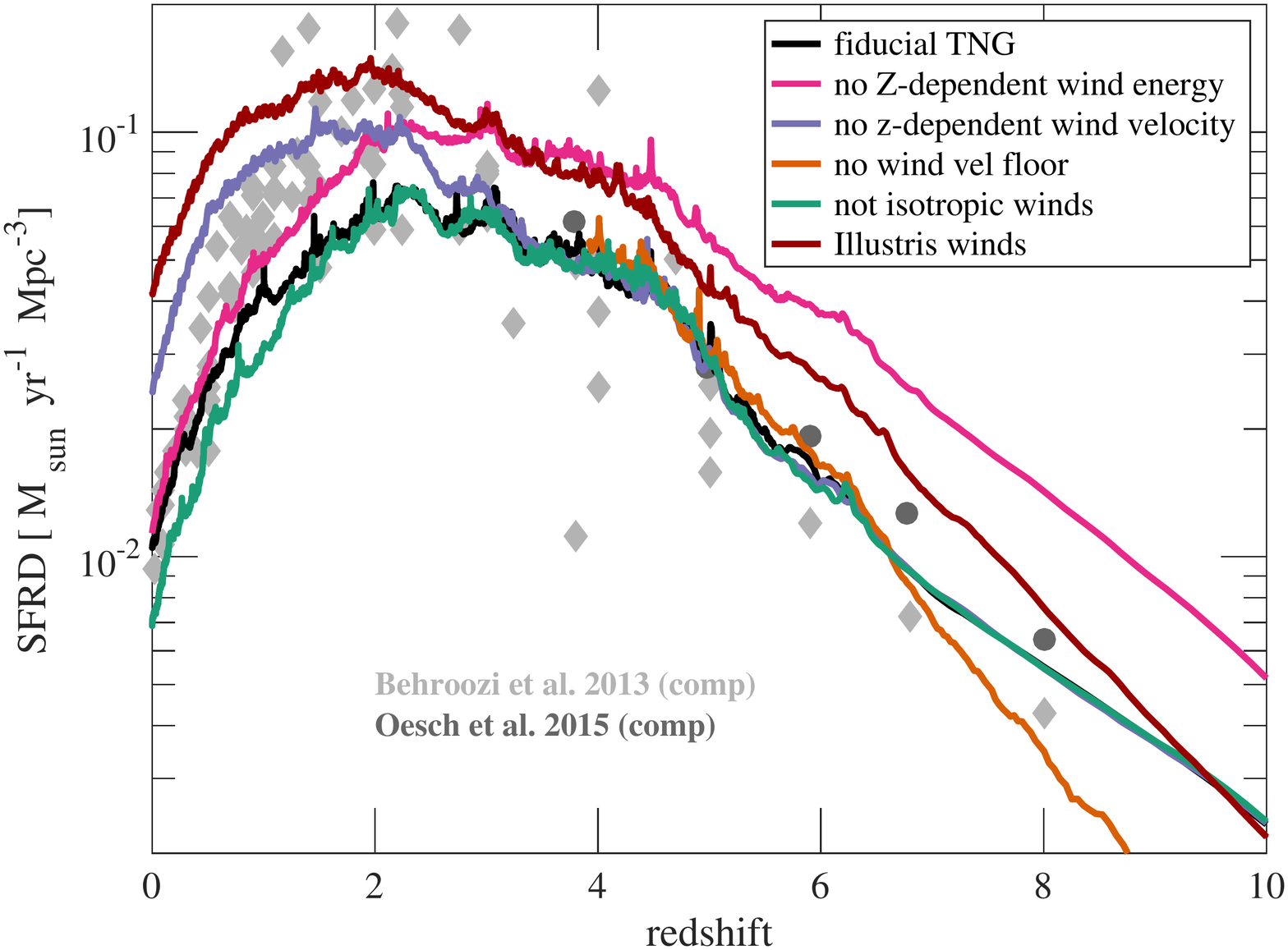}
\includegraphics[width=8.4cm]{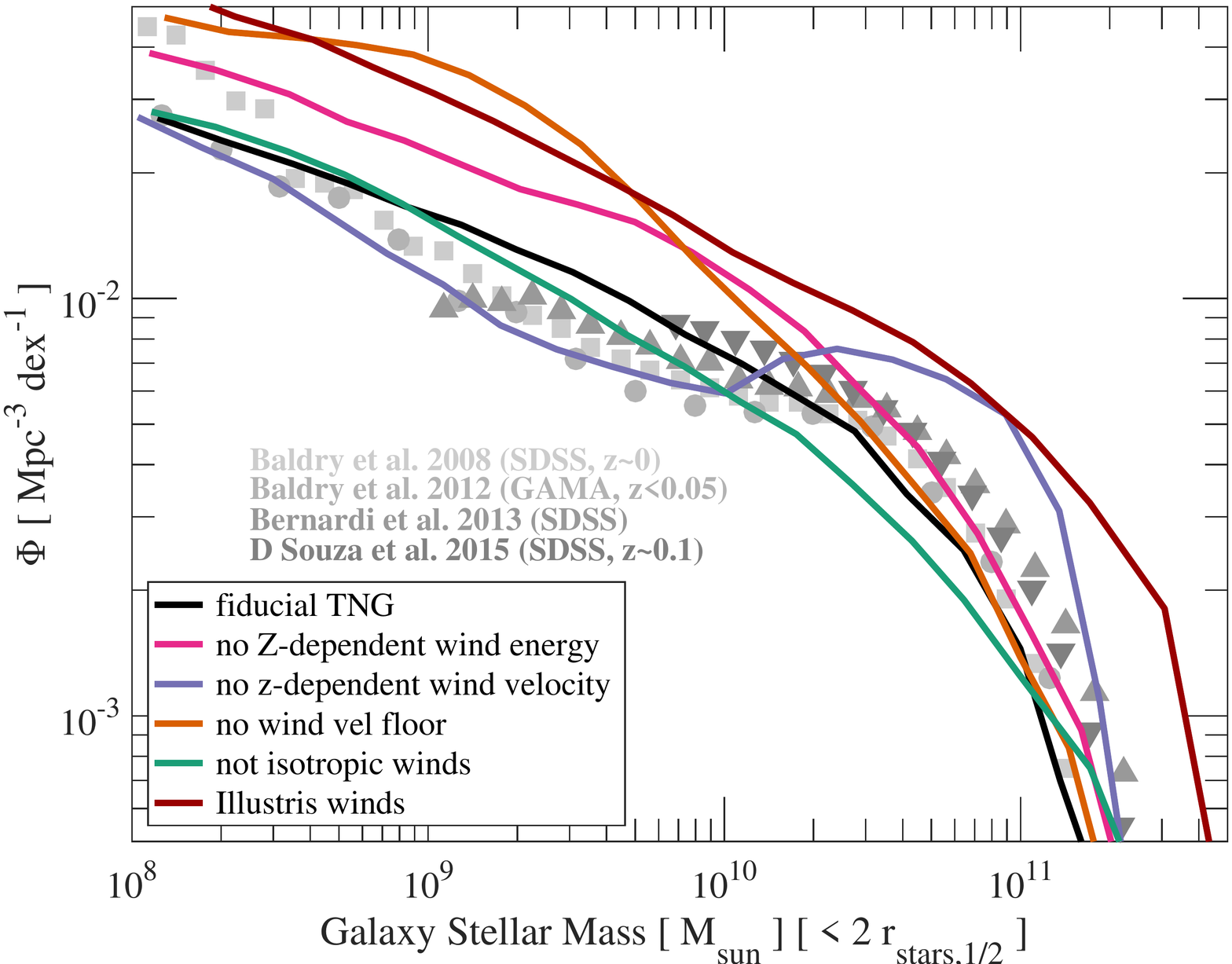}
\includegraphics[width=8.4cm]{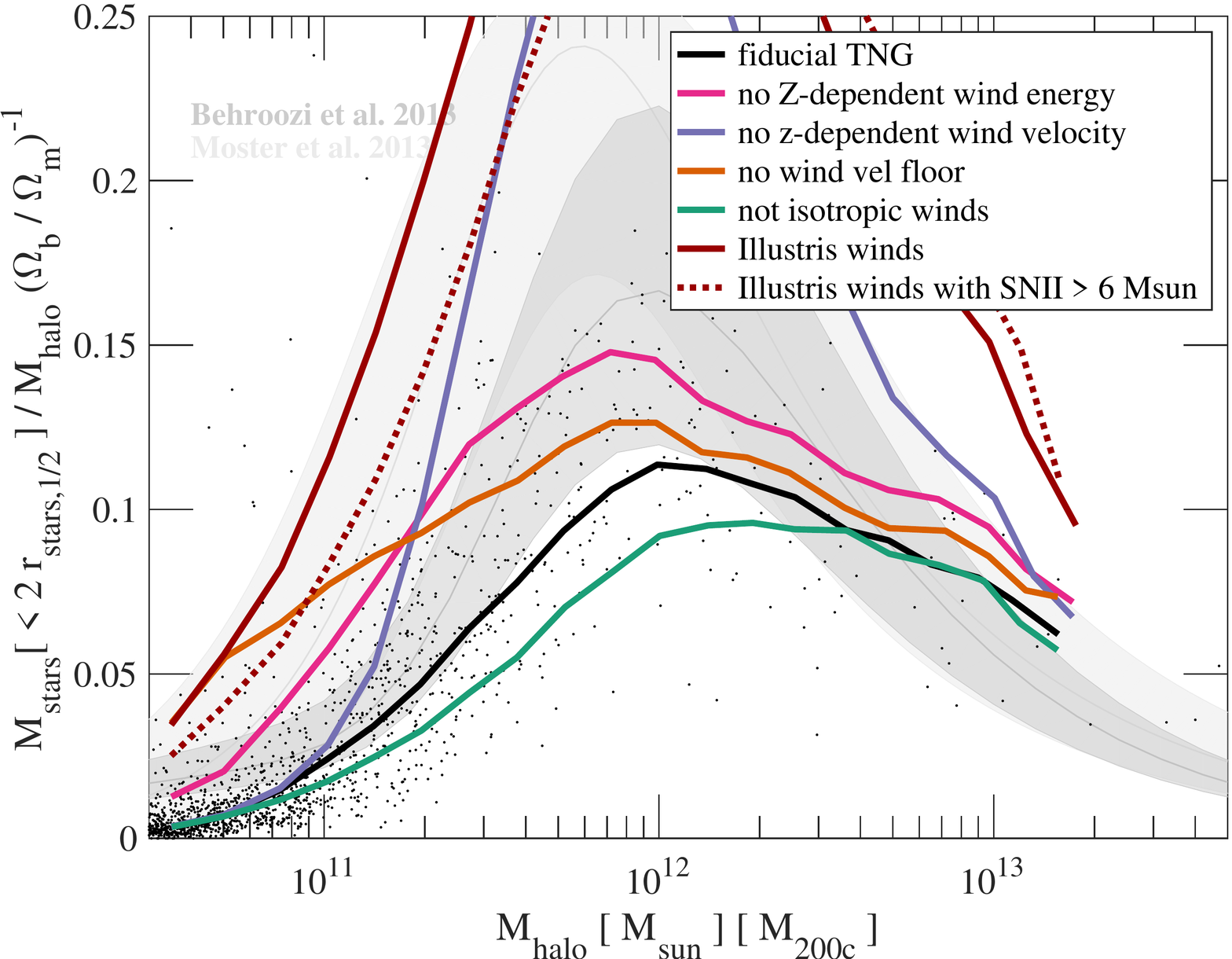}
\caption{Stellar content with the fiducial TNG model (black) and several variations, where we remove each of the new features of the galactic winds one at a time. These are: no metallicity scaling of wind energy (magenta), no redshift scaling of wind velocity (purple), no velocity floor (orange), and no changed/isotropic directionality (green). The brown curves are the TNG model where the galactic winds (only) are exactly as in Illustris. In all solid curves, the minimum mass for SNII remains larger, i.e. is set to the TNG fiducial value of 8 $\MSUN$. For comparison, and only in the bottom panel, the dotted brown curve shows the same `Illustris winds' variation but with the minimum SNII mass also reset to its Illustris value. Parameter choices for all the runs in this Figure are reported in Table \ref{tab:variations}. Symbols and annotations are as in Figure~\ref{fig:galprop_fiducial}, but here stellar masses are measured within twice the stellar half-mass radius.}
\label{fig:galprop_winds_1}
\end{figure}

\subsubsection{Galactic Winds: Scalings}
\label{sec:sims_variations_winds_1}

To better understand how the TNG model functions, we quantitatively demonstrate how the simulated galaxy/halo properties depend on the chosen values of the model parameters, focused specifically on the novel aspects of the galactic winds. In Appendix \ref{sec_appendix2_winds} we show perturbations to the wind total energy, thermal energy, and velocity (`stronger', `faster', `warmer', `cold', and `no velocity floor' winds). These are even more basic changes, some of which can be equally important to those explored below. However, they do not represent new aspects of the TNG model or newly implemented changes \citep[and were explored already for the Illustris model in][although for different galaxy properties]{Vogelsberger:2013b, Bird:2014, Suresh:2015}, so we postpone those results and their discussion to the Appendix.

As summarized in Table \ref{tab:illvstng} and discussed in Section~\ref{sec:sims_fiducial_winds}, the fiducial wind parameters are chosen so that a) the wind velocity in haloes is equivalent to that of the previous Illustris model at redshift $z=5$ (see Eq. \ref{eq:winds_vel}), while they are overall faster at $z<5$; b) the wind energy parameter \egyw  is such that the wind energy in galaxies with average Solar gas metallicity is comparable to the value adopted in Illustris (see Eqs. \ref{eq:winds_eta} and \ref{eq:wind_energy}, and Figure~\ref{fig:galprop_winds_Z}); and c) a floor to the minimum wind velocity is enforced at 350 ${\rm km\,s^{-1}}$ (see Eq. \ref{eq:winds_vel} and Figure~\ref{fig:winds_vel}). Overall, the new winds are faster, warmer, and have smaller mass loading at $z=0$ for intermediate and massive galaxies. In addition, for galaxies less massive than the Milky Way their relative efficiency is higher (Figures \ref{fig:winds_vel}, \ref{fig:winds_eta}).

The effects of these scalings in the velocity and energy of the TNG winds are shown in Figure~\ref{fig:galprop_winds_1}, focusing on the stellar content of galaxies (the corresponding parameter variations are described in Table \ref{tab:variations}): these test cases are designed to recover the Illustris wind parameterization, i.e. by choosing the parameter changes that are most in line with the Illustris model in the absence of each individual scaling. Perturbations on the implementations of the TNG wind scaling are studied in Appendix \ref{sec_appendix2_winds}. In  Figure~\ref{fig:galprop_winds_1}, we also show the outcome of the fiducial TNG model except with the {\it entire} wind model returned to its parameterization in Illustris (``Illustris winds'', brown curve): this in practice corresponds to the case with all the TNG scalings studied in Figure~\ref{fig:galprop_winds_1} turned off.  

We observe that: 1) the redshift-dependent velocity scaling (black vs purple curves) makes the TNG winds faster than Illustris at recent times, producing the different shape of the star formation rate density with redshift and resulting in less massive galaxies at $z=0$ in haloes of masses around $10^{11}\MSUN$ and above. In fact, the shape of the galaxy stellar mass function without the redshift-dependent velocity scaling interestingly resembles the early findings from the {\sc GIMIC} simulation \citep{Crain:2009}. 2) The metallicity-dependent wind mass loading (magenta vs black) and the minimum velocity floor (orange vs black), on the other hand, are responsible for the intended changes at the low-mass end of the galaxy population at the current epoch. They both lead to a higher {\it relative} efficiency of the wind feedback in low-mass systems, by suppressing their star formation more in comparison to Milky Way-sized galaxies. This can be seen in the shape of the $z=0$ galaxy stellar mass function before the knee, and in the steep rise of the TNG fiducial model of the stellar-to-halo mass relation towards the peak. 3) However, these two modifications to the original Illustris winds have starkly different consequences at high redshifts (see top panel), with their net impact on the global star formation at $z \gtrsim 5$ roughly counteracting each other. 4) The directionality of the winds at injection has weaker effects on the stellar content of galaxies, leading to a 10-20 per cent suppression of $M_\star$ around the Milky Way-mass scale for directional winds with respect to the isotropic case. \footnote{Although we do not show it here, the TNG wind energy and velocity adjustments do impact the BH masses and halo gas fractions at fixed halo masses; however, they do not not modify the stellar radii of galaxies.}

It is clear from Figure~\ref{fig:galprop_winds_1} that Illustris-like winds adopted within the TNG model (brown curves vs black) are not effective enough at suppressing star formation in galaxies, across the whole mass and redshift range. In particular, galaxies populating $10^{12}\MSUN$ haloes at $z=0$ (bottom panel) are a factor 2-3 more massive in this case than with the full TNG (or Illustris) fiducial model. One possible culprit is the increase in the minimum mass for core-collapse supernovae. The parameterization of the TNG galactic winds has indeed been chosen to account for this change, in practice by making them faster rather than increasing their energy. However, Illustris-winds adopted within the TNG model with 6 rather than 8 $\MSUN$ for the minimum SNII mass still produce overly massive galaxies, particularly at the peak of stellar efficiency and towards lower masses (dotted brown curve, bottom panel).

%The mass downt to which the BH act: as a whole and in the low accretion state.
%Why did we make the winds stronger? Below the peak, illustris stellar masses are too large, so TNG winds had to be stronger.
%At the peak, 2.2 factor change in Mstars with and without BH, In TNG the net is 1.8 at the peak, but at the peak 
% TNG L25n512: at 1e12: mstars(fiducial) = 0.1136, mstars(no BHs) = 0.1854 => 1.6.
% TNG L25n256: at 1e12: mstars(fiducial) = 0.071, mstars(no BHs) = 0.094 => ratio = 1.3.
% Ill L25n256 (V13 digitized): at 1e12 mstars(no BHs)/mstars(fiducial) ~ 2.5
%We shifted the balance between radio and quasar mode.
It is difficult to adopt the comparison between the Illustris and TNG galactic winds to indirectly compare the relative efficiency of the Illustris and TNG BH models with respect to the suppression of stellar mass growth. Firstly, the various galactic wind parameterizations impact the BH growth differently (see Figures \ref{fig:galprop_misc} and \ref{fig:galprop_winds_2}, top right panels) and hence affect indirectly the efficiency of the concurrent BH feedback. Secondly, the relative effectiveness of the two BH models depends sensitively on the physical parameters adopted for each. Keeping this in mind, we find that the old Illustris BH model was more aggressive than the TNG implementation at suppressing star formation at intermediate masses (including at $L^{\star}$). In the fiducial configurations and with respect to the no black hole cases, Illustris BH feedback suppresses the stellar masses of $10^{12}\MSUN$ haloes by about a factor 2.5 in comparison to about a factor of 1.3-1.5 in the TNG model \citep[from Figure 15 in][accounting for resolution effects in comparison to Figure~\ref{fig:galprop_misc} bottom right panel in this paper]{Vogelsberger:2013b}. The strength of galactic winds in this mass regime is correspondingly larger in TNG in order to counterbalance this change. Both Illustris and the TNG BH models impact the stellar mass of galaxies in haloes as small as $1-2 \times 10^{11}$\,$\MSUN$. Simultaneously, the smallest halo mass above which the low-accretion rate feedback modifies stellar mass is different by about an order of magnitude between the two models: 10$^{11}$\,$\MSUN$ in comparison to about 10$^{12}$\,$\MSUN$ in Illustris and TNG respectively, with the Illustris quasar mode barely having any effect. However, the differential impact between the two models clarifies that this halo mass scale is model dependent. Indeed, we could have shifted this transition point by appropriately redistributing the burden of stellar mass control between the low- and high-accretion rate BH feedback below the $10^{12}\MSUN$ scale or from the winds onto the black holes. It is clear, then, that the optimal choices for wind as well as black hole feedback strongly depend on the {\it whole ensemble} of galaxy formation mechanisms incorporated into the model.

\section{Summary and Discussion}
\label{sec:summary}

In this paper we have described the implementation and performance of the IllustrisTNG model (TNG, The Next Generation), an updated ensemble of numerical routines to form and evolve galaxies in large scale gravo-magneto-hydrodynamical simulations with the moving mesh code {\sc Arepo}. The TNG physics and general approach derive from the galaxy formation model \citep{Vogelsberger:2013b,Torrey:2014a} that has been used to run the Illustris simulation \citep{Vogelsberger:2014a,Vogelsberger:2014b,Genel:2014,Sijacki:2015}. 

Both models therefore include the numerical solution of the gravitational interactions among different resolution elements (gas, dark matter, stars, and black holes); the solution of the (magneto)hydrodynamical equations for the gaseous component; a treatment of the most important radiative cooling and heating processes; a mechanism for the conversion of gas into stars; stellar evolution and subsequent chemical enrichment of the interstellar, circumgalactic and intergalactic medium; outflows of gas from the star-forming regions of galaxies; and the formation, growth, and energetic feedback of supermassive black holes in distinct low- and high-accretion rate states.

In developing the TNG model we have pursued three simultaneous goals: to bring together under a consistent framework the improvements implemented in {\sc Arepo} to the accuracy and robustness of the underlying numerical methods; to introduce new, rich physics; and to address the key deficiencies of the previous model. 

The main numerical updates are summarized in Section~\ref{sec:tng_numerics} and include an improved estimator for spatial gradients across gas cells, a more efficient gravity solver and time integration scheme, and a more consistent advection of the passive scalars. Together, these changes give a more accurate and better convergent hydrodynamical scheme, and will allow us to pursue even more ambitious simulations in the future. 

The principal new physics in the TNG model are the inclusion of MHD (see Section~\ref{sec:mhd} and references therein) and a dual-mode, thermal and kinetic black hole feedback model \citep[see the companion paper by][]{Weinberger:2017}. The new kinetic BH feedback has been shown to alleviate some of the discrepancies identified in Illustris in comparison to observations at the high end of the halo mass function ($\gtrsim 10^{12}-10^{14} \MSUN$), in particular regulating the stellar content of massive galaxies while preserving realistic halo gas fractions.

In fact, in confronting the problems identified in Illustris (see Introduction), we aim to open new physical regimes which were impossible to robustly study in previous simulations. These include not only massive haloes emitting at X-ray wavelengths, but also the galaxies occupying the transitional green valley, and the properties of low-mass dwarfs, including cluster satellite populations. In this paper, we have focused on the intermediate and low mass end of the galaxy population (halo masses of a few $10^{10} - 10^{13}\MSUN$) and have presented an updated model for the galactic winds feedback, by modifying their directionality, thermal content, velocity and energy scalings (Section~\ref{sec:tng_winds}). We have also updated our reference yield tables (Table~\ref{tab:yields} and Figure~\ref{fig:yields}) and adjusted some stellar evolution choices (Section~\ref{sec:tng_stars}).

The TNG galactic winds (whose implementation and scalings largely follow those of \citealt{Springel:2003, Oppenheimer:2006, Oppenheimer:2008, Vogelsberger:2013b}) are now launched isotropically from the star-forming regions of galaxies, bringing a simplification to the modeling that by itself does not significantly affect galaxies, qualitatively (Figure~\ref{fig:winds_patterns}) nor quantitatively (Figure~\ref{fig:galprop_winds_1}). Simultaneously, we better capture the rich complexity of the physical mechanisms governing galactic outflows, allowing the wind energy to depend on the metallicity of the ambient gas (Figures \ref{fig:winds_eta} and \ref{fig:galprop_winds_Z}; and \citealt{Schaye:2015}). Finally, TNG winds are launched with a velocity that depends on the {\it local} velocity dispersion subject to a minimum floor (350 ${\rm km\,s^{-1}}$) and further modulated in redshift following the evolution of the halo virial mass (Figure~\ref{fig:winds_vel}).
%, and so constant at fixed halo mass (Figure~\ref{fig:winds_vel}).

To calibrate the model and determine its free parameters, we have simulated a series of cosmological volumes, exploring a collection of variations in the physical and numerical schemes. In this paper we have given a sense of these explorations by showing perturbations about the fiducial TNG model (Figures \ref{fig:galprop_misc}, \ref{fig:galprop_winds_1}, \ref{fig:galprop_winds_2}, \ref{fig:galprop_winds_Z}, \ref{fig:galprop_velmin}). To arrive at this final parameterization, we have eventually retained the implementation and parameter set that simultaneously alleviated the largest number of examined tensions in comparison to the analogous galaxy population realized with the fiducial Illustris setup, while demonstrating an overall satisfactory agreement with benchmark observational constraints.\footnote{The exploration of the TNG model undertaken here was carried out \textit{after} the model was finalized, and is not meant as a documentation of the actual process undertaken to calibrate the model.} The principal outcomes of the model on our (37\,Mpc)$^3$ test volume are given in Figure~\ref{fig:galprop_fiducial}. A summary of the differences between the TNG and Illustris setups, both schematic and quantitative, is given in Table~\ref{tab:illvstng}. 

As suggested by Figure~\ref{fig:galprop_fiducial}, our comparison between the two models, as well as between TNG and key observational results, has focused on the integral properties of galaxies, particularly their stellar content at $z=0$ (galaxy stellar mass function and stellar to halo mass relation), in addition to the star formation rate density versus time, the current BH mass to galaxy mass relation, the halo gas fraction within $R_{500c}$, and the stellar half-mass radii (sizes) of galaxies. As a result of the new scalings and {\it in combination} with the functioning of the new BH feedback, TNG galactic winds are overall faster and more effective than Illustris winds at preventing star formation at all masses and times (see Figure~\ref{fig:galprop_winds_1} and related discussion). Moreover, we have obtained a higher relative efficiency of wind feedback in low-mass systems, by suppressing their star formation more in comparison to Milky Way-sized galaxies. 
As a result, TNG galaxies exhibit a significantly suppressed $z=0$ galaxy stellar mass function at the low mass end, while the ``knee'' remains unaltered. They also show a more pronounced peak in the present day stellar-to-halo mass relation, although the overall normalization is lower. Simultaneously, the TNG model still appears to produce a reasonable mix of morphological types (shown qualitatively in Figure~\ref{fig:L25n512_box_2}), a reasonable trend of black hole mass with galaxy mass, and total halo gas fractions at the group scale in much better agreement with observations. 

\begin{figure}
\centering
\includegraphics[width=8.4cm]{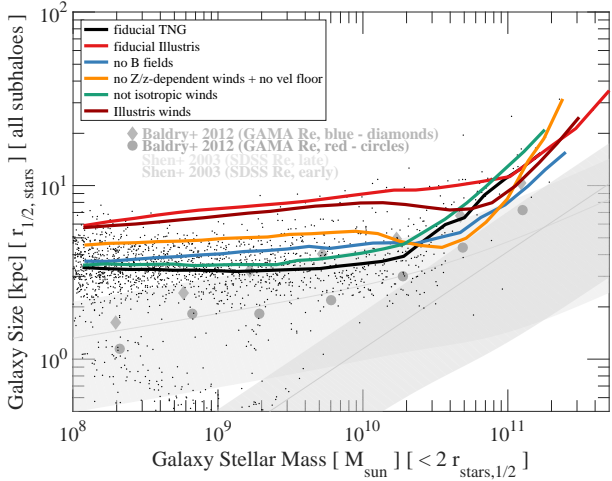}
\caption{Galaxy sizes (stellar half mass radii) as a function of stellar mass at $z=0$. The fiducial TNG model (black) is compared to cases with: no MHD (blue), no metallicity or Hubble related wind scalings (orange), and no change to wind directionality, i.e. reverting to bipolar winds (green). We also include the fiducial TNG model, except with all wind-related aspects equal to Illustris (dark red), and the fiducial Illustris outcome itself (red). While no single change to the winds produces the observed dramatic decrease in galaxy sizes, the new TNG wind model as a whole does.}
\label{fig:galprop_sizes}
\end{figure}

\subsubsection*{Galaxy Sizes}

In Figure~\ref{fig:galprop_fiducial} we have also shown that galaxy sizes in the TNG model are about a factor of two smaller than in the fiducial Illustris model for galaxies with stellar mass below $\sim 10^{10}\MSUN$. This puts them qualitatively at the right magnitude as compared to $z=0$ observations, improving upon the previous result. In Figure~\ref{fig:galprop_misc} we have further demonstrated that the sizes below this mass scale are essentially controlled by our galactic winds. We have not, however, provided an explanation for exactly which components of the new wind model are responsible for this improvement. Indeed, this turns out to arise \textit{only} through a \textit{combination} of several changes.

In Figure~\ref{fig:galprop_sizes} we show galaxy stellar half-mass radii as a function of stellar mass, including several perturbations to the fiducial TNG model which have the largest impact. In particular, both the original Illustris model (red) and the full TNG model except with an `Illustris winds' prescription (dark red) give consistent results: much larger sizes at low mass. None of our four principal modifications to the winds -- directionality, velocity floor, metallicity-dependent energy, and Hubble-dependent velocity -- succeed individually to affect this change (see also bottom right panel of Figures \ref{fig:galprop_winds_2}). The last three of these changes, taken all together (orange line), lead to significant improvement, although indeed this change is larger by far than the sum of their three individual effects. Alone, switching the wind directionality between isotropic and bipolar has negligible impact (green line). Yet, combining the directionality change with the three other adjustments just discussed (by definition, the dark red line) yields the entirety of our improvement in the size-mass relation at low masses. This rather puzzling result points towards a non-trivial coupling between at least three, and possibly all four, of our main changes to the wind model: we postpone a more detailed investigation to future work. For completeness, we also include the impact of excluding MHD on the sizes (blue line), which of all the perturbations explored in this paper is the only additional change to play any noticeable role here.

\subsubsection*{The Impact of Magnetic Fields}

A remarkable addition to the TNG galaxy formation model is the self consistent amplification and subsequent evolution of a primordially-seeded magnetic field, which will open up many new astrophysical regimes and questions for scientific exploration. We have seen early hints that magnetic fields can indeed play an important role in cosmological simulations of galaxy formation, to a sufficient degree that excluding MHD may represent a serious model deficiency. For example, in Figure~\ref{fig:galprop_misc} we have demonstrated that magnetic fields have a non-negligible impact on larger halos, $\gtrsim 10^{12}\MSUN$, suppressing their final stellar content by $z=0$ by up to 30-80 per cent. The inclusion of MHD also suppresses the cosmic SFRD at low redshifts ($z\lesssim 3$), while increasing the gas content in haloes below $10^{12}\MSUN$.

In certain regimes, the impact of MHD can be as large as the differences between the fiducial Illustris and TNG models themselves. 
This is likely driven by the fact, seen in Figure~\ref{fig:galprop_misc} (top right panel), that galaxies of all masses host more massive BHs in the presence of magnetic fields. The feedback associated with this additional accretion energy hampers star formation across cosmic time. Interestingly, the shift in the $z=0$ BH to halo/galaxy mass relation extends across a large range of galaxy masses, i.e. also for haloes below $\sim 10^{12}\MSUN$. This suggests a more general interaction between the magnetic fields and black holes, which we speculate arises from an enhancement of the gas accretion onto the BHs in the presence of a more highly pressurized interstellar medium.
%and which is not related to the dependence of the BH Bondi accretion rate on the . 
A targeted study is needed to properly understand such mechanism in the TNG model and to make a quantitative comparison to the findings of \cite{Marinacci:2015} and \cite{Pakmor:2017} who have quantified the balance among magnetic, thermal and kinetic energy in the ISM and gaseous haloes of samples of galaxies simulated with the Illustris and Auriga \citep{Grand:2017} models, respectively.

\subsubsection*{Open Issues and Future Directions}

By combining the results of this paper with those explored in the companion study, \cite{Weinberger:2017}, we have addressed five of the main six shortcomings of the Illustris model enumerated in the Introduction. These are all related to the overall baryon mass content of galaxies and their haloes. Although Figures \ref{fig:L25n512_box_2} and \ref{fig:winds_patterns} have given a preview of the structural properties of galaxies formed within the TNG model, a quantitative assessment remains needed on sub-galaxy scales. This will enable us also to robustly answer the remaining failure of the previous model: an excess of star forming rings in fragmenting galactic disks. We note that, in initial assessment, the latter phenomenon is reduced in TNG.

Additionally, while we have discussed the resolution convergence of both the fiducial Illustris and TNG models in Appendix \ref{sec_appendix2_winds}, the dependence of some model predictions on numerical resolution remains a fundamental challenge of hydrodynamical cosmological simulations for galaxy formation, and one still wanting for a satisfactory solution. Similarly, we caution against the over-interpretation of specific choices and parameter values adopted in our model. Care is required to connect any model parameter directly to a particular physical value, either in observations or other theoretical models. As in all such simulations, model parameters cannot be separated from the particular details of their numerical implementation. By construction, they have also been defined within the scope of, and only in relation with the other components of, a multi-scale effective theory for galaxy formation, which combines a diverse set of physics.
As a concrete example, we have considered in some detail the properties of TNG galactic winds {\it at injection}. These are the direct consequences of the model inputs convolved with the cosmological context. However, these initial wind properties will differ from the phenomenology of galactic outflows measured at several kilo-parsec or larger scales away from galaxy centers. Similarly, several unexpected aspects of the simulations still need to be understood in detail. These include the impact of MHD on the evolution of massive galaxies and their surrounding gaseous haloes; the emergence of less extended galaxies and their size evolution across redshifts; and finally and rather crucially, the effects of the modifications introduced with the TNG model to the stellar and gas metallicities and elemental compositions of galaxies.

The demonstration of the improvements of the TNG model discussed in this paper constitute only a preliminary analysis.
%are intended only as a first look, and in a relative sense -- namely, in direct  comparison to the realizations of the original Illustris model, as well as to variations of the fiducial TNG model itself, in the same cosmological test volume. 
Much larger simulated volumes are necessary for an adequate assessment. These will give us robust statistical power to quantify trends in the galaxy population as a whole, as well as the ability to make statements about the rarest and/or most massive objects in the Universe. Moving beyond `average' galaxy populations, we can then start to understand the physical origin of galaxy-to-galaxy variations, and so the mechanisms behind different evolutionary pathways and their culmination in the observed galaxy diversity of today. At the same time, more sophisticated analyses of the simulated galaxies are required. Only then can we rigorously assess the validity of the TNG model in comparison to observational data, while also having confidence as we leverage the model in astrophysical regimes over which it has never been directly calibrated, or indeed even ever run at all. To this end a new series of large volume cosmological simulations, the ``IllustrisTNG project'', will be undertaken and introduced in future publications.

\begin{figure*}
\centering
\includegraphics[width=17cm]{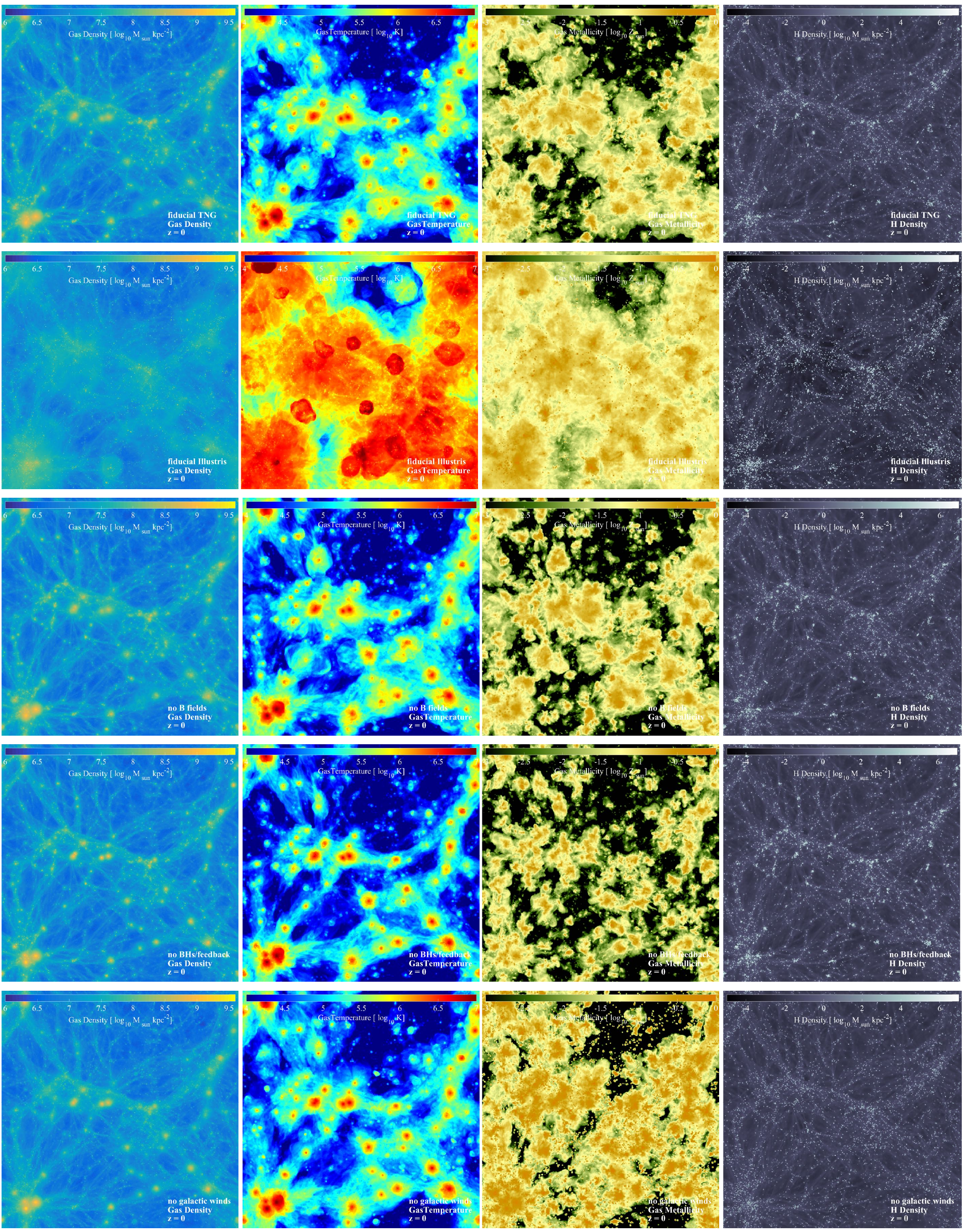}
\caption{Projections of gas properties (column density, mass-weighted temperature, mass-weighted metallicity, and neutral hydrogen column density) on the large scales of the full 25\,Mpc $h^{-1}$ box. We show different choices of the galaxy formation model, one per row, from top to bottom: the TNG fiducial model, the Illustris fiducial model, the TNG model without magnetic fields, the TNG model without BHs and their feedback, and the TNG model without galactic winds. These correspond to the five variations shown quantitatively in Figure~\ref{fig:galprop_misc}. Colour scales are kept fixed across rows, so that a direct comparison is possible. }
\label{fig:box}
\end{figure*}

As a concluding remark, we show in Figure~\ref{fig:box} the gas distribution and its properties on the large scales of the entire box used for all the tests herein. Each row shows a different model variation: fiducial TNG, fiducial Illustris, TNG without MHD, TNG without galactic winds, and TNG without BHs. %In each case the four columns depict: gas column density, temperature, metallicity, and neutral hydrogen (HI) column. 
While our galaxy formation models have always been tested and calibrated by focusing on the integral stellar properties of galaxies, no constraint whatsoever has been applied to the distribution of the cosmic gas. On the one hand, we have already demonstrated that the fraction of total gas mass retained {\it within haloes} can strongly discriminate among models: this is certainly the case for group and cluster size haloes, but even more so for haloes of $10^{12}\MSUN$ and below. Figure~\ref{fig:box} further provides a sense of how much the topology, thermodynamic, and chemical enrichment of Mpc-scale gas and beyond depend on the entire underlying galaxy formation model. The differences between the TNG and Illustris models can be as dramatic, if not more so, than the differences between the fiducial model and the cases when either galactic winds or black holes are entirely excluded. The gas maps in Figure~\ref{fig:box}, justaxposed against the quantitative evaluation of the galaxy properties in Figure~\ref{fig:galprop_misc}, provide a visual demonstration of the multi-scale, multi-physics nature of the problems we have been tackling in this paper: they encapsulate the rationale for ever more comprehensive, self consistent theories for galaxy physics in the full cosmological context.

%-------------------------------------------------------------------------------------------------------------------------------------------------------------

\section*{Acknowledgements}
AP thanks Jan Rybizki and Enrico Ramirez-Ruiz for useful conversations on the yield tables and chemical enrichment. The Flatiron Institute is supported by the Simons Foundation. VS, RW, and RP acknowledge support through the European Research Council under ERC-StG grant EXAGAL-308037 and would like to thank the Klaus Tschira Foundation. MV acknowledges support through an MIT RSC award and the support of the Alfred P. Sloan Foundation. Simulations were run on the Hydra and Draco supercomputers at the Max Planck Computing and Data Facility (MPCDF, formerly known as RZG) in Garching near Munich and on the HazelHen Cray XC40-system at the High Performance Computing Center Stuttgart as part of project GCS-ILLU of the Gauss Centre for Supercomputing (GCS).

\bibliographystyle{mn2eFixed}  
\bibliography{TNG_Methods_cleaned}
%
%-------------------------------------------------------------------------------------------------------------------------------------------------------------

\appendix

\section{Resolution Dependence in TNG}
\label{sec_appendix1}

\begin{figure*}
\centering
\includegraphics[width=8cm]{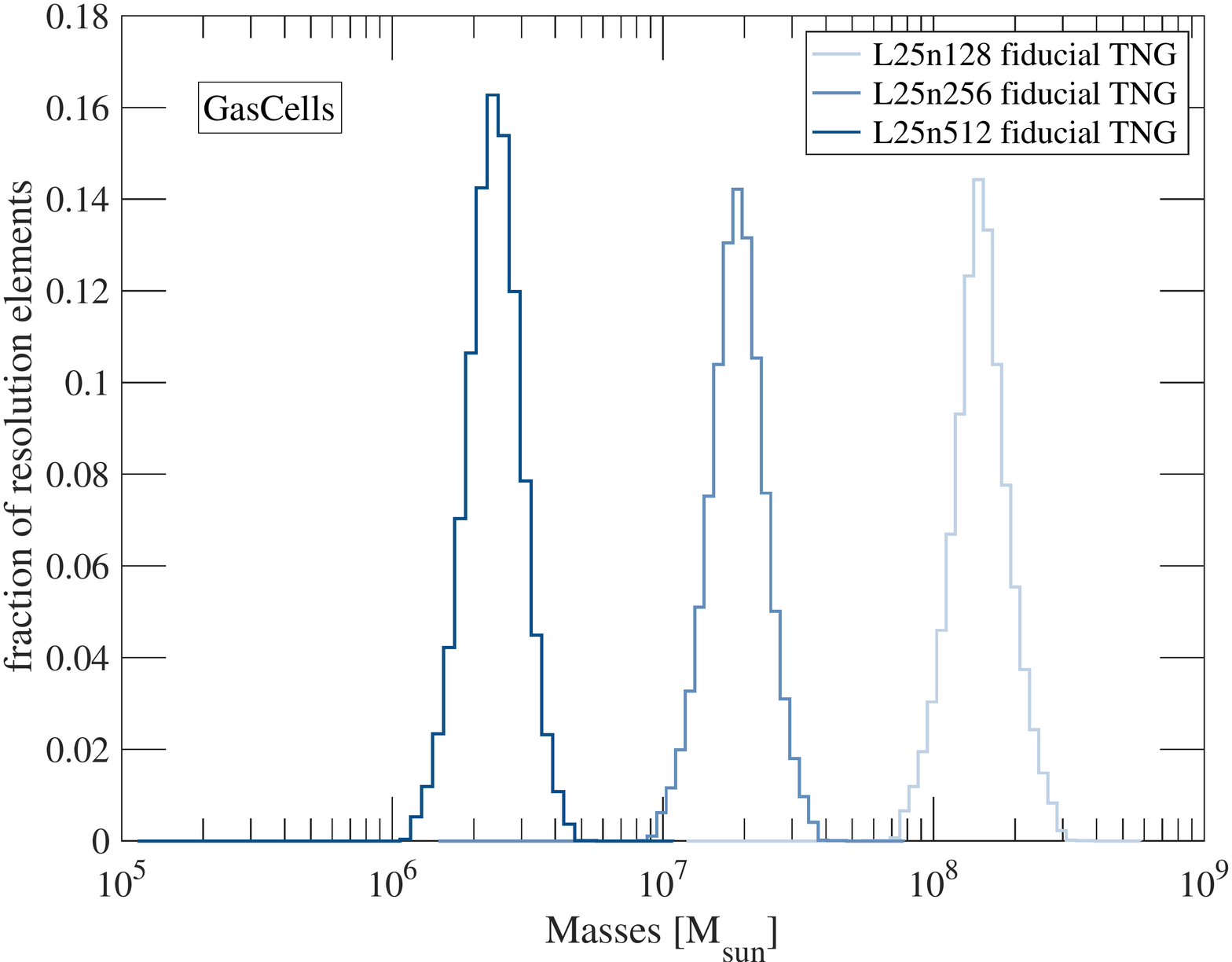}
\includegraphics[width=8cm]{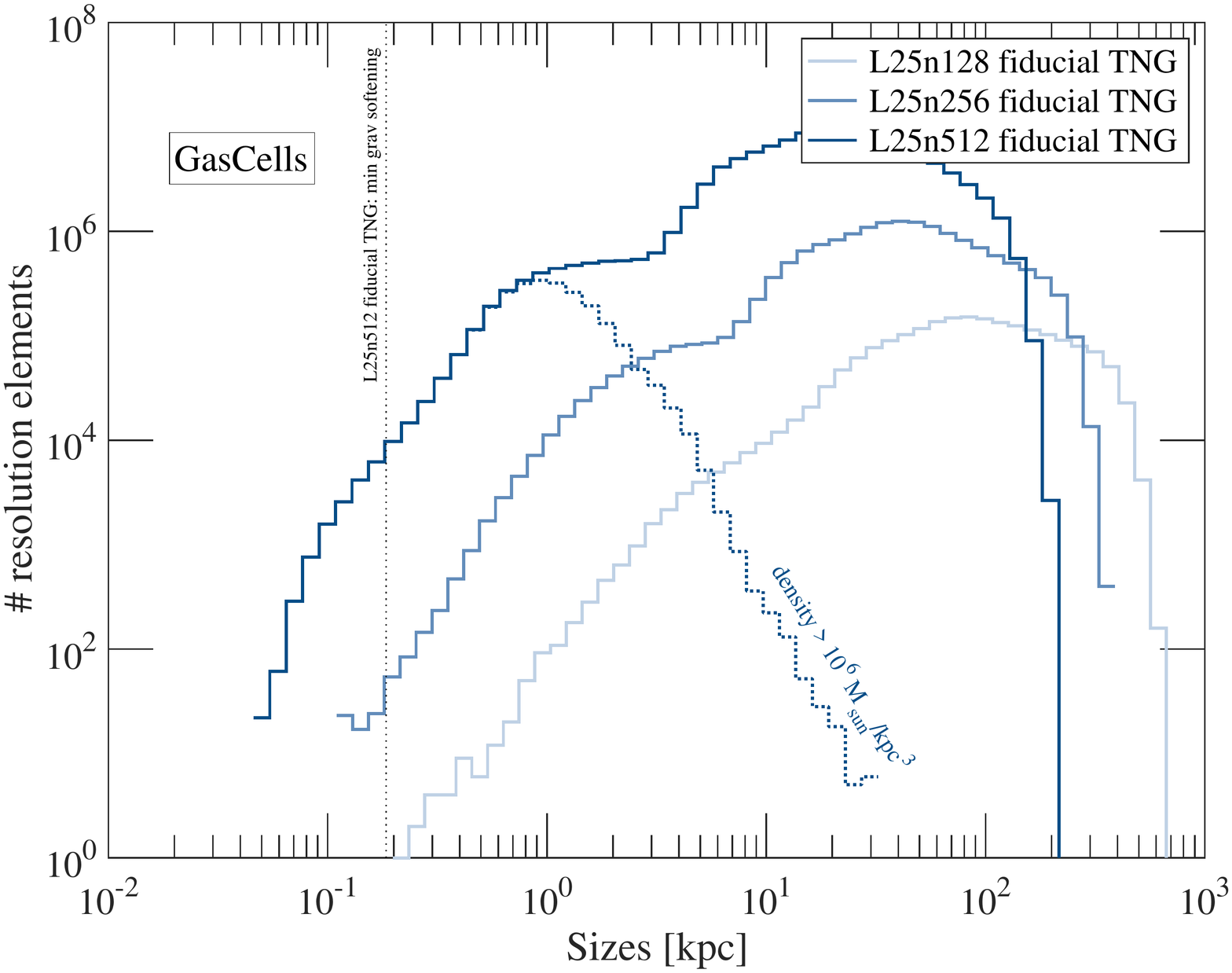}
\caption{Effective resolution of the gas cells in the simulations adopted in this work. Left: distribution of the $z=0$ mass per gas cell, across the simulated volume, for three different resolutions. Gas cells are refined/de-refined such that the mass of the cells is kept within a factor of 2 of a specified target mass $m_{\rm target}$. This is in practice simply the mean baryon mass in the initial conditions. For example, in L25n512, $m_{\rm target} = 2.4 \times 10^6 \MSUN$.  Right: $z=0$ distribution of gas cell radii (sizes), derived from the Voronoi cell volumes assuming each is a sphere. For the L25n512 resolution only, a dotted blue distribution depicts the sizes of gas cells in high density environments ( $> 10^6~ \MSUN$ kpc$^{-3} \sim 0.1 \rm{ cm}^{-3}$, about our star-formation density threshold). For reference, a vertical dotted line denotes the minimum gravitational softening length imposed to the gas cells at L25n512 resolution.}
\label{fig:res_cells}
\end{figure*}
\begin{table*}
  \centering
 % \begin{tabular}{l| cc ccc} 
  %\hline 
  % & $m_{\rm DM}$ & $\bar{m}_{\rm stars}$ & $\epsilon^{\rm com, z > 1}_{\rm DM, stars}$ & $\epsilon^{z \leq 1}_{\rm DM, stars}$ & $\epsilon^{\rm min}_{\rm gas}$\\
  % & $[\MSUN]$ & $[\MSUN]$ & [$h^{-1}$kpc] & [kpc] & [$h^{-1}$kpc]\\ 
  %\hline 
  %L25n1024      & $1.6\times 10^6$ & $2.9\times 10^5$  & 0.5 & 0.37 & 0.0625 \\
  %{\bf L25n512}  & $1.2\times 10^7$ & $1.6\times 10^6$ 	& 1.0 & 0.74 & 0.125 \\
  %L25n256 		 & $9.9\times 10^7$	& $1.3\times 10^7$  & 2.0 & 1.48 & 0.25 \\
  %L25n128 		 & $7.9\times 10^8$	& $1.1\times 10^8$ 	& 4.0 & 2.95 & 0.5 \\ 
  % & & & & & \\
  %Illustris-1    & $6.3\times 10^6$ & $8.3\times 10^5$ 	& 1.0 & 0.71 & 0.5 \\
  %Illustris-2    & $5.0\times 10^7$	& $7.2\times 10^6$	& 2.0 & 1.42 & 1.0 \\
  %Illustris-3    & $4.0\times 10^8$	& $6.0\times 10^7$ 	& 4.0 & 2.84 & 2.0 \\ 
  
  %\hline
  %\end{tabular}
  
    \begin{tabular}{l| cc ccc} 
  \hline 
   & $m_{\rm DM}$ & $\bar{m}_{\rm stars}$ & $\epsilon^{\rm com, z > 1}_{\rm DM, stars}$ & $\epsilon^{z \leq 1}_{\rm DM, stars}$ & $\epsilon^{\rm min}_{\rm gas}$\\
   & $[\MSUN]$ & $[\MSUN]$ & [kpc] & [kpc] & [kpc]\\ 
  \hline 
  %L25n1024      & $1.6\times 10^6$ & $2.9\times 10^5$  & 0.5 & 0.37 & 0.0625 \\
  {\bf L25n512}  & $1.2\times 10^7$ & $1.6\times 10^6$ 	& 1.48 & 0.74 & 0.18 \\
  L25n256 		 & $9.9\times 10^7$	& $1.3\times 10^7$  & 2.95 & 1.48 & 0.37 \\
  L25n128 		 & $7.9\times 10^8$	& $1.1\times 10^8$ 	& 5.90 & 2.95 & 0.74 \\ 
   & & & & & \\
  Illustris-1    & $6.3\times 10^6$ & $8.3\times 10^5$ 	& 1.42 & 0.71 & 0.71 \\
  Illustris-2    & $5.0\times 10^7$	& $7.2\times 10^6$	& 2.84 & 1.42 & 1.42 \\
  Illustris-3    & $4.0\times 10^8$	& $6.0\times 10^7$ 	& 5.68 & 2.84 & 2.84 \\ 
  
  \hline
  \end{tabular}
  
  \caption{\label{tab:res} Numerical characteristics and resolution parameters of the cosmological simulations adopted in this work. All volumes have the same box size, 25 Mpc $h^{-1} \sim 37$ Mpc, and share the same initial random realization of the density field. Unless otherwise stated, we always show results from L25n512 configurations, which have a resolution similar to Illustris-1 (within a factor of 2 in mass resolution). DM particles have all the same mass at fixed resolution, while all other baryonic components exhibit a distribution in particle or cell mass: $\bar{m}_{\rm stars}$ denotes the median stellar particle mass today. Plummer-equivalent gravitational softening lengths are denoted by $\epsilon$. In TNG, the gravitational softenings for all particle types are comoving kpc (with value equal to that of the DM) down to $z=1$, after which they are fixed in physical space to their $z=1$ values (in Illustris, this treatment
is not applied to the DM particles, such that at $z=0$, the DM particles have double the softening length as the reported values for the stars). For comparison, the three lower rows give the same values for the Illustris simulation series, performed in a volume of $\simeq$100 Mpc on a side and with WMAP9 cosmology (\textcolor{blue}{www.illustris-project.org}).}
\end{table*}
\begin{figure*}
\centering
\includegraphics[width=8.2cm]{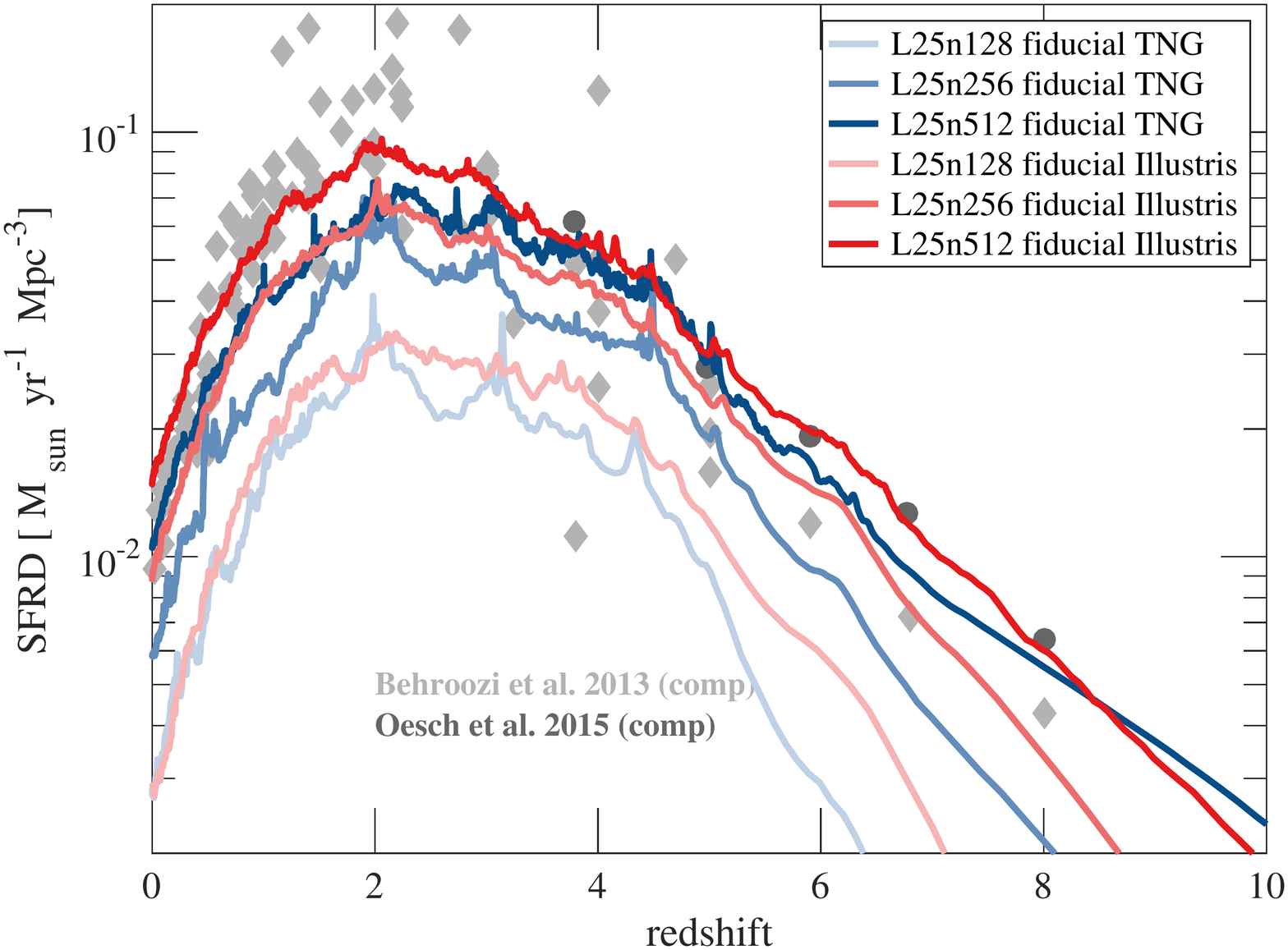}
\includegraphics[width=8.2cm]{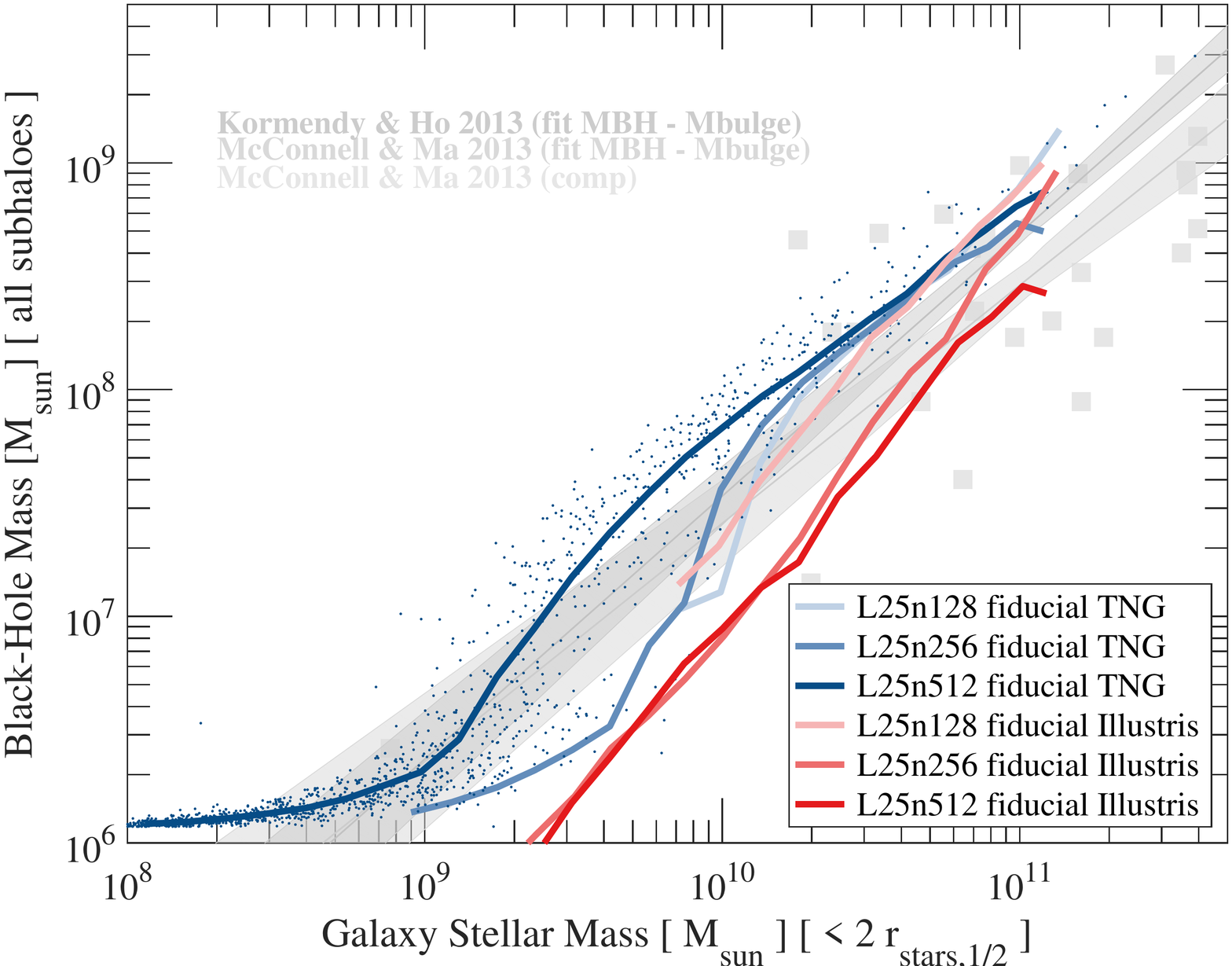}
\includegraphics[width=8.2cm]{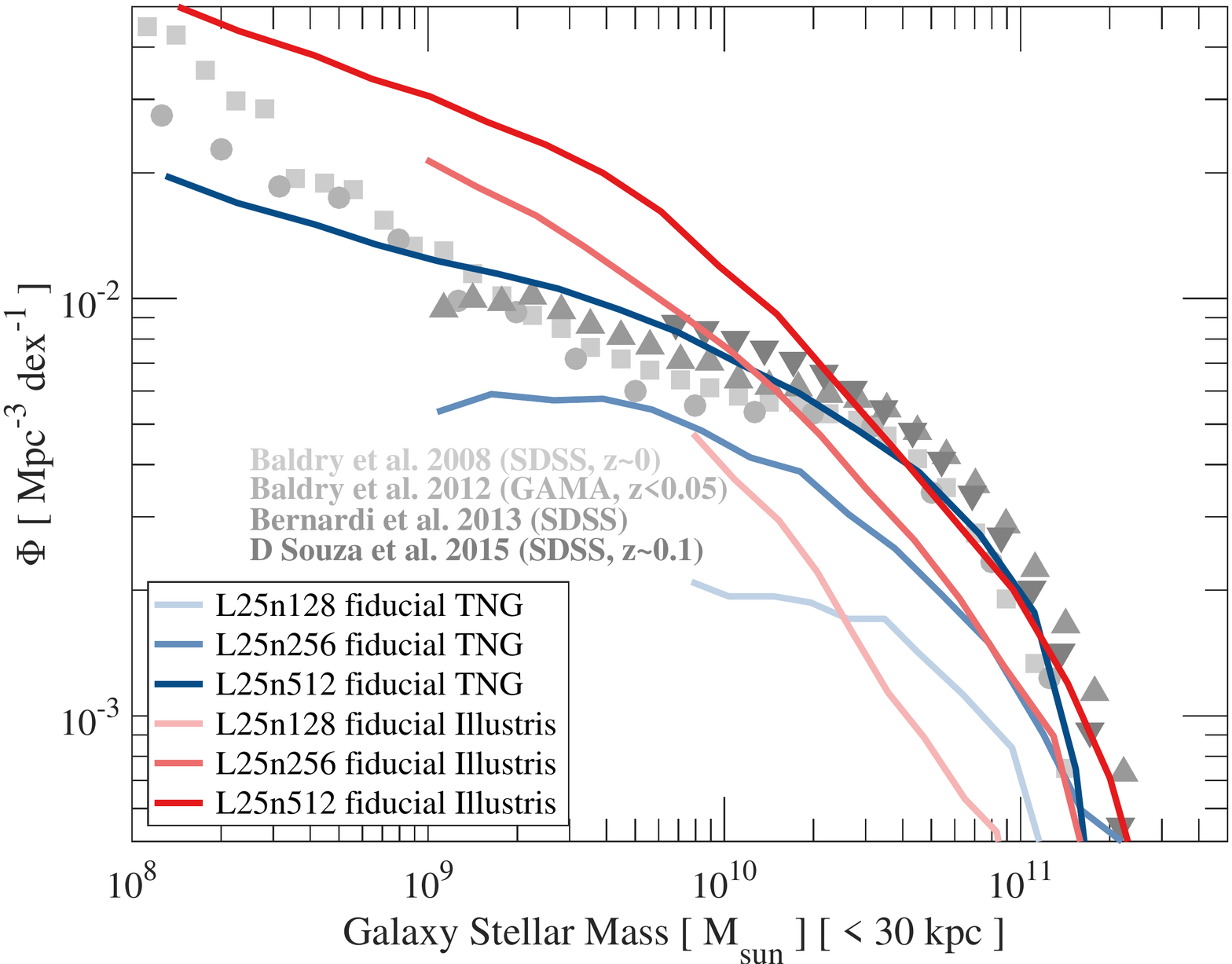}
\includegraphics[width=8.2cm]{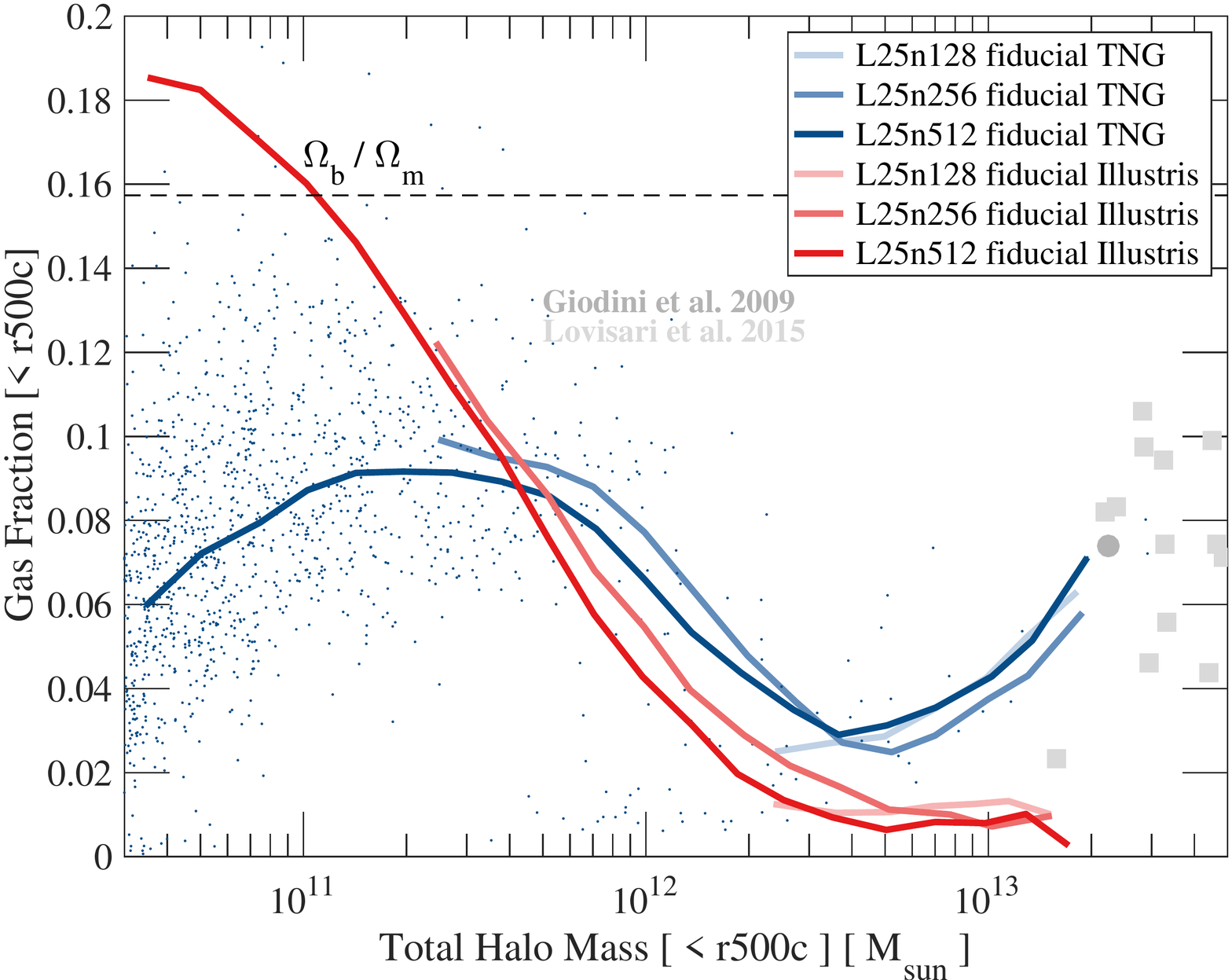}
\includegraphics[width=8.2cm]{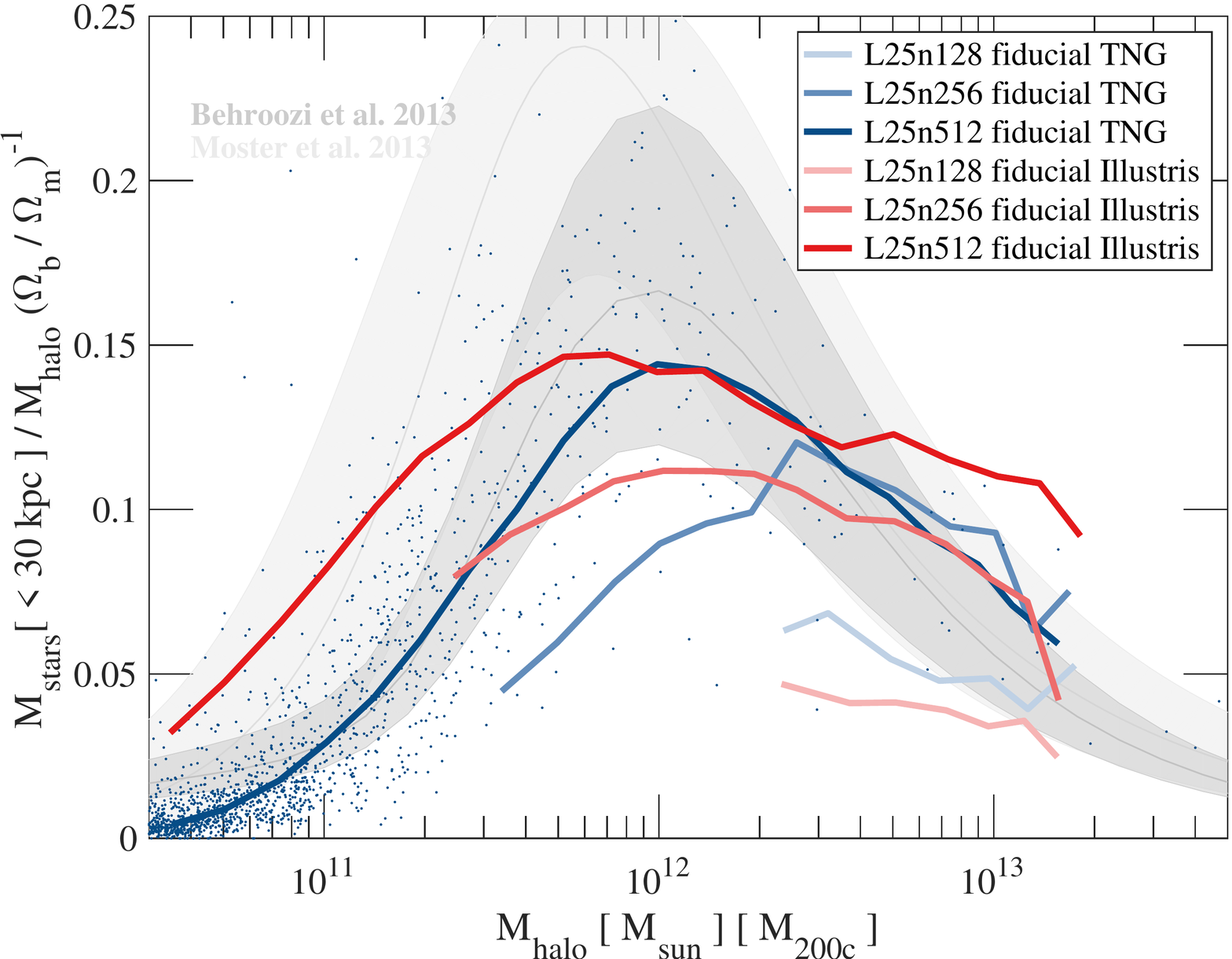}
\includegraphics[width=8.2cm]{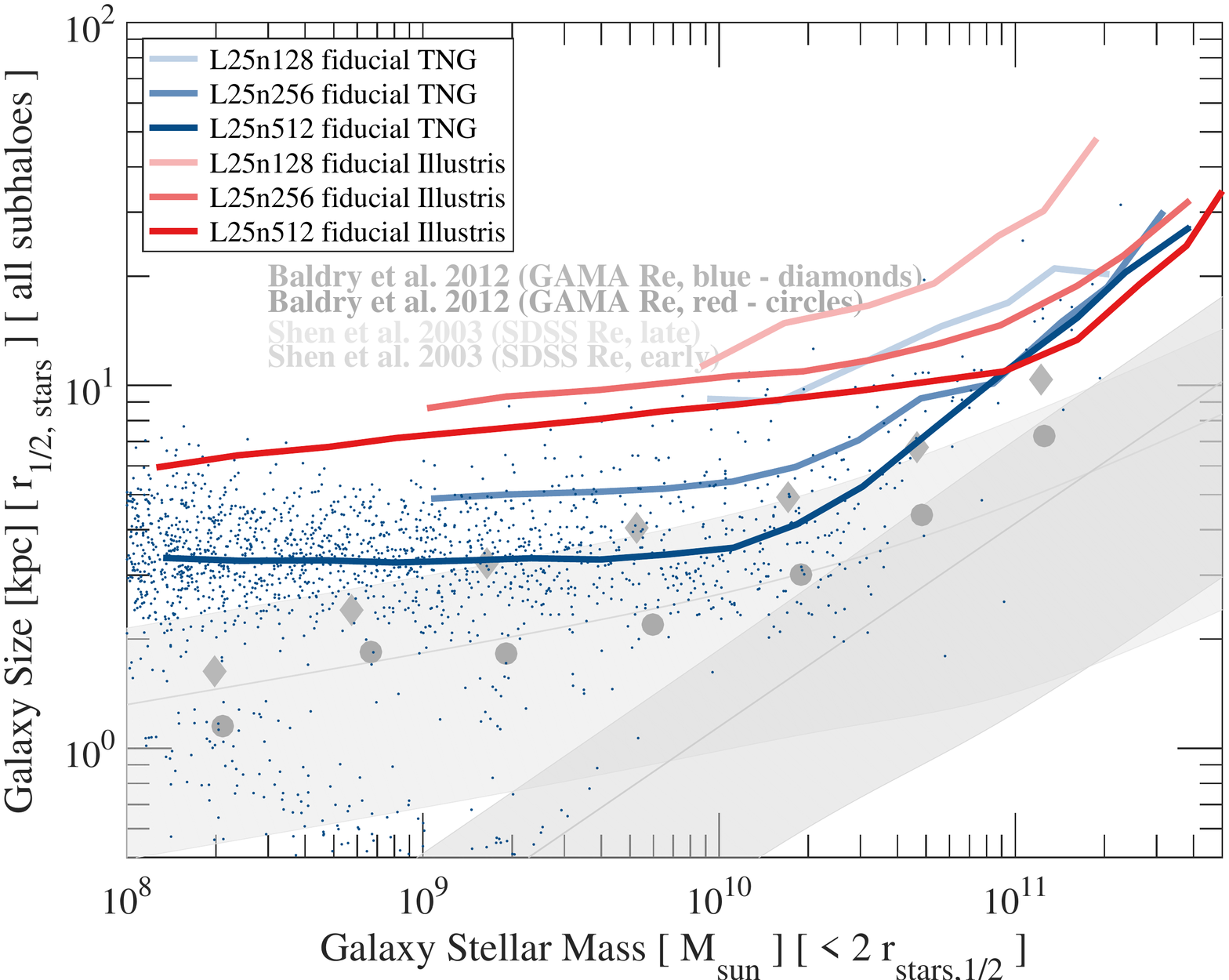}
\caption{Resolution dependence of the TNG and Illustris models in their fiducial implementations. The numerical characteristics of the L25n128, L25n256, and L25n512 boxes are listed in Table \ref{tab:res}. Solid curves denote running medians (but for the cosmic star-formation rate density as a function of cosmic times - top left panel). Individual galaxies are shown as data points only for the high-resolution run L25n512 in the TNG fiducial model. Gray shaded area represent observational data, for reference.}
\label{fig:galprop_res}
\end{figure*}

Throughout this paper we have exclusively shown results from our fiducial L25n512 volume (Illustris-1 equivalent resolution, i.e. within a factor of 2 in particle mass resolution), whereas here we consider the effects of changing resolution. To do so we run realizations with 8 and 64 times lower mass resolution, corresponding to factors of 2 and 4 larger gravitational softening lengths. The numerical parameters of this resolution series are given in Table \ref{tab:res}. The physical properties of the gas component, which in the simulation are Voronoi cells of various masses and sizes, are provided in Figure~\ref{fig:res_cells}. We emphasize that gas cell masses (i.e. the mass resolution of a simulation) and gas cell sizes (i.e. the spatial resolution of a simulation) both form continuous distributions, not necessarily well described by a single number. The gravitational softening of the gas cells is adaptive, cascading down to very small values in high density environments, which we truncate by imposing a floor at a minimum softening value denoted in the table as $\epsilon^{\rm min}_{\rm gas}$. The gas cell masses also form a distribution peaked about their mean value and allowed to vary by a factor of two in either direction.

Using this resolution series, in Figure~\ref{fig:galprop_res} we explore the convergence properties of our model (blue lines), giving also a direct comparison to the convergence behavior of the previous Illustris model (red lines). We see that higher resolution results in larger galaxy stellar masses at fixed halo mass -- the normalization of the stellar mass function increases correspondingly, as does the overall cosmic SFRD. Many of the other galaxy population statistics exhibit a similar qualitative trend, i.e. monotonically larger values with increasing numerical resolution. These include: BH masses for galaxy stellar masses $\lesssim 10^{10}\MSUN$, and galaxy stellar and gas metallicity at fixed galaxy stellar mass (not shown). On the other hand, the halo gas fraction as a function of halo mass is stable against resolution change at essentially all accessible halo masses. Galaxy sizes exhibit an inverse trend at fixed stellar mass: at progressively better spatial and mass resolution galaxies are less extended, and we can demonstrate (although we do not show here) that what sets galaxy sizes is not simply the imposition of a numerical floor due the smaller gravitational softening lengths chosen for the different matter species. Stellar masses at $z=0$ are consistent between L25n256 and L25n512 in haloes with total mass $\gtrsim 2\times 10^{12}\,\MSUN$, i.e. haloes with at least 50,000 resolution elements. The rate at which our method converges, and the regimes in which it is converged at a given resolution, depends on both the observable and on the mass scale.

The dominant effects of resolution can be understood by reference to the implementation of star formation and the treatment of star-forming gas in our model. We use an effective characterization of the inter-stellar medium whereby stars form stochastically on a given time scale ($t_{\rm sfr}$) from gas cells that exceed a given density threshold ($n_{\rm sfr}$, see \citealt{Springel:2003} and \citealt{Vogelsberger:2013b} for details). The SF parameters are fixed across resolutions, with $\rho_{\rm sfr} \simeq 0.1$ neutral hydrogen atoms per cubic cm, and $t_{\rm sfr} = 2.2$ Gyr. However, progressively better spatial resolution leads to the sampling of ever higher gas density regions, allowing more gas mass to become eligible for star formation and to be resolved at higher densities, accelerating the rate at which it is turned into stars. This is an inevitable result of cosmological collapse and galaxy formation. 

Since galactic winds are tied directly to the SFR, feedback increases correspondingly and naturally balances part of the resolution effects of star formation.
%, leading to a self-regulated mechanism \citep{Springel:2003}. This is however only the case at 
Residual resolution trends can however only be avoided at much better spatial and mass resolutions than the ones typically achievable in full cosmological volumes, and only in those cases when all the involved feedback mechanisms are numerially converged. In fact, the self-regulation between star formation and stellar feedback can be compromised by any resolution dependence of the feedback mechanism itself. The interaction of  galactic winds with ambient halo gas and the emergence of galactic fountain flows is a prime example of how the actual way in which a physical feedback mechanism functions, or not, may depend sensitively on numerical resolution.

Our approach is such that the galaxy formation physics choices and parameters are {\it not} changed as a function of mass and spatial resolution; details of our numerical choices related to resolution scalings were given in Section~\ref{sec:resolution_choices}. This approach could be described as ``imposing strong resolution convergence'' in the discussion of \cite{Schaye:2015}, and we prefer it to the alternative ``weak resolution convergence'' where parameters are intentionally re-scaled as a function of numerical resolution in an attempt to obtain less variable results. We prefer the first approach for the following reasons: 1) it allows an in principle arbitrary reduction of numerical truncation errors in the model predictions and thus a separation of the numerical and physical limitations of the modelling; 2) it simplifies the model tuning procedure; and 3) our approach has been shown to provide strongly converged results at sufficiently high resolution \citep[see, e.g.][]{Marinacci:2014, Grand:2017}. While the latter is not yet reached at the resolutions studied in this paper, we are confident that this can be achieved with the present incarnation of our methodology as well.
%\ap{We prefer the first approach not only because it simplifies the model {\it tuning} procedure, but also because we are confident that, although at the resolutions studied in this paper galaxy properties are not perfectly converged, they are converg{\it ing}, which will be demonstrated with future higher resolution runs.}.
%: with future higher resolution runs, we aim at determining whether the model with its fixed parameter values can be intended as truly predictive}.

%-------------------------------------------------------------------------------------------------------------------------------------------------------------

\section{Additional Details on the Galactic Wind Changes}
\label{sec_appendix2_winds}
\begin{figure*}
\centering
\includegraphics[width=8.4cm]{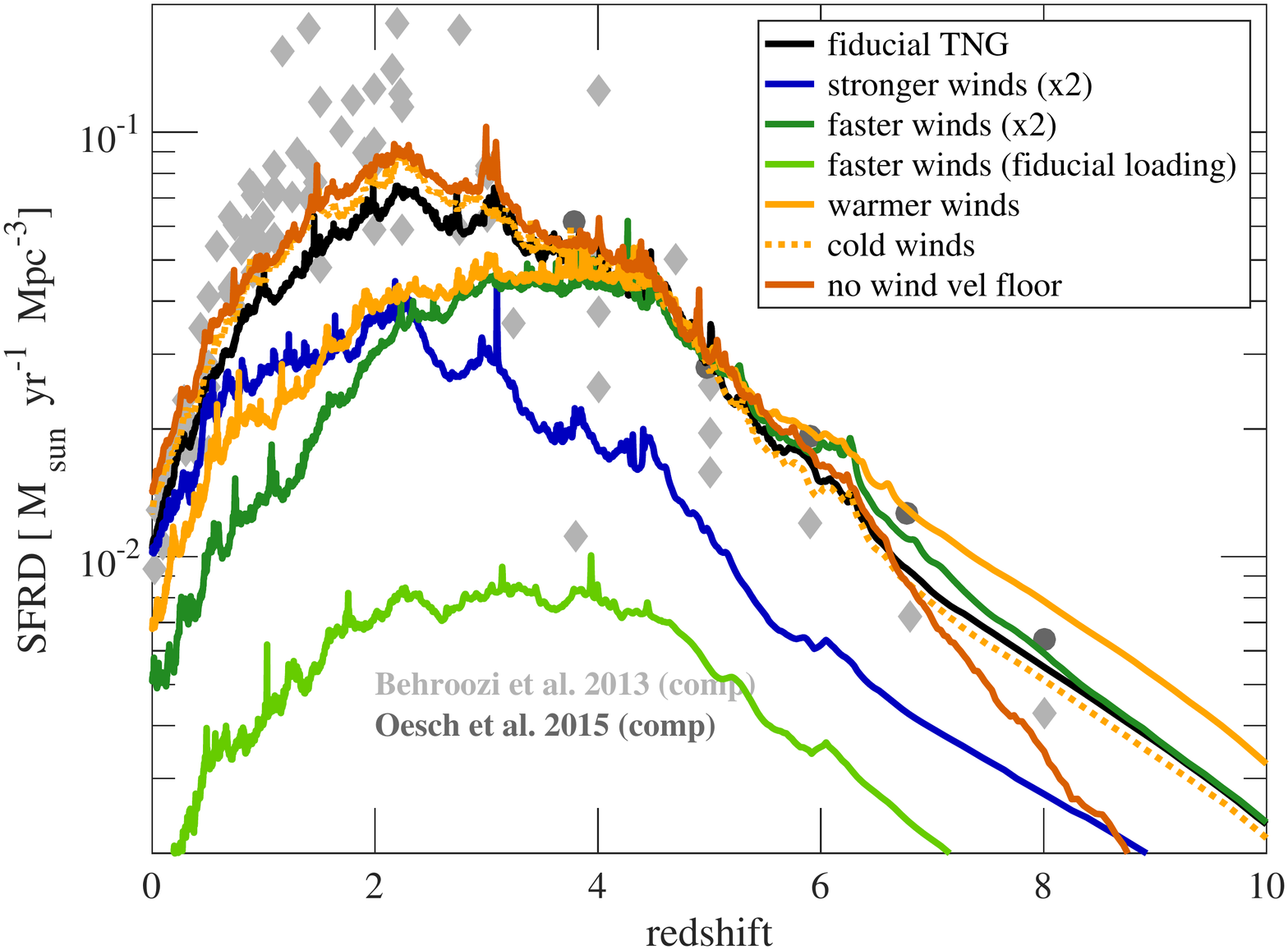}
\includegraphics[width=8.4cm]{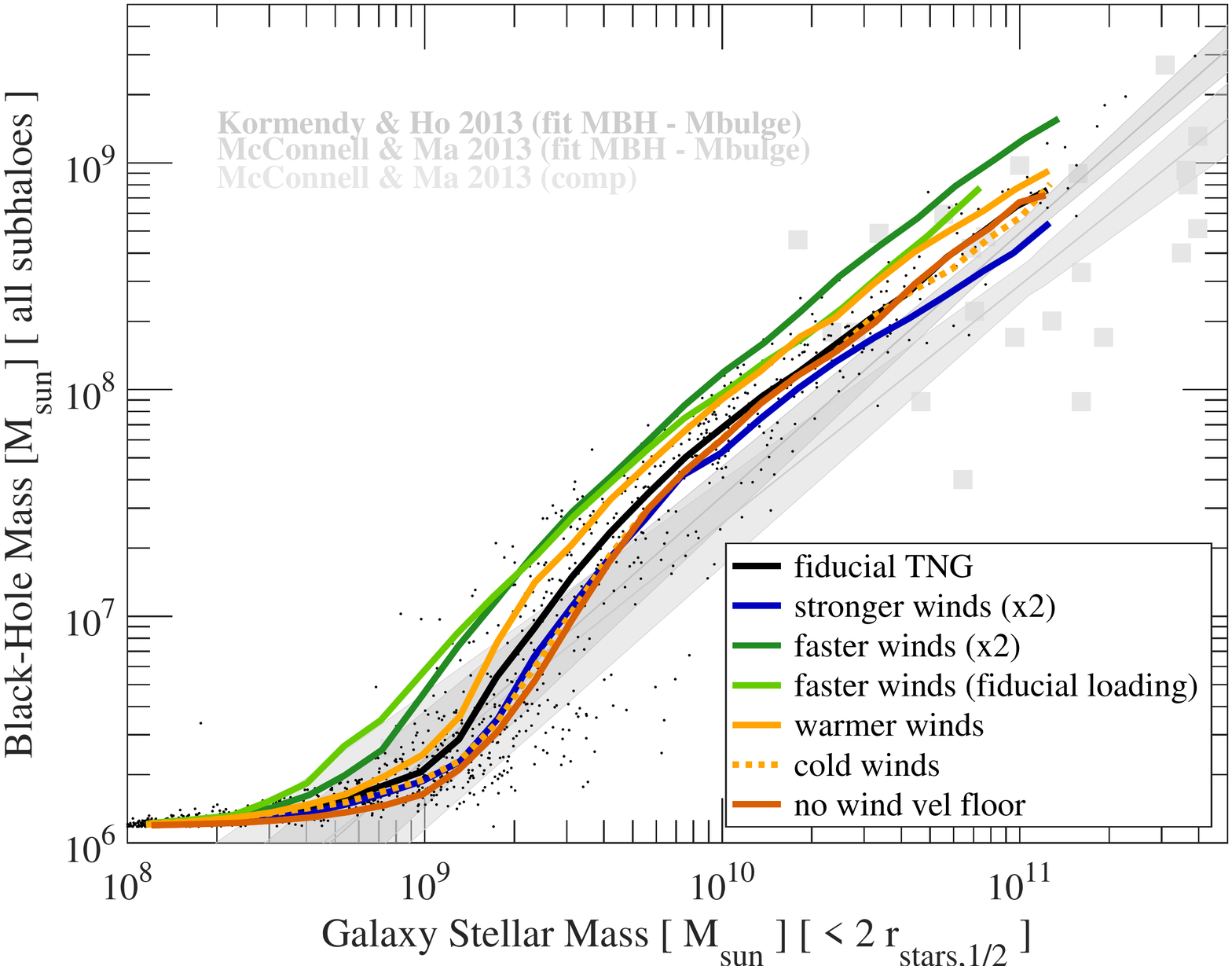}
\includegraphics[width=8.4cm]{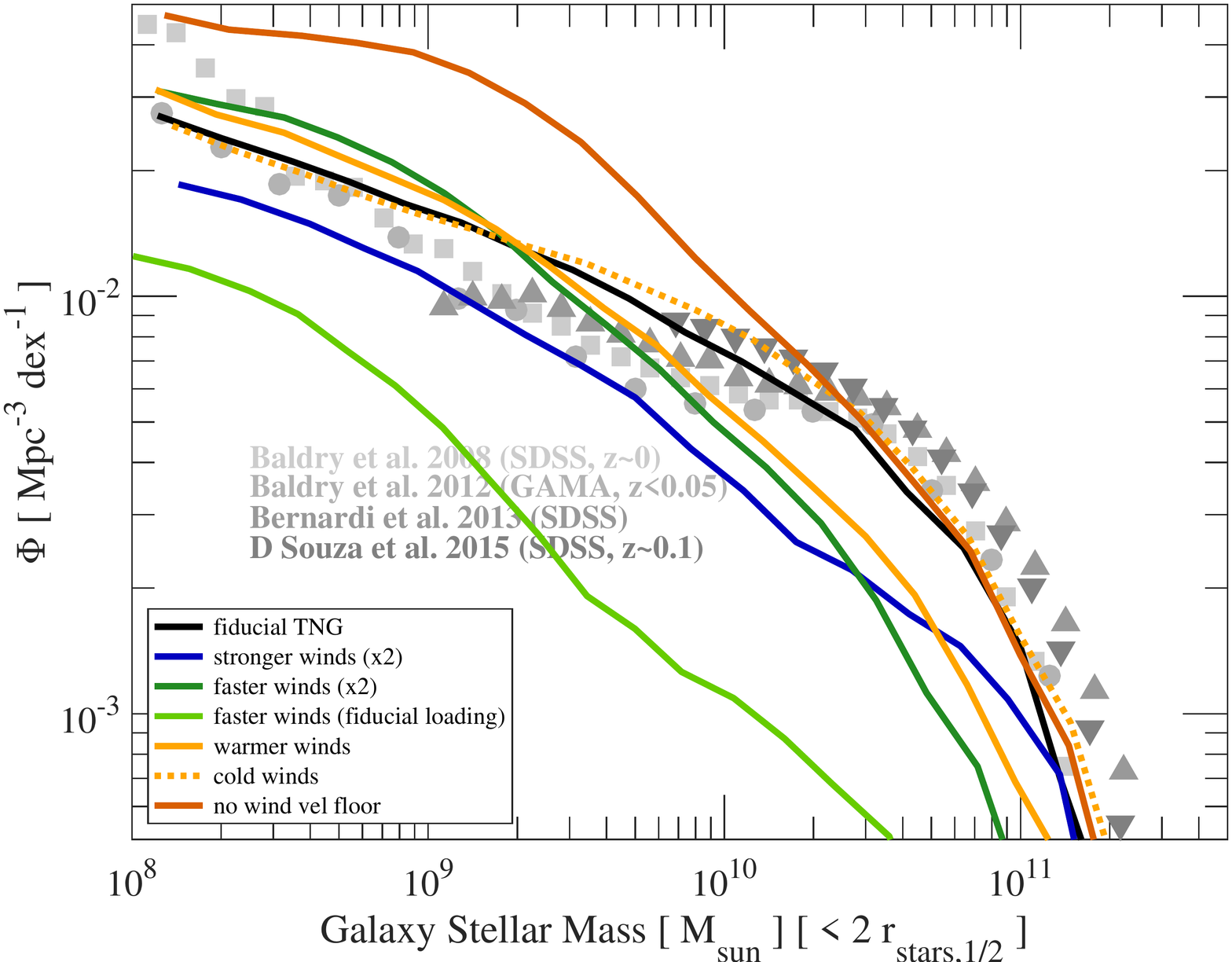}
\includegraphics[width=8.4cm]{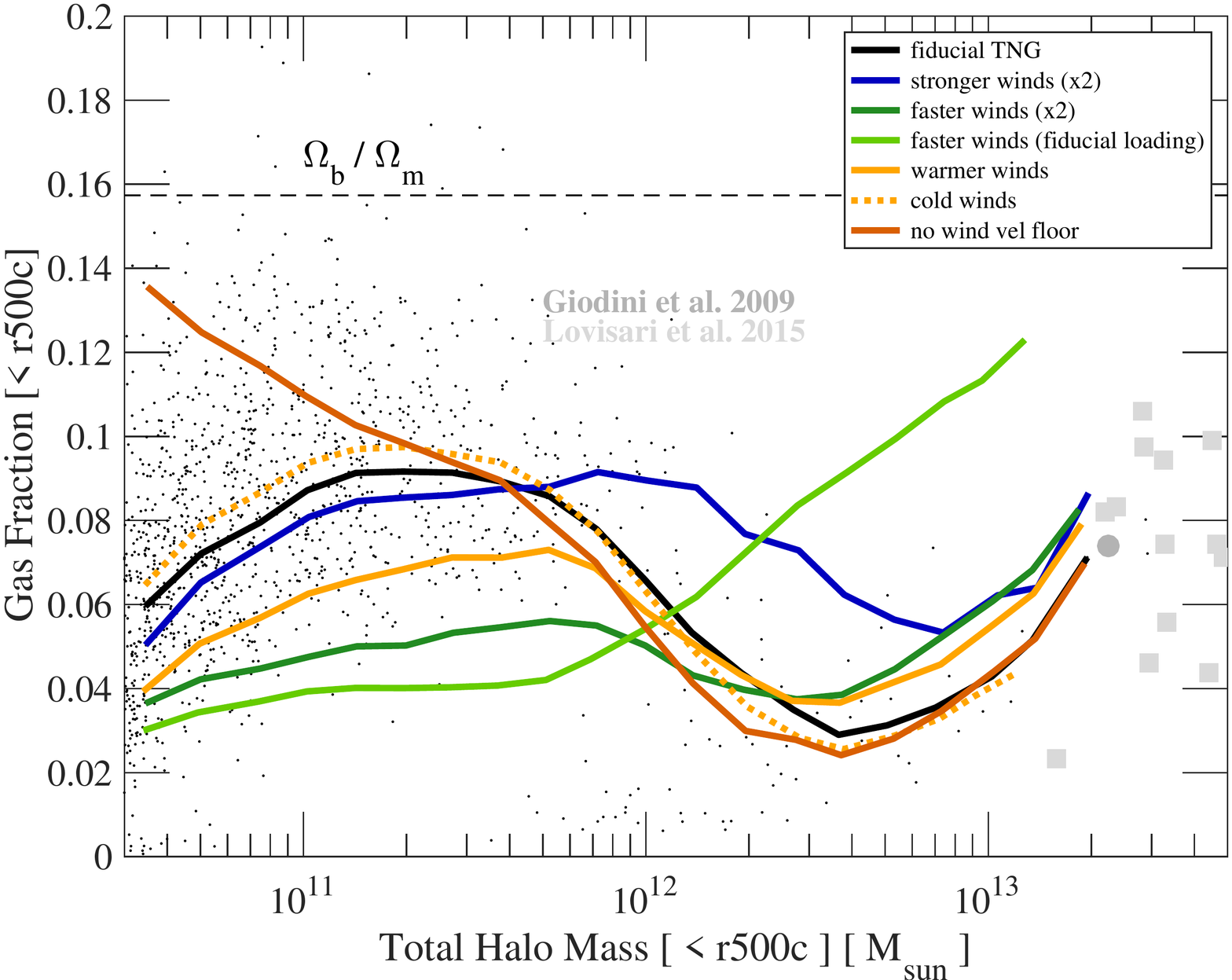}
\includegraphics[width=8.4cm]{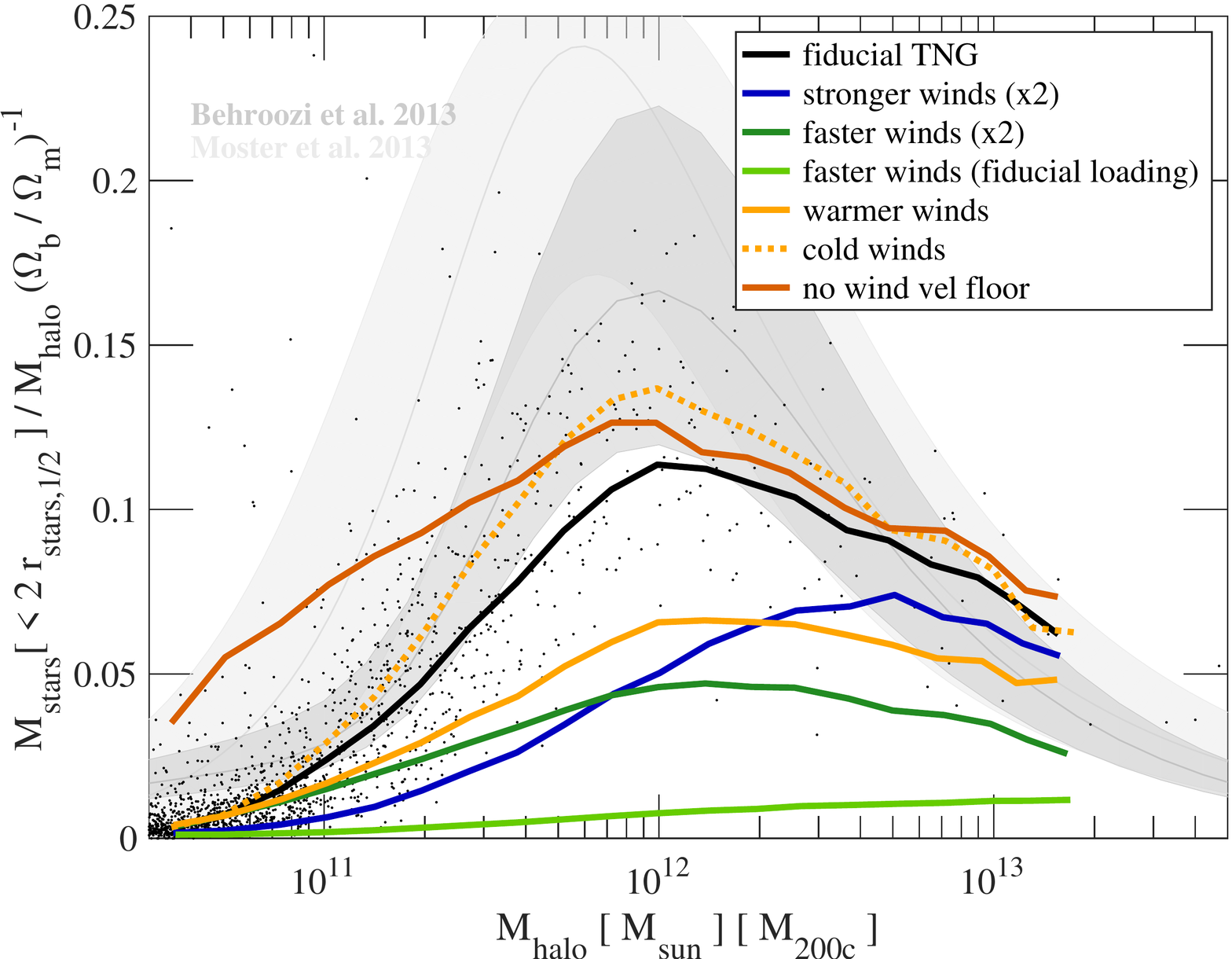}
\includegraphics[width=8.4cm]{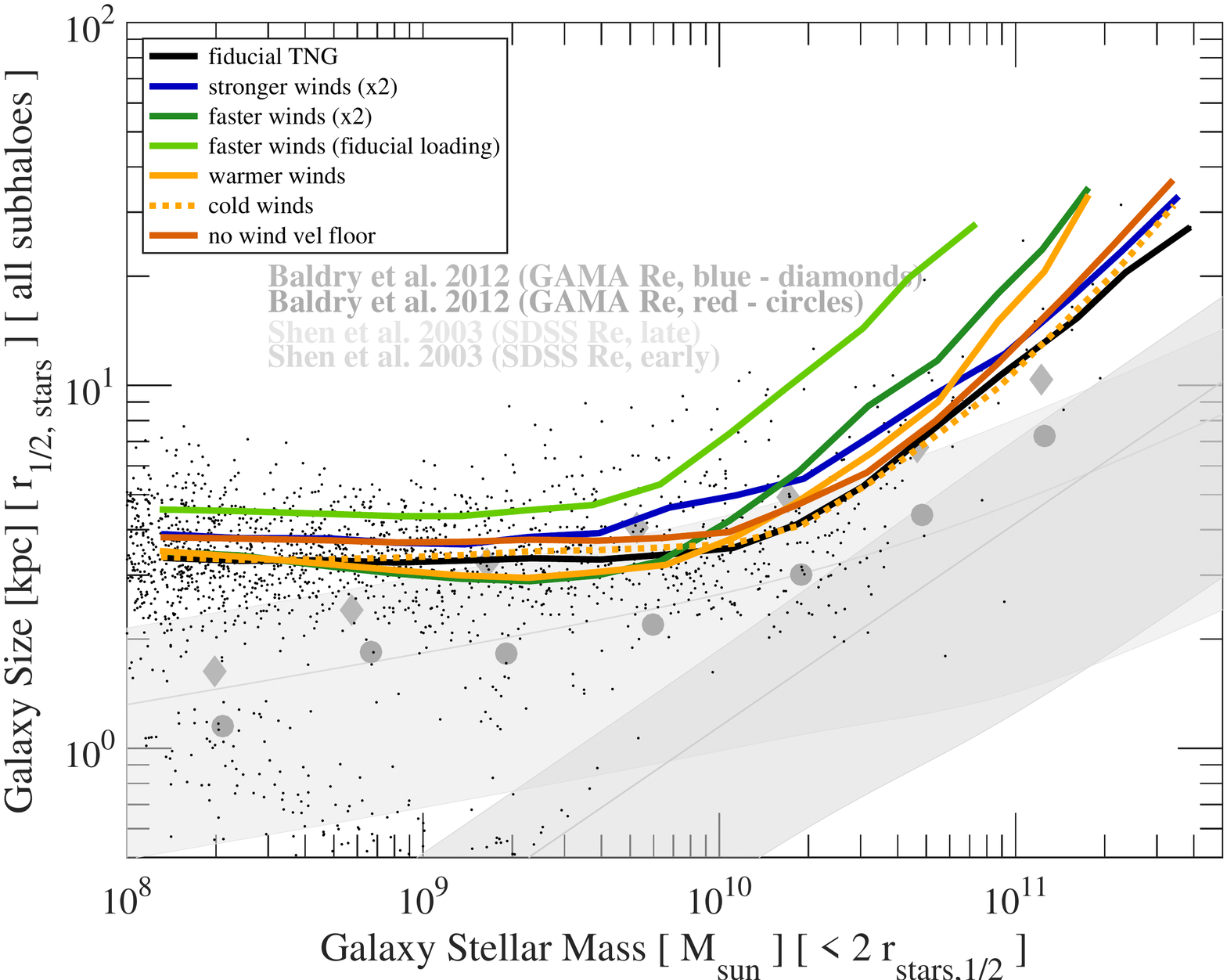}
\caption{The galaxy population at L25n512 resolution for different choices in the TNG wind implementation. Different curves show the outcome of the TNG model for alternative values of model parameters, according to Table \ref{tab:variations}. The ``fiducial TNG'' and the ``no wind vel floor'' curves are the same as in Figure~\ref{fig:galprop_winds_1}, while we also include the impact of either stronger or faster winds, as well as warm versus cold winds.}
\label{fig:galprop_winds_2}
\end{figure*}
\begin{table*}
\centering
\begin{tabular}{llcc}
\hline
&&&\\
Run Name & Parameter(s) of Interest & Fiducial Value & Variation Value \\
&&&\\
\hline
&&&\\

  fiducial TNG                          & all                      & see Table 1 (right column)                 & - \\
  fiducial Illustris                    & all                      & see Table 1 (left column)                  & - \\
  no BHs/feedback                       & all black hole-related   & see Table 1 (second section)               & - \\
  no galactic winds                     & all wind-related         & see Table 1 (third section)                & - \\
  no B fields                           & seed B field strength    & $1.6 \times 10^{-10}$ Gauss at $z=127$     & 0 Gauss \\
  &&& \\
  no $Z$-dependent wind energy            & \egyw, $f_{Z,w}$                                        & 3.6, 0.25   & 0.9, 0.0 \\
  no $z$-dependent wind velocity          & no $H(z)$ dependence in Eq. (\ref{eq:winds_vel}), $k_w$ & 7.4         & 3.7 \\
  no wind vel floor                     & wind velocity floor: $v_{w, \rm min}$                   & 350 ${\rm km\,s^{-1}}$    & 0 ${\rm km\,s^{-1}}$ \\
  no $Z/z$-dependent winds + no vel floor & all three previous modifications together               & (see above) & (see above) \\
  not isotropic winds                   & direction of wind particles at injection                & random      & along to $\vec{v} \times \nabla \Phi$ \\
  Illustris winds                       & many                                                    & see Table 1 & see Table 1 \\
  Illustris winds (SNII $>$ 6 Msun)     & many, also setting $M^{\rm SNII}_{\rm min} = 6\MSUN$    & see Table 1 & see Table 1 \\
  &&& \\
  stronger winds (x2)                       & wind energy factor: \egyw                               & 3.6         & 7.2 \\
  faster winds (x2)                         & wind velocity factor: $\kappa_w$                        & 7.4         & 14.8 \\
  faster winds (fiducial loading)           & wind energy and velocity factors: \egyw, $\kappa_w$                 & 3.6, 7.4    & 14.4, 14.8 \\
  warmer winds                          & thermal fraction: $\tau_w$                       & 0.1         & 0.5 \\
  cold winds                            & thermal fraction: $\tau_w$                       & 0.1         & 0.0 \\
  &&& \\
  $Z$-dependence: smaller $Z_{\rm ref}$        & reference metallicity: $Z_{w, Z}$      & 0.002       & 0.001 \\
  $Z$-dependence: larger $Z_{\rm ref}$         & reference metallicity: $Z_{w, Z}$      & 0.002       & 0.004 \\
  $Z$-dependence: shallower                    & reduction power: $\gamma_{w, Z}$       & 2.0         & 1.0 \\
  $Z$-dependence: steeper                      & reduction power: $\gamma_{w, Z}$       & 2.0         & 3.0 \\
  $Z$-dependence: weaker reduction at high $Z$   & reduction factor: $f_{Z,w}$            & 0.25        & 0.5 \\
  $Z$-dependence: stronger reduction at high $Z$ & reduction factor: $f_{Z,w}$            & 0.25        & 0.125 \\
  min wind vel = 300 ${\rm km\,s^{-1}}$                    & wind velocity floor: $v_{w, \rm min}$  & 350 ${\rm km\,s^{-1}}$    & 300 ${\rm km\,s^{-1}}$ \\
  min wind vel = 400 ${\rm km\,s^{-1}}$                    & wind velocity floor: $v_{w, \rm min}$  & 350 ${\rm km\,s^{-1}}$    & 400 ${\rm km\,s^{-1}}$ \\
&&&\\
\hline
\end{tabular}
\caption{\label{tab:variations}Summary of the different model variations shown in this work to explore the effectiveness of the TNG model. All runs have been performed on the volume L25n512 with the same initial conditions and at the same resolution. Unless otherwise stated, we have always changed one parameter/choice at a time with respect to the TNG fiducial implementation.}
\end{table*}

In the main text we have focused on the dependencies of the galaxy population properties with respect to the inclusion, or not, of various model components (`on'/`off' choices). Here we instead examine the dependence of the galaxy population properties on the specific values of the most important physical parameters. Each of the test runs is listed in Table \ref{tab:variations}, and in every case we evolve the usual L25n512 box to $z=0$.

\begin{figure*}
\centering
\includegraphics[width=8.4cm]{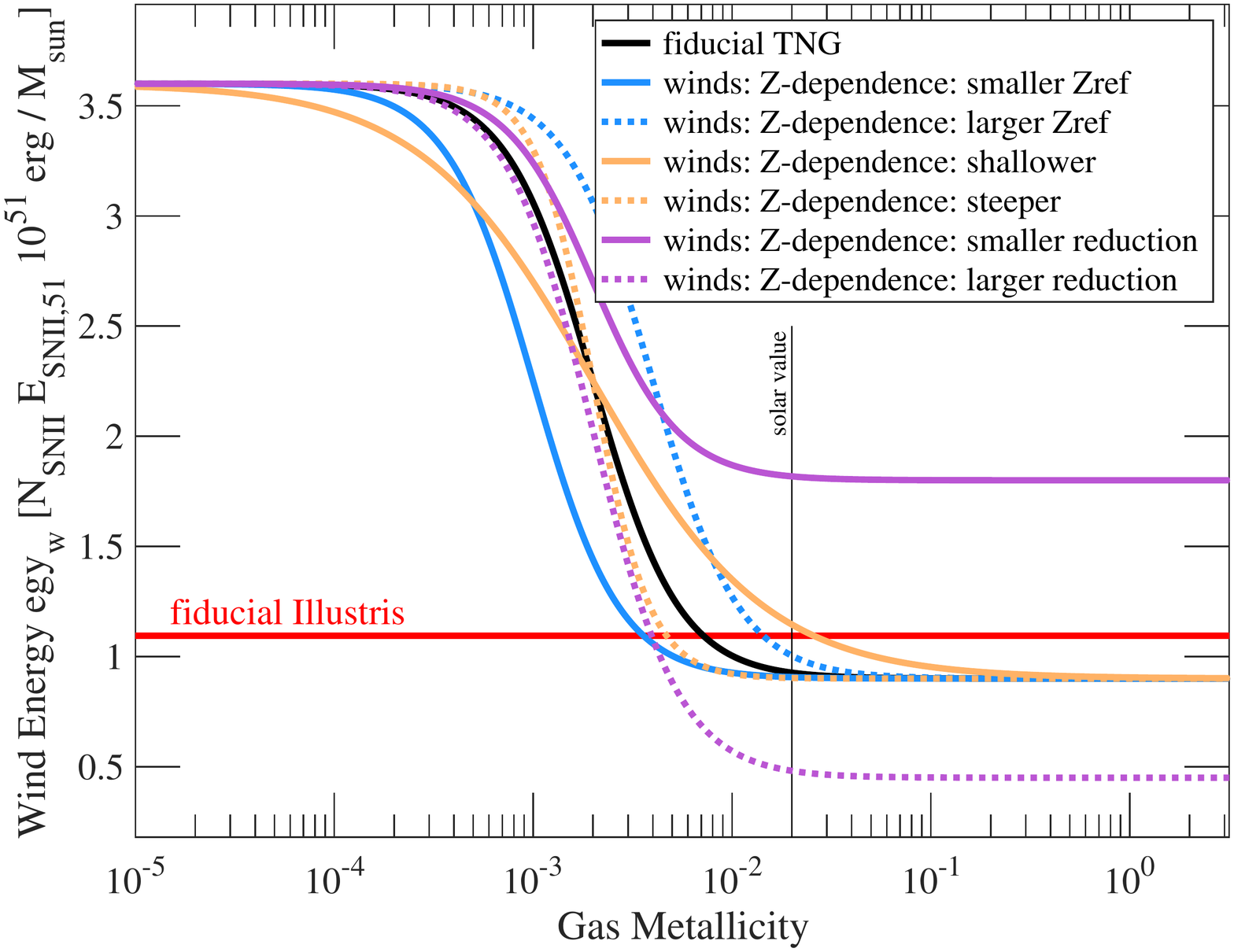}
\includegraphics[width=8.4cm]{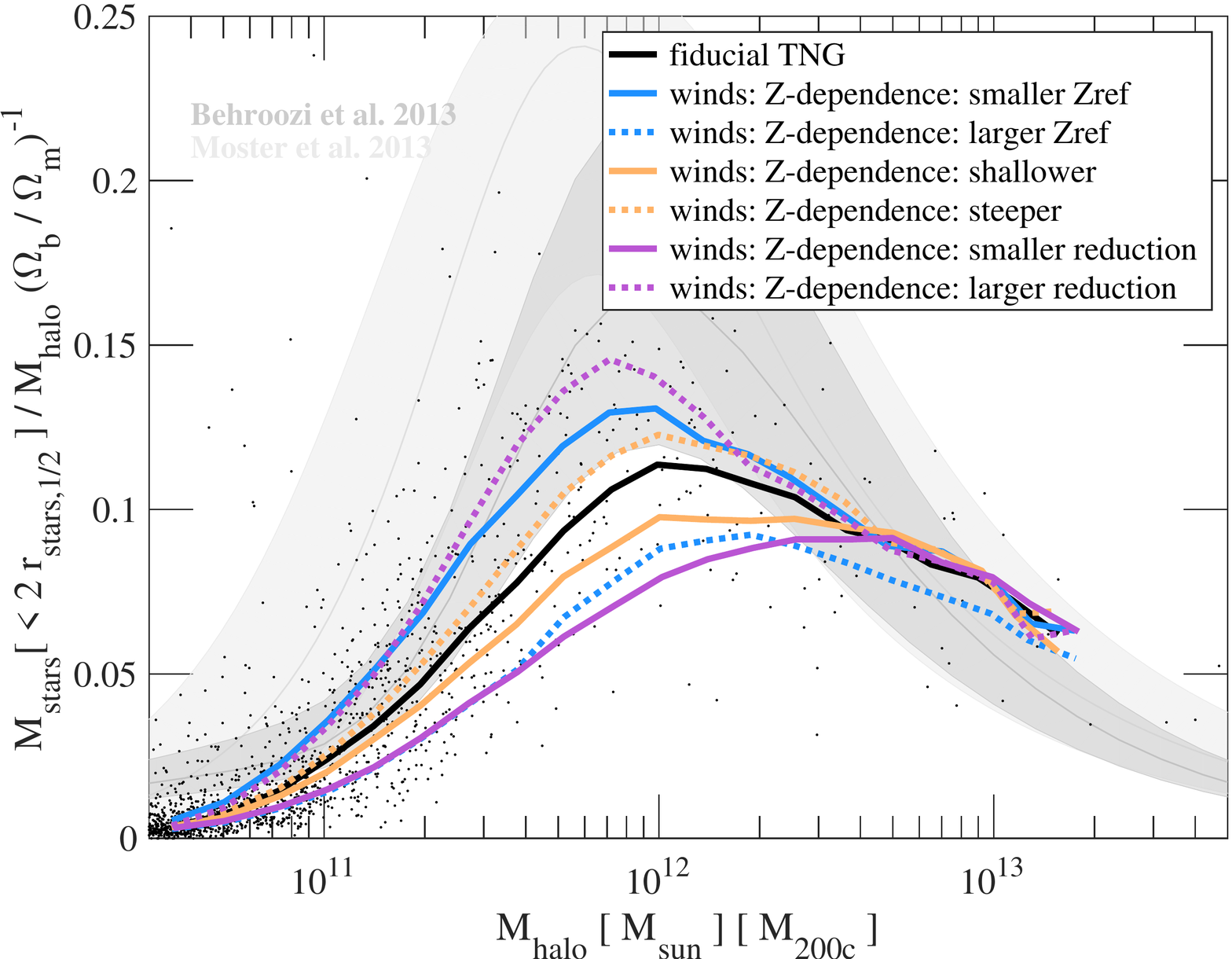}
\caption{The stellar mass to halo mass relation at L25n512 resolution for different choices in the TNG metallicity-dependent modulation of the wind energy. Left panel: the functional form of the metallicity dependence of the $e_w$ parameter of Eq.~\ref{eq:wind_energy}. Right panel: the $z=0$ stellar-to-halo mass relation. We contrast the fiducial model (black) with symmetric changes, to both higher and lower values, of the reference metallicity $Z_{w, Z}$ (blue lines), reduction power $\gamma_{w, Z}$ (orange lines), and high-$Z$ reduction factor $f_{Z,w}$ (purple lines).}
\label{fig:galprop_winds_Z}
\end{figure*}
\begin{figure}
\centering
\includegraphics[width=8.4cm]{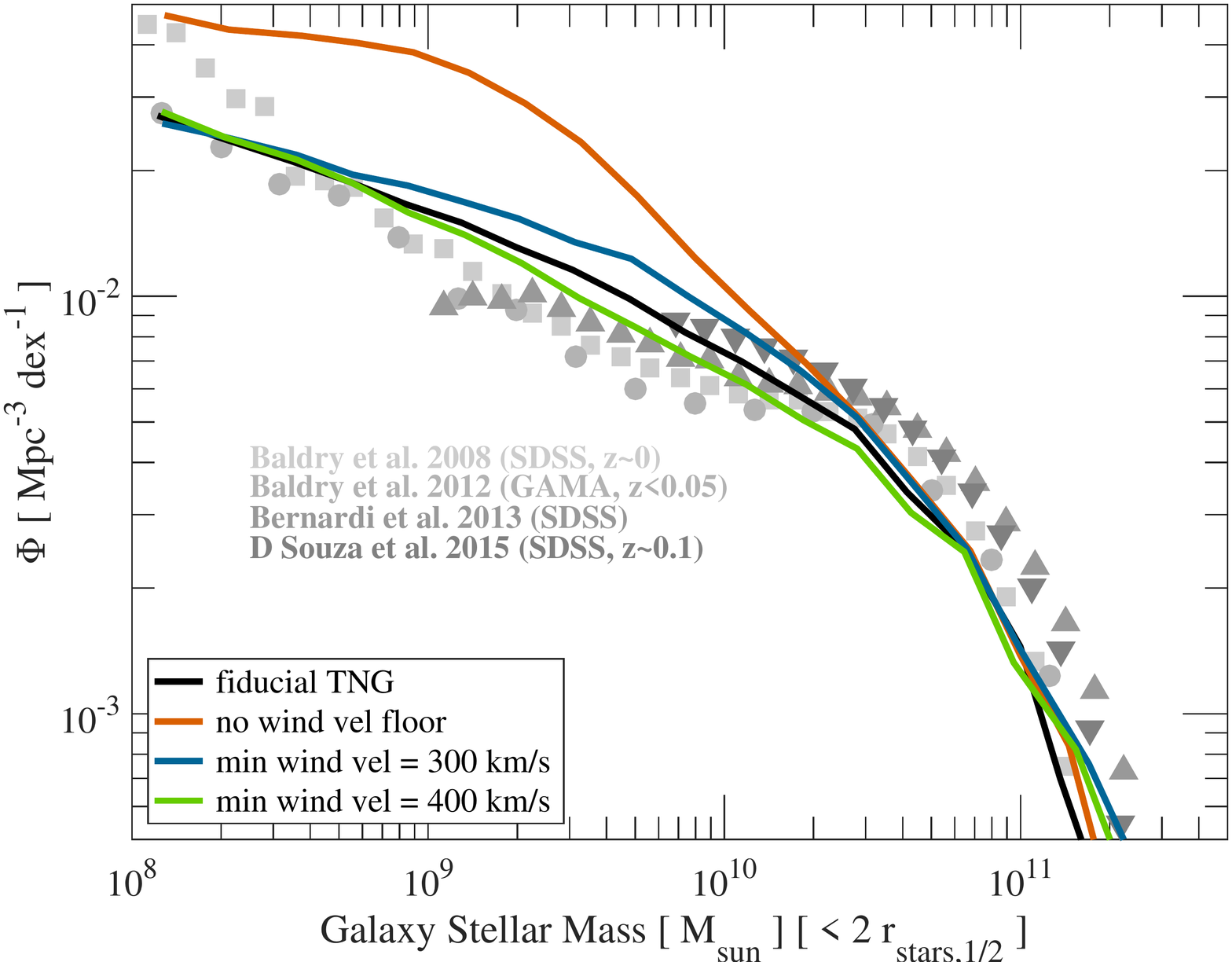}
\caption{Impact of the minimum galactic wind velocity on the stellar mass content of galaxies. Specifically, the $z=0$ stellar mass function (GSMF) is shown in comparison to the same observations as in previous figures. This minimum velocity lowers the normalization of the GSMF left of the knee.}
\label{fig:galprop_velmin}
\end{figure}
Figure~\ref{fig:galprop_winds_2} demonstrates an aspect of the model we have already alluded to: that by increasing either the wind energy (``stronger winds'') or wind velocity (``faster winds''), star formation is efficiently reduced across all masses and redshifts (all three left panels). In fact, injecting faster winds is more effective at preventing SF than injecting more wind mass. Because $\eta_w \propto 1/v_w^2$,  an increase by a factor of two in the velocity corresponds to a reduction by a factor of four in the mass loading (``faster winds'' of Figure~\ref{fig:galprop_winds_2}, green curves). Despite that, faster winds still produce less massive galaxies in haloes larger than $\sim 10^{12}\MSUN$ than winds with a factor of two larger mass loadings. In comparison, winds a factor of 2 faster than fiducial, but where the wind effective mass loading is kept to the fiducial value by raising the wind energy factor (``faster winds (fiducial loading)'' of Figure~\ref{fig:galprop_winds_2}, light green curves) return in all regimes much smaller stellar masses than the fiducial run and the faster and stronger wind cases.

With larger speeds at injection, gas outflows from galaxies can more easily exceed the escape velocity of their host halo. By eventually leaving the host potential well, gas recycling onto the galaxy is reduced, and the available gas reservoir for subsequent star formation is not replenished. This picture is consistent with the measured gas fractions within the halo (middle right panel), which are reduced in the ``faster winds'' case at all masses below $2 \times 10^{12}\MSUN$. Indeed, if we consider the same increase of the wind speed but hold the mass loading fixed, the gas fractions are also suppressed with respect to the fiducial case below $2 \times 10^{12}\MSUN$, and so are the stellar masses in all regimes (green light curves).

The amount of wind thermal energy also impacts the stellar mass content of galaxies: here we show the case of cold winds (as in Illustris, where wind particles have no thermal energy component) and warmer winds with the thermal energy fraction increased from 10 per cent to 50 per cent. The latter is the value adopted for an early generation of MW-zoom simulations \citep{Marinacci:2014} and for the Auriga galaxies \citep{Grand:2017}. From Eq. \ref{eq:winds_eta} it is apparent that giving a non-zero thermal energy to the wind reduces the effective wind mass loading, with the velocity remaining unchanged. Nevertheless, we see that with a 50 per cent wind split between thermal and kinetic, galaxies are substantially less massive at essentially all halo masses. As with faster winds, halo gas fractions are also reduced. Indeed, the impact of 50 per cent thermal `warm' winds is qualitatively similar to a factor of two `faster' winds, and based on the restricted observables we explore here, the two are likely interchangeable for an intermediate case between fiducial and faster. Conversely, entirely cold winds produce slightly more massive galaxies across halo mass together with an increase of gas fractions below $L^\star$. Physically, it may be that the effects of both warmer and faster winds resemble each other because they contribute similarly to an increased buoyancy, or effective pressure, of the outflow with respect to the gravitational potential and the ambient hot halo gas.

Galaxy sizes do not appear to be affected in any significant way by changes to the wind parameter values in galaxies below $10^{10}\MSUN$, even though from Figure~\ref{fig:galprop_misc} it is manifest that galaxy sizes in that same regime are fully due to the functioning of the galactic feedback as a whole. At the high mass end, the largest modifications to the galaxy sizes at fixed stellar mass are due to the underlying modifications to the galaxy stellar masses. 

The metallicity dependent energy modulation is crucial to improve the shape of the $z=0$ galaxy GSMF and SMHM relations (as shown in Figure~\ref{fig:galprop_winds_1}), primarily modifying the low-mass galaxy population, as intended. However, its exact implementation depends strongly on the adopted  parameters, particularly the effective reduction factor achieved at high metallicity. In Figure~\ref{fig:galprop_winds_Z} we explore the new metallicity dependence of the TNG winds further. The shape, location, and asymptotic normalization of the wind energy parameter \egyw as a result of the metallicity modulation are varied. We note that the two changes to the normalization or reduction factor (purple curves) are needed to assess the relative impact of the $Z$-dependence on Milky Way-mass versus lower mass galaxies, which is otherwise fixed. The detailed shape of the stellar-to-halo mass relation is sensitive to the metallicity dependence of the wind energy. The peak height, largely independent of the overall shape, varies weakly with the $Z$-transition shape (broader or narrower, corresponding to smaller or larger values for the power factor $\gamma_{w,Z}$ in Eq. \ref{eq:wind_energy}). The peak position shifts to lower (higher) $M_{\rm halo}$ as the reference metallicity $Z_{w, Z}$ decreases (increases). The overall reduction at high metallicity with respect to the low-$Z$ wind energy has a similar impact. At otherwise fixed fiducial model, the physical processes which we encapsulate in this metallicity dependence are fairly constrained by the abundance matching results.

Finally, Figure~\ref{fig:galprop_winds_2} demonstrates that the minimum wind velocity floor contributes to reduced gas fractions within haloes $\lesssim 10^{12}\MSUN$, one of the most striking features that distinguishes the TNG from the Illustris model. At very early times (high redshifts) the velocity floor also boosts the global SFRD arising from newly forming halos. Its primary impact, however, is the suppression of the $z=0$ stellar mass function at the low-mass end. Here the exact value of $v_{w, \rm min}$ plays a minor role: in Figure~\ref{fig:galprop_velmin} we explore the impact of the velocity floor value in the TNG wind model. Stellar masses below the knee are differentially suppressed, the impact being largest for the smallest galaxies. The exact threshold, 350 ${\rm km\,s^{-1}}$, leads to qualitatively the same behavior as a choice of either 300 or 400 ${\rm km\,s^{-1}}$, although the maximum stellar mass affected shifts correspondingly to lower or higher values. This behavior is driven by competition between the weak dependence of $v_w \propto M_{\rm halo}^{1/3}$ (Eq.~\ref{eq:wind_vel_scaling}) and the strong dependence of $M_\star-M_{\rm halo}$ at these low masses. Any $v_{w, \rm min} \neq 0$ which succeeds to modify stellar masses at all, then has a large impact across much of the GSMF: this effective parameter could be constrained by tight observational constraints on the low-end $z=0$ mass function combined with higher resolution simulations. 

\end{document}